\documentclass[a4paper,11pt]{article}
\pdfoutput=1
\usepackage[T1]{fontenc}
\usepackage[table,xcdraw]{xcolor}
\usepackage{jcappub}
\usepackage{mathtools}
\usepackage{comment}
\usepackage{float}
\usepackage{bm}
\usepackage[normalem]{ulem}
\usepackage{braket}
\usepackage{nameref}
\hypersetup{colorlinks=true}
\newcommand{\eq}[1]{Eq.~(\ref{#1})}
\newcommand{\eqs}[2]{Eqs.~(\ref{#1}) and (\ref{#2})}
\newcommand{\avrg}[1]{\left\langle #1 \right\rangle}
\newcommand{\bv}[1]{\mathbf{#1}} 
\newcommand{\hbv}[1]{\hat{\mathbf{#1}}} 
\newcommand{\h}[1]{\hat{#1}} 
\newcommand{\mcal}[1]{\mathcal{#1}} 
\newcommand{\mrm}[1]{\mathrm{#1}}

\newcommand{\hMpci}{h\mrm{Mpc}^{-1}}
\newcommand{\hiMpc}{h^{-1}\mrm{Mpc}}

\newcommand{\delK}{\delta^\mrm{K}}
\newcommand{\delD}{\delta^\mrm{D}}
\newcommand{\delg}{\delta_\mrm{g}}
\newcommand{\delL}{\delta_\mrm{L}} 
\newcommand{\PL}{P_\mrm{L}} 
\newcommand{\tdelg}{\tilde{\delta}_\mrm{g}}
\newcommand{\Sij}{S_{ij}}
\newcommand{\tSij}{\tilde{S}_{ij}}
\newcommand{\Skl}{S_{kl}}

\newcommand{\tP}{\tilde{P}}
\newcommand{\Tr}[1]{\mrm{Tr}\left[{#1}\right]}
\newcommand{\TF}[1]{\mrm{TF}\left[{#1}\right]}
\title{Redshift-space 3$\times$2pt statistics of galaxy clustering and intrinsic alignments at one-loop order} 
\author[a,b,1]{Toshiki Kurita,}
\affiliation[a]{Max-Planck-Institut f\"ur Astrophysik, Karl-Schwarzschild-Str. 1, 85748 Garching, Germany}
\affiliation[b]{Excellence Cluster ORIGINS, Boltzmannstr. 2, 85748 Garching, Germany}
\emailAdd{ktosh@mpa-garching.mpg.de} 

\author[c,d,e]{Atsushi Taruya}
\affiliation[c]{Center for Gravitational Physics and Quantum Information, Yukawa Institute for Theoretical Physics, Kyoto University, Kyoto 606-8502, Japan}
\affiliation[d]{Kavli Institute for the Physics and Mathematics of the Universe (WPI), The University of Tokyo Institutes for Advanced Study, The University of Tokyo, 5-1-5 Kashiwanoha, Kashiwa, Chiba 277-8583, Japan} 
\affiliation[e]{Korea Institute for Advanced Study, 85 Hoegiro, Dongdaemun-gu, Seoul 02455, Republic of Korea} 

\author[f, d]{and Teppei Okumura}
\affiliation[f]{Academia Sinica Institute of Astronomy and Astrophysics (ASIAA), No. 1, Section 4, Roosevelt Road, Taipei 106216, Taiwan} 

\abstract{ 
Intrinsic alignments of galaxies, long regarded as a systematic in weak lensing, are now recognized as a cosmological probe complementary to galaxy clustering. 
Exploiting them with spectroscopic surveys requires accurate models in redshift space beyond the linear regime. 
In this work, we derive the galaxy density--shape cross-power spectrum and shape auto-power spectra at one-loop order in redshift space. 
We consider two models, one based on the effective field theory (EFT) and the other on the Taruya--Nishimichi--Saito (TNS) prescription. 
They share the bias expansion but differ in how they treat small-scale velocities along the line of sight. 
The EFT model expands the effects of these velocities perturbatively and absorbs their sensitivity to small scales into counterterms. 
We show that, for the sky-projected shape field, this model requires six renormalized bias parameters and two counterterms. 
Up to order $k^2$, the stochastic terms require two parameters for the cross spectrum and four parameters for the auto spectra. 
The TNS-based model keeps the one-point velocity dispersion in an unexpanded damping factor and treats it as a single free parameter. 
We extend our previous linear-bias TNS-based model of the cross spectrum to nonlinear bias and the shape auto spectra. 
We find that all loop integrals reduce to those already used for shape spectra in real space and galaxy clustering in redshift space, and we evaluate them with FFTLog. 
We compare both models with power spectra measured from halos in $N$-body simulations and find good agreement on mildly nonlinear scales. 
These models provide a basis for joint cosmological analyses of galaxy clustering and intrinsic alignments with imaging and spectroscopic galaxy surveys, in which intrinsic alignments are utilized as a cosmological probe. 
}

\begin{document}
\maketitle
\clearpage
\section{Introduction} \label{sec:intro}
Spatial correlations between the unlensed shapes of galaxies, known as intrinsic alignments (IAs), have been studied extensively for their impact on cosmological inference from galaxy surveys \citep{Chisari2025:IA_review}. 
In weak lensing analyses, IAs are a major systematic that can bias cosmological parameter estimates if not properly accounted for \cite{Croft&Metzler2000:IA_dawn,Hirata&Seljak2004:IA_LA}
\citep[see][for reviews]{Joachimi+2015:IA_review,Kiessling+2015:IA_review,Kirk+2015:IA_review,Troxel&Ishak2015:IA_review,Lamman+2024:IAGuide,Chisari2025:IA_review}
This has motivated the development of theoretical models of IA across a wide range of scales, such as 
the linear alignment \citep{Hirata&Seljak2004:IA_LA} 
and nonlinear alignment \citep{Bridle&King2007:IA_NLA} models, quadratic alignment models based on tidal torquing \citep{Crittenden+2002:TidalTorqueAlignment,Mackey+2002:TidalTorqueAlignment}, 
the tidal alignment and tidal torquing model \citep{Blazek+2015:IA_TATT,Blazek+2019:IA_TATT}, 
and halo model approaches \cite{Schneider&Bridle2010:IA_HaloModel,Fortuna+2021:IA_HaloModel,VanAlfen+2024:Halo_model}. 
Observational studies have detected or constrained IA signals in a variety of galaxy samples and surveys \citep{Brown+2002:IA_SuperCOSMOS,Mandelbaum+2006:IA_measurement,Hirata+2007:IA_measurement,Okumura+2009:IA_measurement,Okumura&Jing2009:IA_measurement_letter,Joachimi+2011:AIA_C1,Mandelbaum+2011:WiggleZ_blueIA,Singh+2015:IA_measurement,Johnston+2019:IA_measurement_GAMA,Georgiou+2019:IA_measurement_GAMA+KiDS,Fortuna+2021:IA_measurement_KiDS-1000,Tonegawa&Okumura2021:IA_measurement,Samuroff+2022:IA_measurement_DES&eBOSS,HervasPeters+2024:UNIONS,Georgiou+2025:IA_measurement_KiDS-1000_Bright,Tonegawa+2025:IA_measurement_HSC_OII,Okumura2025:IA_auto_evidence,Navarro-Girones+2026:IA_measurement_PAU,Siegel+2026:IA_measurement_demographics}. 
Cosmological hydrodynamical simulations have also been used to investigate the dependence of IA on galaxy properties and to test theoretical models \citep{Tenneti+2015a:IA_hydro,Chisari+2015:IA_hydro_Horizon-AGN,Velliscig+2015:IA_hydro,Hilbert+2017:IA_hydro,Shi+2021a:IA_hydro,Samuroff+2021:IA_hydro,Delgado+2023:IA_hydro_MTNG,Ferlito+2025:IA_hydro_MTNG_lightcone,Herle+2026:IA_hydro_FLAMINGO}.

More recently, IAs have increasingly been studied as a cosmological probe in their own right. 
On large scales, galaxy shapes respond to the tidal field and can therefore be described as tensor-type biased tracers of large-scale structure \citep{Vlah+2020:IA_EFT}. 
IA correlations can probe signatures imprinted in the initial conditions of structure formation, such as 
anisotropic primordial non-Gaussianity (PNG) \citep{Schmidt+2015:IA_PNG,Kogai+2018:IA_PNG,Akitsu+2021:IA_PNG,Kogai+2021:PNG_spin_detector,Kurita&Takada2023:AniPNGfromIA}, 
parity-violating PNG \citep{Kurita+2026:PVinGS_PNG}, 
primordial gravitational waves and their chirality \citep{Schmidt&Jeong2012:LSSwithGW_2,Schmidt+2014:LSSwithGW_3,Biagetti&Orlando2020:PVGW_IA,Akitsu+2022:SU_sim_GW,Philcox+2024:NewPhys_GalaxyShape,Okumura&Sasaki2024:PVGW_PrjTensor,Mikura+2025:GWPol_PrjTidal}, 
primordial magnetic fields \citep{Saga+2024:IA_with_PMF}, 
and statistical anisotropy \citep{Shiraishi+2023:IA_Stat_Ani,Minato+2026:IA_Dip,Mikura+2026:IA_Dip_WA}. 

At late times, IAs can also probe cosmic expansion and the growth of structure in the late Universe \cite{Chisari&Dvorkin2013:IA_BAO,Taruya&Okumura2020:IA_improvement}. 
The baryon acoustic oscillations (BAO) feature in IAs can be used to measure distances \citep{Chisari&Dvorkin2013:IA_BAO,Okumura+2020:IA_nbody,Kurita+2021:IA_nbody,vanDompseler+2023:IA_BAO_toy_model,Xu+2023IA_BAO,Chan+2026:IA_BAO_GI}. 
The Alcock--Paczynski (AP) effect \citep{Alcock&Paczynski1979,Matsubara&Suto1996:AP} and redshift-space distortions (RSD) provide further information on distances and structure growth \citep{Okumura&Taruya2020:IA_improvement,Zwetsloot&Chisari2022:IA_growth,Inoue+2025:IA_multipoles,Shim+2025:IA_DDE_Letter,Shim+2025:IA_DDE_full}. 
These signatures have been studied in simulations \citep{Okumura+2020:IA_nbody,Kurita+2021:IA_nbody,Okumura+2019:IA_density&velocity,Ishikawa+2025:SHAPE}, 
and analyses of actual data have yielded observational constraints \citep{Xu+2023IA_BAO,Okumura&Taruya2023:IA_constraint_fs8,Okumura2026:IA_constraints_DESI,Chan+2026:IA_BAO_GI}. 
Combining IAs with velocity measurements from the kinetic Sunyaev--Zel'dovich (kSZ) effect can further improve cosmological constraints \citep{Okumura+2019:IA_density&velocity,vanGemeren&Chisari2020:IA_velocity,Okumura&Taruya2022:IA_improvement}. 
The cosmological information from IAs therefore complements that from galaxy clustering and other probes of the late Universe. 

Exploiting IAs as a cosmological probe requires accurate theoretical modeling. 
To include information from mildly nonlinear scales, these models must go beyond the linear alignment model and account for nonlinear gravitational evolution and nonlinear bias. 
In real space, perturbative descriptions of galaxy and dark matter halo shapes have been developed to incorporate these effects \citep{Blazek+2015:IA_TATT,Schmitz+2018:IA_TATT,Blazek+2019:IA_TATT}. 
The effective field theory (EFT) of IAs provides a systematic framework for this modeling, with predictions for one-loop power spectra and tree-level bispectra \citep{Vlah+2020:IA_EFT,Vlah+2021:IA_EFT,Chen&Kokron2024:LEFTofIA,Bakx+2025:IA_Bistectrum_I}. 
Hybrid and field-level approaches have also been developed \citep{Maion+2024:HYMALAIA,Kokron+2026:IA_field_level}. 
$N$-body and hydrodynamical simulations have also been used to test perturbative IA models and measure higher-order shape bias \citep{Samuroff+2021:IA_hydro,Bakx+2023:EFTofIAvsSims,Akitsu+2023:QuadShapeBias,Bakx+2025:IA_Bistectrum_II,Herle+2026:IA_bias_FLAMINGO,Vedder+2026:IA_3P_FLAMINGO}. 
For spectroscopic samples, peculiar velocities affect the galaxy positions inferred from redshifts and thus the three-dimensional IA statistics. 
Hence, analyses on mildly nonlinear scales require an accurate treatment of nonlinear RSD. 
For this purpose, models incorporating velocity damping \citep{Okumura+2024:NLRSDofIA} and perturbative treatments of redshift-space IA correlations \citep{Matsubara2024:IA_iPT_I,Matsubara2024:IA_iPT_II,Taruya+2024:TNSofIA_1} have been developed. 
A model-independent framework for the symmetries of IA statistics in redshift space has also been developed \citep{Akitsu+2026:IAinRS_I}. 

In this work, we derive the one-loop power spectra of galaxy density and projected galaxy shapes in redshift space. 
We start from the real-space bias expansion of Ref.~\cite{Vlah+2020:IA_EFT} and compute the terms generated by redshift-space mapping of the shape field. 
We use this calculation to develop two models. 
The two models share the bias expansion, the one-loop terms, and the stochastic terms, and differ mainly in the treatment of small-scale velocities along the line of sight (LOS), which produce the Fingers-of-God (FoG) effect. 
In the EFT model, we extend the redshift-space treatment of matter and galaxy clustering \citep{Senatore&Zaldarriaga2014:EFTofLSS_RSD,Perko+2016:EFTofLSS_RSD_biased,Lewandowski+2018:EFT_RSD,Desjacques+2018:EF_RSD_Pk_Bk} to the galaxy shape field. 
We renormalize the bias parameters and include counterterms and stochastic contributions to absorb the sensitivity of the loop integrals to small-scale physics. 
We also implement infrared (IR) resummation to account for the damping of the BAO feature by large-scale bulk flows \citep{Senatore&Zaldarriaga2015:IRresum_EFT,Baldauf+2015:IRresum_EP_BAO,Ivanov&Sibiryakov2018:IRresum_RSD}. 
Our second model follows the Taruya--Nishimichi--Saito (TNS) prescription \citep{Taruya+2010:TNS}. 
Instead of expanding the effect of small-scale random velocities, it keeps the one-point velocity dispersion in an unexpanded FoG damping factor with a single free parameter. 
This approach was applied to the density--shape cross-spectrum in Ref.~\cite{Taruya+2024:TNSofIA_1}, assuming linear density and shape bias. 
Here, we include nonlinear bias for both fields and extend the calculation to the $E$- and $B$-mode shape auto-power spectra. 
We compare both models with measurements from $N$-body simulations and find good agreement on mildly nonlinear scales. 

This paper is organized as follows: 
In Section~\ref{sec:ps_1loop_real}, we review the real-space bias expansions for the galaxy density and shape fields and introduce our notation. 
We define the $E$ and $B$ modes of the projected shape field and summarize the real-space EFT power spectra at one-loop order. 
In Section~\ref{sec:ps_1loop_rsd}, we expand the redshift-space mapping and use Eulerian perturbation theory to compute the loop terms. 
We then construct the EFT and TNS-based models from these loop terms. 
In Section~\ref{sec:nbody_sims}, we present the comparison with $N$-body
simulations. 
We conclude in Section~\ref{sec:conclusions}. 
In the Appendices, we give the detailed calculations behind the results in the main text. 
We summarize the real-space calculations, including the helicity decomposition and the degeneracies among the loop integrals in Appendix~\ref{app:details_real}. 
In Appendix~\ref{app:details_rsd}, we give the kernels and loop integrals in redshift space by using some useful azimuthal-integral formulae derived in Appendix~\ref{app:azm_avrg}. 
In Appendix~\ref{app:relation_TNS}, we show how the power spectra of the TNS-based model are related to those in Eulerian perturbation theory. 
In Appendix~\ref{app:summary_table}, we summarize the EFT model as tables of the coefficients of the basis functions.

\section{Recap: Real-space power spectra at one-loop order} \label{sec:ps_1loop_real}
\subsection{Definition of the shape field}
\label{subsec:shape_field}
We describe the three-dimensional shape of galaxy $\mrm{g}$ by a symmetric second-moment tensor $I_{ij,\mrm{g}}$. 
For a galaxy sample, we construct the tensor field 
\begin{align}
    I_{ij}(\bv{x})
    \equiv
    \frac{1}{\bar n_{\mrm{g}}}
    \sum_\mrm{g} I_{ij,\mrm{g}}\delD(\bv{x}-\bv{x}_\mrm{g}),
    \label{eq:def_inertia_tensor_field}
\end{align}
where $\bv{x}_\mrm{g}$ is the position of galaxy $\mrm{g}$ and $\bar n_{\mrm{g}}$ is the mean number density of the sample. 
For a tensor $X_{ij}$, we define its trace as $\Tr{X} = X_{ii}$
and its trace-free part as $\TF{X_{ij}} = X_{ij} - \Tr{X}\delK_{ij}/3$. 
We define the dimensionless, trace-free shape fluctuation as
\begin{align}
    \Sij(\bv{x})
    \equiv
    \frac{\TF{I_{ij}(\bv{x})}}{\avrg{\Tr{I}}}. 
    \label{eq:def_shape_field}
\end{align}
This field is symmetric and trace free, with $\avrg{\Sij}=0$ since $\avrg{I_{ij}}=\delK_{ij}\avrg{\Tr{I}}/3$ under statistical isotropy. 
We will focus only on the trace-free part in this work. 
Since the shapes are sampled at galaxy positions, $\Sij$ is density weighted:
\begin{align}
    \Sij=(1+\delg)S_{ij}^\mrm{unw},
    \label{eq:shape_density_weighting}
\end{align}
where $\delg$ is the galaxy density field and $S_{ij}^\mrm{unw}$ is the underlying unweighted shape field with the same normalization. 
Throughout this paper, $S_{ij}$ denotes this density-weighted shape field.\footnote{The same bias expansion given in Section~\ref{subsec:bias_expansion} also applies to the unweighted field, with different bias coefficients. \citep{Vlah+2020:IA_EFT}}

\subsection{Bias expansion of the galaxy shape field} 
\label{subsec:bias_expansion} 
At leading order in spatial derivatives, a complete set of bias operators for the shape field up to third order in real space \citep{Vlah+2020:IA_EFT} is given by 
\begin{alignat}{2}
    &\mrm{1st:} &\quad&\TF{\Pi_{ij}^{[1]}}, \label{eq:1st_bias_op}\\
    &\mrm{2nd:} &\quad&\TF{\Pi_{ij}^{[2]}}, \TF{\Pi_{ik}^{[1]}\Pi_{kj}^{[1]}}, \TF{\Pi_{ij}^{[1]}}\Tr{\Pi^{[1]}}, \label{eq:2nd_bias_op}\\
    &\mrm{3rd:} &\quad&\TF{\Pi_{ij}^{[3]}}, \TF{\Pi_{ik}^{[1]}\Pi_{kj}^{[2]} + \Pi_{ik}^{[2]}\Pi_{kj}^{[1]}}, \TF{\Pi_{ij}^{[2]}}\Tr{\Pi^{[1]}}, \nonumber\\
    & &\quad&\TF{\Pi_{ik}^{[1]}\Pi_{kl}^{[1]}\Pi_{lj}^{[1]}}, \TF{\Pi_{ik}^{[1]}\Pi_{kj}^{[1]}}\Tr{\Pi^{[1]}}, \TF{\Pi_{ij}^{[1]}}\left(\Tr{\Pi^{[1]}}\right)^2, \label{eq:3rd_bias_op}
\end{alignat}
where $\Pi_{ij}^{[1]} \equiv \partial_i\partial_j\Phi$ is the Hessian matrix of the rescaled gravitational potential $\Phi$, with $\mrm{Tr}[\Pi^{[1]}] = \nabla^2\Phi = \delta$ following from the Poisson equation. 
We denote the tidal field as $K_{ij} \equiv \mrm{TF}[\Pi_{ij}^{[1]}]$. 
$\Pi_{ij}^{[n]}~(n>1)$ is the higher-generation tensor recursively defined by the convective time derivative as \citep{Mirbabayi:2015,Desjacques+2018:bias_review}:
\begin{align}
    \Pi_{ij}^{[n]} = \frac{1}{(n-1)!}\left[(\mcal{H}f)^{-1}\frac{D}{D\tau}\Pi_{ij}^{[n-1]} - (n-1)\Pi_{ij}^{[n-1]}\right], 
\end{align}
with the conformal Hubble parameter $\mcal{H}$, the logarithmic growth rate $f$, and the convective time derivative: 
$D/D\tau \equiv \partial/\partial\tau + v^i \partial_i$. 
This basis contains one, three, and six independent operators that \textit{start} at first, second, and third order in perturbations, respectively.\footnote{In Ref.~\cite{Vlah+2020:IA_EFT}, seven third-order operators are listed: the six operators in \eq{eq:3rd_bias_op} and the additional operator $\mrm{TF}[\Pi_{ij}^{[1]}]\mrm{Tr}[(\Pi^{[1]})^2]$. 
However, since the four cubic operators constructed solely from $\Pi^{[1]}$ are related by a Cayley--Hamilton identity (see Refs.~\cite{Ansari+2025:NonLocalInTime,Bakx+2026:TwoLoopRenom_GalBias}), only six operators are independent.}

Following Ref.~\cite{Vlah+2020:IA_EFT}, we perform a linear transformation of the quadratic operator basis to one that is convenient for calculating the one-loop power spectrum: 
\begin{align}
    Q_{1,ij} &\equiv \TF{\Pi_{ik}^{[1]}\Pi_{kj}^{[1]}}, \label{eq:def_Q1ij}\\
    Q_{2,ij} &\equiv \frac{7}{20}\TF{\Pi_{ij}^{[2]}-\Pi_{ik}^{[1]}\Pi_{kj}^{[1]}}, \label{eq:def_Q2ij}\\
    Q_{3,ij} &\equiv \TF{\Pi_{ij}^{[1]}\Tr{\Pi^{[1]}} -\frac{7}{20}\Pi_{ij}^{[2]} - \frac{13}{20}\Pi_{ik}^{[1]}\Pi_{kj}^{[1]}}. \label{eq:def_Q3ij}
\end{align}
The advantage of this choice will shortly become clear from the Fourier-space kernels. 
As shown in Ref.~\cite{Vlah+2020:IA_EFT}, the cubic operators give only two independent contributions to the $(13)$ part of the one-loop power spectrum after renormalization. 
We choose the following operators to represent these contributions: 
\begin{align}
    C_{1,ij} &\equiv 4\TF{\Pi_{ik}^{[1]}\Pi_{kj}^{[2]} + \Pi_{ik}^{[2]}\Pi_{kj}^{[1]} - \Pi_{ij}^{[2]}\Tr{\Pi^{[1]}}}, \label{eq:def_C1ij}\\
    C_{2,ij} &\equiv \TF{\Pi_{ik}^{[1]}\Pi_{kj}^{[2]} + \Pi_{ik}^{[2]}\Pi_{kj}^{[1]}}. \label{eq:def_C2ij}
\end{align} 
With this choice of cubic operators, we write the bias expansion for the one-loop power spectrum as 
\begin{align}
    \Sij
    =
    b_K K_{ij}
    +\sum_{a=1}^{3}b_{2,a}Q_{a,ij}
    +\sum_{a=1}^{2}b_{3,a}C_{a,ij}
    + b_{\nabla^2K} \nabla^2K_{ij} 
    + S_{ij}^\mrm{stoch}. \label{eq:bias_exp_Sij}
\end{align}
For the galaxy density field, we also retain only the bias operators needed for the one-loop power spectrum: 
\begin{align}
    \delg 
    = 
    b_1 \delta 
    + \frac{b_2}{2}\delta^2 + b_{\mcal{G}_2}\mcal{G}_2 
    + b_{\Gamma_3}\Gamma_3 
    + b_{\nabla^2\delta}\nabla^2\delta 
    + \delg^\mrm{stoch}, \label{eq:bias_exp_delg}
\end{align}
where $\mcal G_2(\Phi)=(\partial_i\partial_j\Phi)^2-(\nabla^2\Phi)^2$ is the second-order Galileon operator, and 
$\Gamma_3=\mcal G_2(\Phi)-\mcal G_2(\Phi_v)$, with the rescaled velocity potential 
$\nabla^2\Phi_v=-\nabla\cdot\bv{v}/(\mcal H f)$ \citep{Chan+2012:Galileon,Assassi+2014:Ren_halo_bias}. 
In this basis, $\Gamma_3$ is the only cubic bias operator needed for the one-loop galaxy density power spectrum after renormalization. 
The operators $\nabla^2 K_{ij}$ and $\nabla^2\delta$ are the leading higher-derivative terms. 
We discuss their contributions in Section~\ref{subsec:ps_1loop}, where we introduce the EFT counterterms. 
The last terms in \eqs{eq:bias_exp_Sij}{eq:bias_exp_delg} denote the stochastic parts of the shape and density fields, respectively. 
They account for small-scale fluctuations that the large-scale fields
do not determine. 
These real-space bias expansions, including the higher-derivative and stochastic terms, are the common starting point of the two redshift-space models in Section~\ref{sec:ps_1loop_rsd}.
The models differ in how the redshift-space mapping of these fields is treated. 

We now expand the deterministic terms (at leading order in spatial derivatives) in powers of the linear matter density field $\delL$ \citep{Desjacques+2018:bias_review,Vlah+2020:IA_EFT}. 
We denote the Fourier-space kernels at $n$th order in $\delL$ as $K_n$ and $K_{n,ij}$, respectively: 
\begin{align}
    \delg^{(n)}(\bv{k}) 
    &= \int_{\bv{q}_1,\cdots,\bv{q}_n} (2\pi)^3\delD_{\bv{k}-\bv{q}_{1\cdots n}}
    K_n(\bv{q}_1,\cdots,\bv{q}_n) \prod_{m=1}^n\delL(\bv{q}_m), \label{eq:bias_exp_delg_real}\\
    \Sij^{(n)}(\bv{k}) &= \int_{\bv{q}_1,\cdots,\bv{q}_n} (2\pi)^3\delD_{\bv{k}-\bv{q}_{1\cdots n}} 
    K_{n,ij}(\bv{q}_1,\cdots,\bv{q}_n) \prod_{m=1}^n\delL(\bv{q}_m). \label{eq:bias_exp_Sij_real}
\end{align} 
We give the first- and second-order real-space bias kernels here.
As we will see in Section~\ref{subsec:redshift-space_mapping}, the additional terms generated by the redshift-space mapping up to third order involve only these lower-order bias kernels. 
The third-order kernels are given in Appendix~\ref{subapp:3rd_kernel_real}. 
For the density field, $K_1=b_1$ and 
\begin{align}
    K_2(\bv{q}_1,\bv{q}_2)
    = b_1 F_2(\bv{q}_1,\bv{q}_2)
    + \frac{b_2}{2} + b_{\mcal{G}_2}\sigma^2(\bv{q}_1,\bv{q}_2), 
    \label{eq:def_K2}
\end{align} 
where $F_2$ is the standard second-order matter kernel \citep{Bernardeau+2002:PT_review} and 
$\sigma^2(\bv{q}_1,\bv{q}_2) \equiv (\mu_{12}^2-1)$ with $\mu_{12}\equiv\hbv{q}_1\cdot\hbv{q}_2$. 
For the shape field, 
$K_{1,ij}(\bv{k}) = b_K \mcal{D}_{ij}(\bv{k}) $ with $\mcal{D}_{ij}(\bv{k}) \equiv \h{k}_i\h{k}_j - \delK_{ij}/3$, and 
\begin{align} 
    K_{2,ij}(\bv{q}_1,\bv{q}_2) 
    = b_K F_2(\bv{q}_1,\bv{q}_2)\mcal{D}_{ij}(\bv{k}) 
    + \sum_{a=1}^3b_{2,a}K_{Q_a,ij}(\bv{q}_1,\bv{q}_2).  
    \label{eq:shape_kernel_2nd}
\end{align} 
The second-order kernels are defined as 
\begin{align}
    K_{Q_1,ij}(\bv{q}_1,\bv{q}_2) &= \mu_{12}\left(\frac{1}{2} \left(\h{q}_{1i}\h{q}_{2j}+\h{q}_{2i}\h{q}_{1j}\right) - \frac{\mu_{12}}{3}\delK_{ij}\right), \label{eq:def_KQ1ij} \\
    K_{Q_2,ij}(\bv{q}_1,\bv{q}_2) &= -\frac{1}{4} \sigma^2(\bv{q}_1,\bv{q}_2) \mcal{D}_{ij}(\bv{k}), \label{eq:def_KQ2ij} \\
    K_{Q_3,ij}(\bv{q}_1,\bv{q}_2) &= \frac{k^2}{2q_1^2q_2^2}\left(q_{1\perp i} q_{1\perp j} - \frac{q_{1\perp}^2}{2}\mcal{P}_{ij}(\bv{k})\right), \label{eq:def_KQ3ij}
\end{align} 
where
$\mcal{P}_{ij}(\bv{k}) \equiv \delK_{ij}-\h{k}_i\h{k}_j$ projects onto the plane perpendicular to $\bv{k}$, and $q_{\perp i} \equiv \mcal{P}_{ij}(\bv{k})q_j$. 
Note that $K_{Q_3,ij}$ is symmetric under the exchange $\bv{q}_1\leftrightarrow\bv{q}_2$, since $\bv{q}_{1\perp}=-\bv{q}_{2\perp}$.

Since $K_{Q_2,ij}$ is proportional to $\mcal{D}_{ij}(\bv{k}) $, it contains only the longitudinal scalar mode (helicity 0) with respect to $\bv{k}$. 
For $K_{Q_3,ij}$, we have rewritten the corresponding kernel in Refs.~\cite{Vlah+2020:IA_EFT,Vlah+2021:IA_EFT,Bakx+2023:EFTofIAvsSims} in terms of $\bv{q}_{1\perp}$. 
This form makes its transverse and trace-free structure explicit (helicity $\pm2$ with respect to $\bv{k}$): $\h{k}_iK_{Q_3,ij}=0$. 
The helicity decomposition is described in Appendix~\ref{subapp:helicity_22}.

\subsection{Projected galaxy shapes on the sky} 
\label{subsec:projection} 
In galaxy imaging surveys, the two spin-2 components of galaxy shapes projected onto the sky are observable rather than the full three-dimensional shapes.
In practice, these components are measured with respect to the local spherical-coordinate basis $\{\bv{e}_\theta,\bv{e}_\phi\}$ at the angular position $\hbv{x}$: 
\begin{align}
    {}_{\pm2}\gamma(\bv{x})
    &= m_i^\mp m_j^\mp
    \Sij(\bv{x}), 
    \label{eq:def_proj}
\end{align}
where we defined the complex unit vectors 
$m_i^\pm \equiv \left[e_{\theta,i}\mp i e_{\phi,i}\right]/\sqrt{2}$.%
\footnote{In observations, the ellipticities of galaxies are normalized by the trace
of the second-moment tensor on the sky.
Because this trace depends on the LOS, the corrections to Eq.~\eqref{eq:def_proj} can introduce additional angle-dependent terms starting at one-loop order (see Appendix~B of Ref.~\cite{Bakx+2023:EFTofIAvsSims}). 
These corrections could be important for some shape estimators.
We ignore them in this analysis and leave a detailed study of their effects to future work.}
In this work, we adopt the plane-parallel approximation, assuming a single, fixed LOS direction $\hbv{x}\simeq\hbv{n}$ for all galaxies.
Since the projection basis is then independent of position, the Fourier transform of Eq.~\eqref{eq:def_proj} becomes 
\begin{align}
    {}_{\pm2}\gamma(\bv{k})
    &= m_i^\mp m_j^\mp
    \Sij(\bv{k}).
    \label{eq:def_proj_F}
\end{align}
The Fourier-space $E$- and $B$-mode fields are defined by \citep{Kurita&Takada2022:AnalysisIAPS} 
\begin{align}
    E(\bv{k})\pm iB(\bv{k})
    &\equiv
    {}_{\pm2}\gamma(\bv{k})
    e^{\mp2i\phi}, 
    \quad 
    e^{\pm i\phi}
    \equiv \frac{\sqrt{2}m_i^\mp\h{k}_i}{\sqrt{1-\mu^2}}, 
    \label{eq:def_EB_modes}
\end{align}
where $\phi$ is the azimuthal angle of the transverse component of $\bv{k}$, measured from $\bv{e}_\theta$ toward $\bv{e}_\phi$, with $\mu\equiv\hbv{k}\cdot\hbv{n}$. 
The phase rotation removes the dependence on the orientation of the transverse coordinate basis, so that $E$ and $B$ modes are invariant under rotations of this basis about $\hbv{n}$. 
Note that $\bv{k}$ denotes the three-dimensional wave vector. 

Here, we rewrite \eq{eq:def_EB_modes} in a more convenient form by absorbing the phase factors into the projection basis \citep{Bakx+2025:IA_Bistectrum_I,Akitsu+2026:IAinRS_I}. 
From \eqs{eq:def_proj_F}{eq:def_EB_modes} with 
$m_i^\pm m_j^\mp = \mcal{P}_{ij}(\hbv{n})/2 \pm i\varepsilon_{ijk}\h{n}_k/2$, we obtain
\begin{align}
    E(\bv{k})\pm iB(\bv{k})
    &=
    M_i^\mp M_j^\mp \Sij(\bv{k}),
    \label{eq:def_EB_modes_phase-free}
\end{align}
where 
\begin{align}
    M_i^\pm
    \equiv m_i^\pm e^{\pm i\phi} 
    = \frac{1}{\sqrt{2}} \left[\h{a}_i\mp i \h{b}_i\right], \quad
    \h{a}_i \equiv \frac{\h{k}_i-\mu\h{n}_i}{\sqrt{1-\mu^2}},\quad 
    \h{b}_i \equiv \varepsilon_{ijk}\h{n}_j\h{a}_k.  
    \label{eq:def_M_pm}
\end{align}
These vectors depend only on $\hbv{k}$ and $\hbv{n}$, which eliminates any dependence on the choice of transverse basis.%
\footnote{The resulting basis is equivalent to that used in Refs.~\cite{Bakx+2025:IA_Bistectrum_I,Akitsu+2026:IAinRS_I}, up to overall signs of $ M_i^\pm$, which cancel in the spin-2 projection.} 
It is useful to collect the real and imaginary parts of the same projection into two real tensors, 
\begin{align}
    M_{ij}^E
    &\equiv \mrm{Re}\left[M_i^-M_j^-\right]
    = \frac{\h{a}_i\h{a}_j-\h{b}_i\h{b}_j}{2}
    = \frac{(\h{k}_i - \mu\h{n}_i)(\h{k}_j - \mu\h{n}_j)}{1-\mu^2} - \frac12 \mcal{P}_{ij}(\bv{n}), \label{eq:def_M_E}\\
    M_{ij}^B
    &\equiv \mrm{Im}\left[M_i^-M_j^-\right]
    = \frac{\h{a}_i\h{b}_j+\h{b}_i\h{a}_j}{2}
    = \frac{(\h{k}_i - \mu\h{n}_i) \varepsilon_{jkl}\h{n}_k\h{k}_l + \varepsilon_{ikl}\h{n}_k\h{k}_l(\h{k}_j - \mu\h{n}_j)}{2(1-\mu^2)}.\label{eq:def_M_B}
\end{align}
Thus $E=M_{ij}^E S_{ij}$ and $B=M_{ij}^B S_{ij}$. 

We use these projectors at every stage of the perturbative calculation. 
For the $n$th-order kernel in \eq{eq:bias_exp_Sij_real}, we define
\begin{align}
    K_n^{E,B}(\bv{q}_1,\cdots,\bv{q}_n)
    \equiv 
    M_{ij}^{E,B}(\bv{k},\bv{n})
    K_{n,ij}(\bv{q}_1,\ldots,\bv{q}_n), 
    \quad \bv{k}=\bv{q}_{1\cdots n}.
    \label{eq:def_EB_kernel}
\end{align} 
In particular, since $M_{ij}^E\mcal{D}_{ij}(\bv{k})=(1-\mu^2)/2$ and $M_{ij}^B\mcal{D}_{ij}(\bv{k})=0$, we have 
\begin{align}
    K_1^{E}&=\frac{(1-\mu^2)}{2}b_K,\quad K_1^{B}=0, \label{eq:kernel_1st_prj}\\
    K_2^{E}&=\frac{(1-\mu^2)}{2}b_KF_2 + \sum_{a=1}^3 b_{2,a}K_{Q_a}^E, \quad
    K_2^{B}=b_{2,1}K_{Q_1}^B + b_{2,3}K_{Q_3}^B. \label{eq:kernel_2nd_prj}
\end{align} 
Note that the linear alignment produces only the $E$ mode, i.e., the $B$-mode arises only from the nonlinear mode couplings.
Also, the nonlinear evolution of the matter field in the first-order operator and one of the quadratic operators, $Q_{2,ij}$, do not contribute to the $B$-mode signal.

\subsection{Power spectra at one-loop order} \label{subsec:ps_1loop} 
Within the EFT framework, the power spectra up to one-loop order can be schematically decomposed as 
\begin{align}
    P_{XY} = P_{XY}^\mrm{tree+1\mathchar`-loop} + P_{XY}^\mrm{ctr} + P_{XY}^\mrm{stoch}, \label{eq:PXY_decomp_real}
\end{align} 
where $XY\in\{\mrm{gg},\mrm{g}E,EE,BB\}$ labels the power spectra considered in this work. 
We assume parity invariance, so that $P_{\mrm{g}B}=P_{EB}=0$ (see Ref.~\cite{Kurita+2026:PVinGS_PNG} for the parity-violating case). 
The first term contains the tree-level and one-loop contributions computed from the bias expansion above using Eulerian perturbation theory. 
The second and third terms contain the deterministic counterterms and the stochastic contributions, respectively. 
The coefficients of both terms also absorb UV-sensitive contributions from the loop integrals, as discussed below.

\subsubsection{Power spectrum of galaxy clustering} 
\label{subsubsec:pgg_real} 
First, we briefly review the results of the galaxy density auto-power spectrum and introduce the notation that will subsequently be used for the galaxy-shape power spectra. 
The tree-level and one-loop parts of the power spectrum are given by 
\begin{align}
    P_\mrm{gg}^\mrm{tree+1\mathchar`-loop}(k) 
    &= \left[K_1^\mrm{g}(\bv{k})\right]^2\PL(k) 
    + 2 \int_\bv{q} \left[K_2^\mrm{g}(\bv{q},\bv{k}-\bv{q})\right]^2 \PL(q) \PL(|\bv{k}-\bv{q}|)\nonumber\\
    &\quad+ 
    6 \PL(k) K_1^\mrm{g}(\bv{k}) \int_\bv{q} K_3^\mrm{g}(\bv{k},\bv{q},-\bv{q}) \PL(q) \nonumber\\ 
    &\equiv P_\mrm{gg}^{(11)}(k) + P_\mrm{gg}^{(22)}(k) + P_\mrm{gg}^{(13)}(k). 
\end{align} 
The tree-level part is 
$P_\mrm{gg}^{(11)} = b_1^2 \PL$, where $\PL$ denotes the linear matter power spectrum. 

To give the loop contributions explicitly, we follow the notation used in Refs.~\citep{Assassi+2014:Ren_halo_bias,Chudaykin+2020:CLASS-PT}. 
The (22) contribution can be written as 
\begin{align}
    P_\mrm{gg}^{(22)}
    = b_1^2 P_m^{(22)}
    +
    b_1b_2 \mcal{I}_{\delta,\delta^2}
    +
    2b_1b_{\mcal{G}_2} \mcal{I}_{\delta,\mcal{G}_2}
    +
    \frac{b_2^2}{4} \mcal{I}_{\delta^2,\delta^2}
    +
    b_2b_{\mcal{G}_2} \mcal{I}_{\delta^2,\mcal{G}_2}
    +
    b_{\mcal{G}_2}^2 \mcal{I}_{\mcal{G}_2,\mcal{G}_2}, \label{eq:pgg_22_real}
\end{align} 
where we define the (22)-type loop integrals as 
\begin{align}
    \mcal{I}_{X,Y}(k)
    \equiv
    2\int_{\bv{q}}
    K_X(\bv{q},\bv{k}-\bv{q})K_Y(\bv{q},\bv{k}-\bv{q})
    \PL(q)\PL(|\bv{k}-\bv{q}|).  \label{eq:def_mcalI_scalar}
\end{align} 
Here, $X,Y\in\{\delta,\delta^2,\mcal{G}_2\}$ label the second-order scalar operators, with second-order kernels $K_\delta=F_2$, $K_{\delta^2}=1$, and $K_{\mcal{G}_2}=\sigma^2$. 
We denote the $(22)$ contribution to the one-loop matter power spectrum in standard perturbation theory (SPT) by $P_m^{(22)}\equiv\mcal{I}_{\delta,\delta}$ \citep{Bernardeau+2002:PT_review}. 
Throughout this paper, any constant ($k$-independent) parts of these loop integrals are understood to be subtracted and absorbed into the leading stochastic contribution. 
Among the integrals appearing in Eq.~\eqref{eq:pgg_22_real}, only $\mcal{I}_{\delta^2,\delta^2}$ has a constant limit before this subtraction: 
$\mcal{I}_{\delta^2,\delta^2}(0) = 2\int_\bv{q}\PL^2(q)$. 

The (13) contribution is given by 
\begin{align}
    P_\mrm{gg}^{(13)}
    = 
    b_1^2P_m^{(13)} + 2b_1\left(b_{\mcal{G}_2}+\frac25b_{\Gamma_3}\right)\mcal{F}_{\mcal{G}_2}, 
    \label{eq:pgg_13_real}
\end{align}
where the first term corresponds to the (13) contribution to the one-loop matter power spectrum in SPT, and $\mcal{F}_{\mcal{G}_2}$ is the non-degenerate function from the renormalized third-order bias contributions. 
The explicit forms of these functions are given in Appendix~\ref{subapp:3rd_kernel_real}. 
We note that, in our notation, we use six basis functions, $\mcal{I}_{X,Y}$, for the (22) contribution in \eq{eq:pgg_22_real}, while two basis functions, $P_m^{(13)}$ and $\mcal{F}_{\mcal{G}_2}$, for the (13) contribution in \eq{eq:pgg_13_real}, respectively. 

The leading UV-sensitive parts of $P_m^{(13)}$ and $\mcal{F}_{\mcal{G}_2}$ both scale as $k^2\PL(k)$. 
The EFT counterterm that absorbs this dependence has the same form as the contribution from $b_{\nabla^2\delta}\nabla^2\delta$ in \eq{eq:bias_exp_delg}. 
Therefore, we use a single effective coefficient $c_{\mrm{g},0}$ for their combined contribution: 
\begin{align}
    P_\mrm{gg}^\mrm{ctr}(k)
    =
    -2b_1c_{\mrm{g},0}k^2\PL(k). 
\end{align}

Finally, since stochastic correlations are short-ranged, their large-scale power spectrum can be expanded in even powers of $k$. 
We keep terms up to $\mcal{O}(k^2)$ in this work: 
\begin{align}
    P_\mrm{gg}^\mrm{stoch}(k)
    =
    N_{\mrm{gg},0}
    +
    N_{\mrm{gg},2}k^2. \label{eq:Pgg_stoch_real}
\end{align}
For purely Poisson noise, $N_{\mrm{gg},0}=1/\bar{n}_\mrm{g}$ and $N_{\mrm{gg},2}=0$.

\subsubsection{Power spectra of projected galaxy shapes} 
\label{subsubsec:pgE_pEE_pBB_real} 
We next consider the galaxy density--shape cross-power spectrum and the galaxy-shape auto-power spectrum in real space. 
Analytic expressions within the EFT framework for these spectra up to one-loop order were derived in Ref.~\cite{Vlah+2020:IA_EFT}. 
Here we rewrite these results in notation closely aligned with that used for galaxy clustering. 
This notation makes clear which loop functions can be reused from the scalar-tracer calculation. 
The correspondence with the notation of Refs.~\cite{Vlah+2020:IA_EFT,Bakx+2023:EFTofIAvsSims} is summarized in Appendix~\ref{subapp:degeneracy_22}. 
The tree-level and one-loop contributions for the cross-power spectrum are written as 
\begin{align}
    P_{\mrm{g}E}^\mrm{tree+1\mathchar`-loop}(k,\mu) 
    &= K_1^\mrm{g}(\bv{k}) K_1^{E}(\bv{k}) \PL(k)
    + 2 \int_\bv{q} K_2^\mrm{g}(\bv{q},\bv{k}-\bv{q}) K_2^{E}(\bv{q},\bv{k}-\bv{q}) \PL(q) \PL(|\bv{k}-\bv{q}|)\nonumber\\
    &\quad+ 
    3 \PL(k) \left[K_1^\mrm{g}(\bv{k}) \int_\bv{q} K_3^{E}(\bv{k},\bv{q},-\bv{q}) \PL(q) + K_1^E(\bv{k}) \int_\bv{q} K_3^\mrm{g}(\bv{k},\bv{q},-\bv{q}) \PL(q)\right] \nonumber\\
    &\equiv P_{\mrm{g}E}^{(11)}(k,\mu) + P_{\mrm{g}E}^{(22)}(k,\mu) + P_{\mrm{g}E}^{(13)}(k,\mu).  
\end{align}
For the galaxy-shape auto-power spectra, we write
\begin{align}
    P_{EE}^\mrm{tree+1\mathchar`-loop}(k,\mu) 
    &= \left[K_1^{E}(\bv{k})\right]^2\PL(k) 
    + 2 \int_\bv{q} \left[K_2^{E}(\bv{q},\bv{k}-\bv{q})\right]^2 \PL(q) \PL(|\bv{k}-\bv{q}|)\nonumber\\
    &\quad+ 
    6 \PL(k) K_1^E(\bv{k}) \int_\bv{q} K_3^{E}(\bv{k},\bv{q},-\bv{q}) \PL(q), \nonumber\\
    &\equiv P_{EE}^{(11)}(k,\mu) + P_{EE}^{(22)}(k,\mu) + P_{EE}^{(13)}(k,\mu) \\
    P_{BB}^\mrm{tree+1\mathchar`-loop}(k,\mu) 
    &= 
    2 \int_\bv{q} \left[K_2^{B}(\bv{q},\bv{k}-\bv{q})\right]^2 \PL(q) \PL(|\bv{k}-\bv{q}|) 
    \equiv P_{BB}^{(22)}(k,\mu).  
\end{align} 
Notice that there are no tree-level and (13) contributions to the $B$-mode power spectrum since $K_1^B=0$ in \eq{eq:kernel_1st_prj}. 
The tree-level contributions to the $E$-mode power spectra, corresponding to the linear alignment model \citep{Hirata&Seljak2004:IA_LA}, are given by 
\begin{align}
    P_{\mrm{g}E}^{(11)} = \frac{1-\mu^2}{2}b_1b_K\PL, \quad 
    P_{EE}^{(11)} = \frac{(1-\mu^2)^2}{4}b_K^2\PL. 
\end{align} 
Due to the projection of galaxy shapes, the power spectra are already anisotropic ($\mu$-dependent) at linear scales \citep{Kurita+2021:IA_nbody}. 

The (22) contributions are given by
\begin{align}
    P_{\mrm{g}E}^{(22)} 
    &= \frac{1-\mu^2}{2}b_K \left[b_1P_m^{(22)} + \frac{b_2}{2} \mcal{I}_{\delta,\delta^2} + b_{\mcal{G}_2} \mcal{I}_{\delta,\mcal{G}_2} \right]
    + 
    \sum_{a=1}^2 b_{2,a}\left[b_1 \mcal{I}_{\delta,Q_a}^E + \frac{b_2}{2}\mcal{I}_{\delta^2,Q_a}^E + b_{\mcal{G}_2}\mcal{I}_{\mcal{G}_2,Q_a}^E\right], \label{eq:PgE_22}\\
    P_{EE}^{(22)} 
    &= \frac{(1-\mu^2)^2}{4}b_K^2P_m^{(22)}
    + (1-\mu^2)b_K \sum_{a=1}^2 b_{2,a} \mcal{I}_{\delta,Q_a}^E 
    + \sum_{a,b=1}^3b_{2,a}b_{2,b} \mcal{I}_{Q_a,Q_b}^{EE}, \label{eq:PEE_22}\\
    P_{BB}^{(22)} 
    &= b_{2,1}^2 \mcal{I}_{Q_1,Q_1}^{BB} + 2b_{2,1}b_{2,3} \mcal{I}_{Q_1,Q_3}^{BB} + b_{2,3}^2 \mcal{I}_{Q_3,Q_3}^{BB}. 
    \label{eq:PBB_22}
\end{align} 
Here, we define the $(22)$-type loop integrals involving the quadratic shape kernels in \eq{eq:kernel_2nd_prj}, following the scalar case in \eq{eq:def_mcalI_scalar} 
\begin{align}
    \mcal{I}_{X,Q_a}^{E}(k,\mu)
    &\equiv
    2\int_{\bv{q}}
    K_X(\bv{q},\bv{k}-\bv{q})K_{Q_a}^E(\bv{q},\bv{k}-\bv{q})
    \PL(q)\PL(|\bv{k}-\bv{q}|),
    \label{eq:def_mcalI_OQaE} \\
    \mcal{I}_{Q_a,Q_b}^{AA}(k,\mu)
    &\equiv
    2\int_{\bv{q}}
    K_{Q_a}^A(\bv{q},\bv{k}-\bv{q}) K_{Q_b}^A(\bv{q},\bv{k}-\bv{q})
    \PL(q)\PL(|\bv{k}-\bv{q}|), 
    \label{eq:def_mcalI_QaQbAA}
\end{align}
where $X\in\{\delta,\delta^2,\mcal{G}_2\}$ labels the scalar operator and $A\in\{E,B\}$ labels the projected shape components. 
Following Ref.~\cite{Vlah+2020:IA_EFT}, we rewrite these loop integrals in a more convenient form as sums of scalar functions of $k$ multiplied by geometrical factors depending on $\mu$ using the helicity decomposition of the operators. 
The detailed calculation is given in Appendix~\ref{subapp:helicity_22}. 
Using the reduction relations in Appendix~\ref{subapp:degeneracy_22}, we need only nine basis functions for the real-space $(22)$ contributions to $P_{\mrm{g}E}$, $P_{EE}$, and $P_{BB}$. 
Six already appear in the galaxy density power spectrum in \eq{eq:pgg_22_real}.
The three additional functions are given in  Eqs.~\eqref{eq:def_mcalI_delta2Q1}--\eqref{eq:def_mcalI_Q1Q1_lam0}.%
\footnote{We find an additional relation that reduces the four shape-specific $(22)$-type integrals in Refs.~\cite{Vlah+2020:IA_EFT,Bakx+2023:EFTofIAvsSims} to three. 
See Appendix~\ref{subapp:degeneracy_22} for details.} 

The (13) contributions to the $E$-mode power spectra are given by
\begin{align}
    P_{\mrm{g}E}^{(13)} 
    &= \frac{1-\mu^2}{2}
    \left[
    b_1b_K P_m^{(13)}
    +
    \left(b_Kb_{\mcal{G}_2}+\frac25b_Kb_{\Gamma_3} + b_1\bar{b}_{3,1}\right)\mcal{F}_{\mcal{G}_2}
    + 
    b_1\bar{b}_{3,2}\mcal{F}_{Q_1}
    \right], \label{eq:PgE_13}\\
    P_{EE}^{(13)} 
    &= \frac{(1-\mu^2)^2}{4}
    \left[b_K^2 P_m^{(13)} + 2 b_K\left(\bar{b}_{3,1}\mcal{F}_{\mcal{G}_2} + \bar{b}_{3,2}\mcal{F}_{Q_1}\right)
    \right], \label{eq:PEE_13}
\end{align}
where $\mcal{F}_{Q_1}$ is the additional $(13)$-type loop function required for tensor tracers.
We give its explicit expression in Appendix~\ref{subapp:3rd_kernel_real}.
The third-order parameters $\bar{b}_{3,1}$ and $\bar{b}_{3,2}$ differ from those in \eq{eq:bias_exp_Sij} because the former include the third-order contributions from the quadratic operators.
The relation between them is also given in Appendix~\ref{subapp:3rd_kernel_real}. 

Since all three basis functions are proportional to $k^2\PL$ at leading order in the UV limit, the corresponding counterterms are degenerate with the leading higher-derivative bias operator, $\nabla^2K_{ij}$. 
Therefore, we write the contributions in the same way as for the density power spectrum by introducing an effective parameter $c_{S,0}$:
\begin{align}
    P_{\mrm{g}E}^\mrm{ctr}(k,\mu)
    &=
    -\frac{1-\mu^2}{2}
    \left(b_1c_{S,0}+b_Kc_{\mrm{g},0}\right)
    k^2\PL(k),\\
    P_{EE}^\mrm{ctr}(k,\mu)
    &=
    -\frac{(1-\mu^2)^2}{2}
    b_Kc_{S,0}k^2\PL(k). 
\end{align}

Lastly, we parametrize the stochastic power spectra up to order $k^2$ as 
\begin{align}
    P_{\mrm{g}E}^\mrm{stoch}(k,\mu)
    &=
    \frac{1-\mu^2}{2} N_{\mrm{g}S,2} k^2, \label{eq:PgE_stoch_real}\\
    P_{EE}^\mrm{stoch}(k,\mu)
    &=
    N_{SS,0} 
    + 
    \left(
    N_{SS,2} + N_{SS,2\mu}\mu^2 
    \right)k^2
    + 
    \frac{(1-\mu^2)^2}{4}N_{SS,2\Delta}k^2, \label{eq:PEE_stoch_real}\\
    P_{BB}^\mrm{stoch}(k,\mu)
    &=
    N_{SS,0} 
    + 
    \left(
    N_{SS,2} + N_{SS,2\mu}\mu^2 
    \right)k^2. \label{eq:PBB_stoch_real}
\end{align} 
The cross spectrum has no constant term due to statistical isotropy. 
For the auto-power spectra, $N_{SS,0}$ denotes the renormalized constant noise amplitude and can differ from the Poisson shape noise. 
At order $k^2$, we treat the three coefficients, $\{N_{SS,2}, N_{SS,2\mu},N_{SS,2\Delta}\}$, as independent. 
Note that, for a strictly local stochastic tensor, only two of these coefficients are independent \citep{Chen&Kokron2024:LEFTofIA}, which corresponds to $N_{SS,2\Delta}=0$. 
In this work, we keep $N_{SS,2\Delta}$ as an independent parameter to absorb
UV contributions from $(22)$-type loop integrals with the inverse Laplacian in the shape operators. 
See Appendix~\ref{subapp:stochasticity} for details. 

\section{Redshift-space power spectra at one-loop order} \label{sec:ps_1loop_rsd} 
In this section, we construct the two redshift-space models from the real-space bias expansions in Section~\ref{subsec:bias_expansion}. 
Sections~\ref{subsec:redshift-space_mapping} and \ref{subsec:loops} are common to both models: we introduce the redshift-space mapping and compute the tree-level and one-loop contributions in Eulerian perturbation theory (EPT). 
The two models differ in how they treat the effects of small-scale LOS velocities that enter through the mapping.
In Sections~\ref{subsec:ps_1loop_rsd_ctr_stoch} and \ref{subsec:ps_1loop_rsd_summary}, we construct the EFT model, in which these velocity effects are expanded, and their UV sensitivity is absorbed into counterterms. 
In Section~\ref{subsec:tns_model}, we construct the TNS-based model, in which the one-point velocity dispersion is kept in an unexpanded damping factor. 

\subsection{Redshift-space mapping} \label{subsec:redshift-space_mapping} 
The mapping from real space to redshift space along the fixed LOS direction $\hbv{n}$ is given by 
\begin{align}
    \bv{s} = \bv{x} + \frac{v_\parallel(\bv{x})}{\mcal{H}} \hbv{n}, \label{eq:rsd_map}
\end{align}
where $v_\parallel \equiv \bv{v}\cdot\hbv{n}$ is the LOS component of the dark matter peculiar velocity. 
Here, we assume that galaxies follow the dark matter velocity field. 
For convenience, we define the rescaled LOS velocity 
$u_\parallel \equiv v_\parallel/\mcal{H}$. 

From number conservation, we have the relation between the real-space density field $\delg$ and the redshift-space density field $\tdelg$ as
\begin{align}
    \left[1+\tdelg(\bv{s})\right]\mrm{d}\bv{s}=\left[1+\delg(\bv{x})\right]\mrm{d}\bv{x}.  
    \label{eq:number_conservation}
\end{align}
Hereafter, we use tildes to distinguish fields and statistics in redshift space from those in real space. 
Taking the Fourier transform of \eq{eq:number_conservation}
using \eq{eq:rsd_map}, we obtain
\begin{align}
    \tdelg(\bv{k})
    &=
    \int \mrm{d}\bv{x}
    \left[
    \left[1+\delg(\bv{x})\right]
    e^{-ik_\parallel u_\parallel(\bv{x})}-1
    \right] e^{-i\bv{k}\cdot\bv{x}},
    \label{eq:delg_rsd_formal_Fourier}
\end{align}
where $k_\parallel \equiv \bv{k}\cdot\hbv{n}=k\mu$. 
For the shape field, we assume that the mapping changes their positions without changing their assigned shape tensors. 
The real-space and redshift-space \textit{density-weighted} shape fields, $\Sij$ and $\tSij$, then satisfy\footnote{
At corresponding positions, we assume that the unweighted fields satisfy 
$\tSij^\mrm{unw}(\bv{s})=\Sij^\mrm{unw}(\bv{x})$. 
Combining this with \eq{eq:number_conservation} and the definitions
$\Sij=(1+\delg)\Sij^{\mrm{unw}}$ and $\tSij=(1+\tdelg)\tSij^{\mrm{unw}}$ gives \eq{eq:shape_conservation}.}
\begin{align}
    \tSij(\bv{s})\mrm{d}\bv{s} = \Sij(\bv{x})\mrm{d}\bv{x}.
    \label{eq:shape_conservation}
\end{align} 
The corresponding Fourier-space expression is
\begin{align}
    \tSij(\bv{k})
    &=
    \int \mrm{d}\bv{x} 
    \Sij(\bv{x})e^{-ik_\parallel u_\parallel(\bv{x})} e^{-i\bv{k}\cdot\bv{x}}.
    \label{eq:Sij_rsd_formal_Fourier}
\end{align} 

Expanding the exponential factors and transforming back to configuration space, we write the terms needed for the one-loop power spectra as 
\begin{align}
    \tdelg &= \delg 
    - \partial_\parallel \left[ u_\parallel \left[1+\delg\right] \right] 
    + \frac12 \partial_\parallel^2 \left[ u_\parallel^2 \left[1+\delg\right] \right] 
    - \frac16 \partial_\parallel^3 \left[ u_\parallel^3 \left[1+\delg\right] \right] +\cdots, \label{eq:tdelg_exp_cubic}\\
    \tSij &= \Sij 
    - \partial_\parallel \left[ u_\parallel \Sij \right] 
    +  \frac12 \partial_\parallel^2 \left[ u_\parallel^2 \Sij \right] +\cdots, \label{eq:tSij_exp_cubic}
\end{align} 
where $\partial_\parallel\equiv\hbv{n}\cdot\nabla$.
These terms have not yet been expanded in $\delL$. 
Since the products of fields at the same position are sensitive to short-scale modes that are not under perturbative control, we renormalize them as in the real-space bias expansion.  
We discuss this in Section~\ref{subsec:ps_1loop_rsd_ctr_stoch}. 
Note that since $\Sij$ has no homogeneous component and starts at first order, we only need terms up to $\Sij u_\parallel^2$ for a third-order expansion in $\delL$. 

We now expand \eqs{eq:tdelg_exp_cubic}{eq:tSij_exp_cubic} in powers of $\delL$, as in real space, and denote the $n$th-order density and shape kernels by $Z_n$ and $Z_{n,ij}$, respectively: 
\begin{align}
    \tdelg^{(n)}(\bv{k}) 
    &= \int_{\bv{q}_1,\cdots,\bv{q}_n} (2\pi)^3\delD_{\bv{k}-\bv{q}_{1\cdots n}}
    Z_n(\bv{q}_1,\cdots,\bv{q}_n) \prod_{m=1}^n\delL(\bv{q}_m), \label{eq:bias_exp_delg_redshift}\\
    \tSij^{(n)}(\bv{k}) &= \int_{\bv{q}_1,\cdots,\bv{q}_n} (2\pi)^3\delD_{\bv{k}-\bv{q}_{1\cdots n}} 
    Z_{n,ij}(\bv{q}_1,\cdots,\bv{q}_n) \prod_{m=1}^n\delL(\bv{q}_m). \label{eq:bias_exp_Sij_redshift}
\end{align} 
Explicit expressions of these kernels are listed in Appendix~\ref{subapp:PT_kernel_rsd}. 
For the shape field, the $E$ and $B$ modes are defined using the same projection tensors as in real space: 
$\tilde{E} = M_{ij}^E \tSij$, $\tilde{B} = M_{ij}^B \tSij$. 
Also, following \eq{eq:def_EB_kernel}, we define the projected shape kernels in redshift space as 
\begin{align}
    Z_n^{E,B}(\bv{q}_1,\cdots,\bv{q}_n)
    \equiv 
    M_{ij}^{E,B}(\bv{k},\bv{n})
    Z_{n,ij}(\bv{q}_1,\ldots,\bv{q}_n), 
    \quad \bv{k}=\bv{q}_{1\cdots n}.
    \label{eq:def_EB_kernel_rsd}
\end{align} 

As in the real-space case in \eq{eq:PXY_decomp_real}, we decompose the redshift-space power spectra within the EFT framework into three parts: 
\begin{align}
    \tP_{XY} = \tP_{XY}^\mrm{tree+1\mathchar`-loop} + \tP_{XY}^\mrm{ctr} + \tP_{XY}^\mrm{stoch}, \label{eq:PXY_decomp_rsd}
\end{align} 
where $XY\in\{\mrm{gg},\mrm{g}E,EE,BB\}$. 
We derive each contribution below and summarize the final expressions in Section~\ref{subsec:ps_1loop_rsd_summary}.

\subsection{Tree-level and one-loop contributions} \label{subsec:loops} 
The deterministic part of the power spectrum is computed by replacing the real-space kernels, $K_n^\mrm{g}$ and $K_{n,ij}$, with the corresponding redshift-space kernels, $Z_n^\mrm{g}$ and $Z_{n,ij}$. 
The tree-level contributions are given by 
\begin{align}
    \tP_\mrm{gg}^{(11)}(k,\mu) 
    &= 
    \left[Z_1^\mrm{g}(\bv{k})\right]^2\PL(k) 
    = 
    (b_1+f\mu^2)^2\PL(k), \\
    \tP_{\mrm{g}E}^{(11)}(k,\mu) 
    &= Z_1^\mrm{g}(\bv{k}) Z_1^E(\bv{k}) \PL(k) = \frac{1-\mu^2}{2}(b_1+f\mu^2)b_K\PL(k), \\
    \tP_{EE}^{(11)}(k,\mu) 
    &= \left[Z_1^E(\bv{k})\right]^2 \PL(k) = \frac{(1-\mu^2)^2}{4}b_K^2\PL(k), 
\end{align}
where we used $Z_{1,ij} = K_{1,ij}$ since the redshift-space correction to the shape field starts at second order. 
Note that there is no $B$-mode at the tree level: $\tP_{BB}^{(11)} = 0$.

\subsubsection{The (22)-type loops} \label{subsubsec:ps_1loop_rsd_22loop} 
The second-order kernel of the galaxy density field in redshift space can be writtern as
\begin{align}
    Z_2^\mrm{g}(\bv{q}_1,\bv{q}_2) 
    &= b_1 K_\delta(\bv{q}_1,\bv{q}_2) 
    + \frac{b_2}{2}K_{\delta^2}(\bv{q}_1,\bv{q}_2) 
    + b_{\mcal{G}_2}K_{\mcal{G}_2}(\bv{q}_1,\bv{q}_2) \nonumber\\
    &\quad+ 
    fb_1 K_{u\delta}(\bv{q}_1,\bv{q}_2) 
    + f\mu^2 K_\theta(\bv{q}_1,\bv{q}_2) 
    + f^2\mu^2 K_{uu}(\bv{q}_1,\bv{q}_2) , \label{eq:def_Z2}
\end{align}
where $K_\theta=G_2$, and the two LOS-dependent kernels are 
\begin{align}
    K_{u\delta}(\bv{q}_1,\bv{q}_2) 
    \equiv \frac{k\mu}{2}\left(\frac{q_{1\parallel}}{q_1^2} + \frac{q_{2\parallel}}{q_2^2}\right),\quad
    K_{uu}(\bv{q}_1,\bv{q}_2) 
    \equiv \frac{k^2q_{1\parallel} q_{2\parallel}}{2q_1^2q_2^2}, \label{eq:def_Kud_Kuu}
\end{align} 
where $q_\parallel \equiv \bv{q}\cdot\hbv{n}$. 
From these definitions, the (22) contribution to the redshift-space density auto-power spectrum becomes 
\begin{align}
    &\tP_{\mrm{g}\mrm{g}}^{(22)}(k,\mu) 
    \equiv 2 \int_\bv{q} \left[Z_2^\mrm{g}(\bv{q},\bv{k}-\bv{q})\right]^2 \PL(q) \PL(|\bv{k}-\bv{q}|) \nonumber\\
    &= P_{\mrm{g}\mrm{g}}^{(22)}(k) 
    + 2b_1
    \left[
    fb_1\mcal{I}_{\delta,u\delta} + f\mu^2\mcal{I}_{\delta,\theta} + f^2\mu^2\mcal{I}_{\delta,uu}
    \right] 
    + b_2 
    \left[
    fb_1\mcal{I}_{\delta^2,u\delta} + f\mu^2\mcal{I}_{\delta^2,\theta} + f^2\mu^2\mcal{I}_{\delta^2,uu}
    \right] \nonumber\\
    &\quad+ 2b_{\mcal{G}_2} 
    \left[
    fb_1\mcal{I}_{\mcal{G}_2,u\delta} + f\mu^2\mcal{I}_{\mcal{G}_2,\theta} + f^2\mu^2\mcal{I}_{\mcal{G}_2,uu}
    \right] \nonumber\\
    &\quad+ f^2
    \left[ b_1^2 \mcal{I}_{u\delta,u\delta} + \mu^4 \mcal{I}_{\theta,\theta} + f^2\mu^4 \mcal{I}_{uu,uu} + 2b_1\mu^2 \mcal{I}_{\theta,u\delta} + 2f\mu^4 \mcal{I}_{\theta,uu} + 2fb_1\mu^2 \mcal{I}_{u\delta,uu}
    \right].  
    \label{eq:Pgg_22_rsd} 
\end{align} 
We use the definition of $\mcal{I}_{X,Y}$ in \eq{eq:def_mcalI_scalar}, extended to
$X,Y\in\{\delta,\delta^2,\mcal{G}_2,\theta,u\delta,uu\}$.
Note that the integrals involving $K_{u\delta}$ or $K_{uu}$ depend on $k$ and $\mu$.
As shown in Appendix~\ref{subsubapp:22_integral_density_auto}, all the additional integrals reduce to the six real-space density integrals in \eq{eq:pgg_22_real} and one additional function of $k$, $\mcal{I}_\mrm{RSD}$, defined in \eq{eq:def_I_rsd}. 

For the shape field, the second-order redshift-space kernels can be written in the reduced form (see Appendix~\ref{subsubapp:22_integral_shape}) 
\begin{align}
    Z_2^E(\bv{q}_1,\bv{q}_2) 
    &= 
    \frac{1-\mu^2}{2}b_K K_\delta(\bv{q}_1,\bv{q}_2) 
    + \sum_{a=1}^3\tilde{b}_{2,a}K_{Q_a}^E 
    + \frac{1-\mu^2}{2}fb_K K_{uu}(\bv{q}_1,\bv{q}_2), 
    \label{eq:def_Z2_E_red} \\
    Z_2^B(\bv{q}_1,\bv{q}_2) 
    &= \tilde{b}_{2,1}K_{Q_1}^B(\bv{q}_1,\bv{q}_2) 
    + \tilde{b}_{2,3}K_{Q_3}^B(\bv{q}_1,\bv{q}_2),
    \label{eq:def_Z2_B_red}
\end{align}
where $K_{Q_a}^{E,B}$ are the projected real-space kernels in \eq{eq:kernel_2nd_prj}, and we define the ``shifted'' bias parameters as  
\begin{align}
    \tilde{b}_{2,1}=b_{2,1},\quad
    \tilde{b}_{2,2}=b_{2,2}+fb_K,\quad
    \tilde{b}_{2,3}=b_{2,3}+fb_K. 
    \label{eq:shape_bias_shift_rsd} 
\end{align}
Thus, part of the redshift-space correction is accounted for by shifting the quadratic shape-bias coefficients.
The remaining term in $Z_2^E$ involves $K_{uu}$, which already appeared in the density case. 
From \eqs{eq:def_Z2}{eq:def_Z2_E_red}, we obtain the (22) contributions to the density--shape cross spectrum in redshift space: 
\begin{align}
    &\tP_{\mrm{g}E}^{(22)}(k,\mu)
    \equiv 2 \int_\bv{q} Z_2^\mrm{g}(\bv{q},\bv{k}-\bv{q}) Z_2^E(\bv{q},\bv{k}-\bv{q}) \PL(q) \PL(|\bv{k}-\bv{q}|) \nonumber\\
    &= P_{\mrm{g}E}^{(22)}(k,\mu;\tilde{b}_{2,a})\nonumber\\
    &\quad+ 
    \frac{1-\mu^2}{2}fb_K
    \left[
    b_1\mcal{I}_{\delta,uu} + \frac{b_2}{2}\mcal{I}_{\delta^2,uu} + b_{\mcal{G}_2}\mcal{I}_{\mcal{G}_2,uu} 
    + fb_1\mcal{I}_{u\delta,uu} + f\mu^2\mcal{I}_{\theta,uu} + f^2\mu^2\mcal{I}_{uu,uu}
    \right] \nonumber\\
    &\quad+ 
    \frac{1-\mu^2}{2}b_K
    \left[
    fb_1\mcal{I}_{\delta,u\delta} + f\mu^2\mcal{I}_{\delta,\theta} + f^2\mu^2\mcal{I}_{\delta,uu}
    \right]
    + 
    \sum_{a=1}^3 \tilde{b}_{2,a} 
    \left[
    fb_1\mcal{I}_{u\delta,Q_a}^E + f\mu^2\mcal{I}_{\theta,Q_a}^E + f^2\mu^2\mcal{I}_{uu,Q_a}^E
    \right], 
    \label{eq:PgE_22_rsd}
\end{align}
and the shape auto-power spectra: 
\begin{align}
    &\tP_{EE}^{(22)}(k,\mu)
    \equiv 2 \int_\bv{q} \left[Z_2^E(\bv{q},\bv{k}-\bv{q})\right]^2 \PL(q) \PL(|\bv{k}-\bv{q}|) \nonumber\\
    &= P_{EE}^{(22)}(k,\mu;\tilde{b}_{2,a}) 
    + \frac{(1-\mu^2)^2}{2}fb_K^2  \mcal{I}_{\delta,uu} 
    + (1-\mu^2)fb_K  \sum_{a=1}^3 \tilde{b}_{2,a} \mcal{I}_{uu,Q_a}^E 
    + \frac{(1-\mu^2)^2}{4} f^2b_K^2\mcal{I}_{uu,uu}, \label{eq:PEE_22_rsd} \\
    &\tP_{BB}^{(22)}(k,\mu) 
    \equiv 2 \int_\bv{q} \left[Z_2^B(\bv{q},\bv{k}-\bv{q})\right]^2 \PL(q) \PL(|\bv{k}-\bv{q}|) \nonumber\\
    &= P_{BB}^{(22)}(k,\mu;\tilde{b}_{2,a}). \label{eq:PBB_22_rsd}
\end{align}
Here, the real-space spectra in Eqs.~\eqref{eq:PgE_22}--\eqref{eq:PBB_22} are evaluated with the shifted coefficients in \eq{eq:shape_bias_shift_rsd}. 
We extend the definition of $\mcal{I}_{X,Q_a}^E$ in \eq{eq:def_mcalI_OQaE} to $X\in\{\theta,u\delta,uu\}$. 
As shown in Appendix~\ref{subsubapp:22_integral_shape}, all the additional integrals reduce to the real-space loop integrals and $\mcal{I}_\mrm{RSD}$, i.e., no new shape-specific integrals need to be introduced.

\subsubsection{The (13)-type loops} \label{subsubsec:ps_1loop_rsd_13loop} 
The third-order redshift-space kernels are explicitly listed in Appendix~\ref{subapp:PT_kernel_rsd}. 
Combining these with the renormalized real-space contributions, we obtain the following spectra (see Appendix~\ref{subapp:13_integral} for a derivation). 
For the density auto-power spectrum, 
\begin{align}
    \tP_\mrm{gg}^{(13)}(k,\mu) 
    &\equiv 
    6 \PL(k) Z_1^\mrm{g}(\bv{k}) \int_\bv{q} Z_3^\mrm{g}(\bv{k},\bv{q},-\bv{q}) \PL(q) \nonumber\\
    &= 
    (b_1+f\mu^2)^2(1+2f\mu^2)P_m^{(13)}(k) 
    -
    (b_1+f\mu^2)^2f^2\mu^2 \sigma_{v,\mrm{L}}^2k^2\PL(k) \nonumber\\
    &\quad+
    2(b_1+f\mu^2)
    \left[
    b_{\mcal{G}_2}+\frac25b_{\Gamma_3}
    + 
    \frac{f\mu^2}{60} \left\{8(1-b_1)+f(9+\mu^2)\right\}
    \right] 
    \mcal{F}_{\mcal{G}_2}(k). 
    \label{eq:Pgg_13_rsd}
\end{align}
For the density--shape cross-power spectrum, 
\begin{align} 
    \tP_{\mrm{g}E}^{(13)}(k,\mu) 
    &\equiv 
    3 \PL(k) \left[Z_1^\mrm{g}(\bv{k}) \int_\bv{q} Z_3^{E}(\bv{k},\bv{q},-\bv{q}) \PL(q) + Z_1^E(\bv{k}) \int_\bv{q} Z_3^\mrm{g}(\bv{k},\bv{q},-\bv{q}) \PL(q)\right] \nonumber\\
    &= 
    \frac{1-\mu^2}{2}
    \Bigg[
    b_K(b_1+f\mu^2)(1+2f\mu^2)P_m^{(13)}(k) 
    -
    b_K(b_1+f\mu^2)f^2\mu^2 \sigma_{v,\mrm{L}}^2k^2\PL(k) \nonumber\\
    &\quad\quad\quad\quad\quad+
    \Bigg\{
    b_K\left(b_{\mcal{G}_2}+\frac25b_{\Gamma_3}\right)
    + (b_1+f\mu^2)\bar{b}_{3,1} \nonumber\\
    &\quad\quad\quad\quad\quad\quad\quad+ 
    \frac{f\mu^2}{60} 
    \left(
    b_K(8-9b_1+9f) - \frac74(b_1+f\mu^2)(b_{2,2}-b_{2,3})
    \right)
    \Bigg\} \mcal{F}_{\mcal{G}_2}(k) \nonumber\\
    &\quad\quad\quad\quad\quad+
    (b_1+f\mu^2)
    \Bigg\{
    \bar{b}_{3,2}
    + \frac{f\mu^2}{3}
    \left(b_K + \frac78(b_{2,2}-b_{2,3})\right)
    \Bigg\} \mcal{F}_{Q_1}(k)
    \Bigg], \label{eq:PgE_13_rsd}
\end{align}
and for the shape auto-power spectrum, we obtain 
\begin{align}
    \tP_{EE}^{(13)}(k,\mu) 
    &\equiv 
    6 \PL(k) Z_1^E(\bv{k}) \int_\bv{q} Z_3^E(\bv{k},\bv{q},-\bv{q}) \PL(q) \nonumber\\
    &= 
    \frac{(1-\mu^2)^2}{4}b_K
    \Bigg[
    b_K(1+2f\mu^2)P_m^{(13)}(k) 
    -
    b_Kf^2\mu^2
    \sigma_{v,\mrm{L}}^2k^2\PL(k) \nonumber\\ 
    &\quad\quad\quad\quad\quad\quad\quad\quad+
    2\left\{
    \bar{b}_{3,1}
    - 
    \frac{f\mu^2}{60}
    \left(b_K + \frac74 (b_{2,2}-b_{2,3})\right)
    \right\} \mcal{F}_{\mcal{G}_2}(k) \nonumber\\
    &\quad\quad\quad\quad\quad\quad\quad\quad+
    2\left\{
    \bar{b}_{3,2}
    + 
    \frac{f\mu^2}{3}
    \left(b_K + \frac78 (b_{2,2}-b_{2,3})\right)
    \right\} \mcal{F}_{Q_1}(k) 
    \Bigg].
    \label{eq:PEE_13_rsd}
\end{align}
These expressions involve only the real-space loop functions 
$P_m^{(13)}$, $\mcal{F}_{\mcal{G}_2}$, and $\mcal{F}_{Q_1}$, with the moment $\sigma_{v,\mrm{L}}^2$ defined in \eq{eq:13_zero_lag_moments}. 
The reductions are derived in Appendix~\ref{subapp:13_integral}.
Since $Z_1^B=K_1^B=0$, the $(13)$-type contribution to the $B$-mode auto-power spectrum in redshift space also vanishes as it does in real space: 
\begin{align}
    \tP_{BB}^{(13)}(k,\mu)=0.
    \label{eq:PBB_13_rsd}
\end{align}

We use the FFTLog-based method in Ref.~\cite{Simonovic+2018:FFTLog} to compute the loop integrals in Sections~\ref{subsubsec:ps_1loop_rsd_22loop} and \ref{subsubsec:ps_1loop_rsd_13loop}.
We expand the linear matter power spectrum as a sum of power laws with complex exponents and evaluate the corresponding loop integrals analytically. 
This reduces the calculation to operations with cosmology-independent matrices and vectors for the $(22)$- and $(13)$-type loops, respectively.

\subsection{EFT model: counterterms and stochastic contributions} \label{subsec:ps_1loop_rsd_ctr_stoch} 
We first construct the EFT model. 
Here we discuss counterterms and stochastic contributions that account for small-scale effects on large scales. 
We first consider the counterterms linear in $\delL$ that contribute at order $k^2\PL$ to the one-loop power spectra.
We derive the counterterms from the redshift-space mapping in Appendix~\ref{subsubapp:UVlim_3rd_kernels}. 
Combining them with the real-space counterterms, we obtain\footnote{
The counterterm for $\partial_\parallel^2[u_\parallel^2\Sij]$ in \eq{eq:tSij_exp_cubic} can also contain the following terms allowed by symmetry: 
\begin{align*}
    \mcal{D}_{ij}(\hbv{n})\delL,\qquad
    \TF{\h{n}_k\h{n}_{(i}K_{\mrm{L},j)k}}.
\end{align*}
However, the contraction at one-loop order generates only the tensor structure $\partial_\parallel^2 K_{\mrm{L},ij}$ (see Appendix~\ref{subsubapp:UVlim_3rd_kernels}). 
Furthermore, even if the two tensor structures above were present, both would vanish under $E/B$ projection. 
} 
\begin{align}
    \tdelg^\mrm{ctr} 
    &= 
    c_{\mrm{g},0} \nabla^2\delL + c_{\mrm{g},2} \partial_\parallel^2 \delL + c_{\mrm{g},4} \frac{\partial_\parallel^4}{\nabla^2} \delL, \label{eq:tdelg_str}\\
    \tSij^\mrm{ctr} 
    &= 
    c_{S,0} \nabla^2 K_{\mrm{L},ij} 
    + c_{S,2} \partial_\parallel^2 K_{\mrm{L},ij} 
    + c_{S,2}^\parallel \TF{\h{n}_{(i} \partial_{j)}} \partial_\parallel \delL. \label{eq:tSij_str}
\end{align} 
The last term in $\tSij^\mrm{ctr}$ vanishes after $E/B$ projection, since
$M_{ij}^{E,B}\h{n}_i=M_{ii}^{E,B}=0$. 
Thus, three counterterm parameters are needed for the density
field and two for the projected shape field at this order. 
Cross-correlating these fields with the linear fields, we obtain 
\begin{align}
    \tP_\mrm{gg}^\mrm{ctr}(k,\mu) 
    &= 
    -2(b_1+f\mu^2)\left(c_{\mrm{g},0} + c_{\mrm{g},2}\mu^2 + c_{\mrm{g},4}\mu^4 \right) k^2 \PL(k), \\
    \tP_{\mrm{g}E}^\mrm{ctr}(k,\mu) 
    &= 
    -\frac{1-\mu^2}{2}\big[
    b_K\left(c_{\mrm{g},0} + c_{\mrm{g},2} \mu^2 + c_{\mrm{g},4} \mu^4 \right) \nonumber\\
    &\qquad\qquad\qquad+ 
    (b_1+f\mu^2)(c_{S,0} + c_{S,2}\mu^2)
    \big]k^2 \PL(k), \\
    \tP_{EE}^\mrm{ctr}(k,\mu) 
    &= -\frac{(1-\mu^2)^2}{2} b_K (c_{S,0} + c_{S,2}\mu^2) k^2 \PL(k), \\
    \tP_{BB}^\mrm{ctr}(k,\mu) 
    &= 0. 
\end{align}

For the stochastic power spectra, we keep contributions up to $k^2$ as in the real-space case. 
Assuming the locality of correlation functions with a characteristic scale in redshift space $\tilde{R_*}$, we obtain the power spectrum by a Taylor expansion of its Fourier transform for $k\tilde{R_*}\ll 1$: 
\begin{align}
    \tP_{XY}^\mrm{local}(\bv{k},\hbv{n}) 
    = \int \mrm{d}\bv{r}~ \xi_{XY}^\mrm{local}(\bv{r},\hbv{n}) e^{-i\bv{k}\cdot\bv{r}} 
    \simeq C_0^{XY}(\hbv{n}) + k_ak_b C_{2;ab}^{XY}(\hbv{n}) + \cdots, 
    \label{eq:PXY_local_exp}
\end{align}
where $X,Y\in\{\mrm{g},S\}$ and we suppressed the tensor indices of the shape field. 
The terms odd in $\bv{k}$ vanish by symmetry. 
The coefficient tensors are built from both $\delK_{ij}$ and $\h{n}_i$ in redshift space, but from $\delK_{ij}$ alone in real space.  
For the density auto-power spectrum, since $C_{2;ab}^\mrm{gg}$ is a linear combination of $\delK_{ab}$ and $\h{n}_{a}\h{n}_{b}$, we obtain 
\begin{align}
    \tP_\mrm{gg}^\mrm{stoch}(k,\mu) 
    = N_{\mrm{gg},0} + \left(N_{\mrm{gg},2} + N_{\mrm{gg},2\mu}\mu^2 \right)k^2. \label{eq:Pgg_stoch_rsd}
\end{align} 
These three coefficients absorb the UV dependence of the $(22)$-type loops of the redshift-space density field through order $k^2$ \citep{Senatore&Zaldarriaga2014:EFTofLSS_RSD,Perko+2016:EFTofLSS_RSD_biased}. 

For power spectra involving projected shapes, the coefficient tensors can be written as 
$C_{0,ij}^{\mrm{g}S}$ and $C_{2,ij;ab}^{\mrm{g}S}$ for the
cross spectrum, and $C_{0,ijkl}^{SS}$ and $C_{2,ijkl;ab}^{SS}$ for the auto spectrum. 
Since the $E$- and $B$-mode projectors in \eqs{eq:def_M_E}{eq:def_M_B}  are traceless and transverse to the LOS, only structures built from $\delK_{ij}$ survive on the shape indices. 
Therefore, the strictly local power spectra with projected shapes have the same functional forms as in real space. 

The UV contributions of the $(22)$-type loop integrals in redshift space are derived in Appendix~\ref{subsubapp:UVlim_2nd_kernels}. 
For the auto spectra, the redshift-space mapping only shifts the coefficients of the real-space terms, so we use \eqs{eq:PEE_stoch_real}{eq:PBB_stoch_real} without modification, i.e., $\tP_{AA}^\mrm{stoch}=P_{AA}^\mrm{stoch}$ for $A\in\{E,B\}$.  
However, for the cross power spectrum, the (22) terms generate an additional contribution proportional to $k_\parallel^2 \mcal{D}_{ij}(\bv{k})$, which is non-analytic in $\bv{k}$. 
Therefore, we write 
\begin{align}
    \tP_{\mrm{g}E}^\mrm{stoch}(k,\mu)
    =
    \frac{1-\mu^2}{2} \left(N_{\mrm{g}S,2} + N_{\mrm{g}S,2\mu} \mu^2\right) k^2, \label{eq:PgE_stoch_rsd}
\end{align} 
which adds one parameter, $N_{\mrm{g}S,2\mu}$, to the real-space parametrization. 

\subsection{EFT model: summary and IR resummation} \label{subsec:ps_1loop_rsd_summary} 
Before IR resummation, the one-loop EFT power spectra can be written as finite sums of radial basis functions $\mcal{B}_i(k)$ and even powers of $\mu$: 
\begin{align}
    \tP_X(k,\mu)
    &= A_X(\mu)\sum_i\sum_{n=0,2,\ldots,n_X}
    c_X^{(i,n)}\mcal{B}_i(k)\mu^n,
    \label{eq:PX_eft_tot}
\end{align}
where $X\in\{\mrm{gg},\mrm{g}E,-,BB\}$, and $\tP_- \equiv (\tP_{EE} - \tP_{BB})/2$. 
For these four spectra, the angular prefactors $A_X(\mu)$ are
$\{1,(1-\mu^2)/2,(1-\mu^2)^2/4,1\}$ and the maximum powers $n_X$ are $\{8,6,4,2\}$, in the same order. 
The basis functions and their coefficients are listed in Tables~\ref{tab:galaxy_rsd_coefficients}--\ref{tab:ia_rsd_BB_coefficients}.
The $EE$ power spectrum can be obtained by 
$\tP_{EE} = \tP_{BB} + 2\tP_-$. 
The angular prefactors in $\tP_{\mrm{g}E}$ and $\tP_{-}$ arise from the geometric projection of the shape field and remain unchanged under the mapping from real to redshift space \citep{Akitsu+2026:IAinRS_I}. 
In total, the EFT model for the four spectra has 24 free parameters:
10 bias parameters, 5 counterterm parameters, and 9 stochastic parameters. 
Using the FFTLog method described in Section~\ref{subsec:loops}, we combine the resulting radial basis functions with the tabulated coefficients to obtain the power spectra. 

Finally, we account for the damping of BAO in these power spectra by resumming the leading effect of large-scale bulk flows \citep{Senatore&Zaldarriaga2015:IRresum_EFT,Baldauf+2015:IRresum_EP_BAO}. 
For the IR resummation, we follow Ref.~\cite{Chudaykin+2020:CLASS-PT} and write the one-loop power spectrum as 
\begin{align}
    \tP_X^\mrm{IR}
    = 
    \tP_{X,\mrm{nw}}
    + e^{-D} \left[\tP_{X,\mrm{full}} - \tP_{X,\mrm{nw}}\right] 
    + 
    D e^{-D} \left[\tP_{X,\mrm{full}}^\mrm{tree} - \tP_{X,\mrm{nw}}^\mrm{tree}\right],  
\end{align}
where $\tP_{X,\mrm{full}}$ and $\tP_{X,\mrm{nw}}$ are the one-loop power spectra in \eq{eq:PX_eft_tot} computed with the full and no-wiggle linear matter power spectra, respectively. 
We obtain the no-wiggle power spectrum with the discrete sine transform method of Refs.~\citep{Hamann+2010:PlinWiggle_split,Baumann+2018:LightRelics_LSS}. 
The damping exponent is given by 
\begin{align}
    D(k,\mu) = k^2
    \left[
    \{1+f(2+f)\mu^2\} \Sigma^2 
    + 
    f^2\mu^2 (\mu^2-1) \delta\Sigma^2 
    \right], 
\end{align}
where 
\begin{align}
    \Sigma^2 &= \int_0^{k_\mrm{IR}} \frac{\mrm{d}q}{6\pi^2} P_{\mrm{L,nw}}(q) 
    \left[1 - j_0(qr_\mrm{BAO}) + 2j_2(qr_\mrm{BAO})\right], \\
    \delta\Sigma^2 &= \int_0^{k_\mrm{IR}} \frac{\mrm{d}q}{2\pi^2} P_{\mrm{L,nw}}(q) 
    j_2(qr_\mrm{BAO}). 
\end{align}
Here, $j_\ell$ is the spherical Bessel function of order $\ell$. 
We set the cutoff scale to $k_\mrm{IR}=0.2 \hMpci$ and the BAO scale to $r_\mrm{BAO} = 105 \hiMpc$.

\subsection{TNS-based model} \label{subsec:tns_model} 
Our second model is based on the TNS prescription \citep{Taruya+2010:TNS}.
In contrast to the EFT model, it does not expand the effect of small-scale LOS velocities on the power spectra in powers of $k_\parallel$, but keeps part of it in a nonperturbative form. 
This model builds on the one developed in our previous work for the density--shape cross spectrum with linear bias \citep{Taruya+2024:TNSofIA_1}. 
We start from the expression for the power spectra given by the redshift-space mappings in \eqs{eq:delg_rsd_formal_Fourier}{eq:Sij_rsd_formal_Fourier},
\begin{align}
    \tP_{XY}(\bv{k})
    &=
    \int \mrm{d}\bv{r}~ e^{-i\bv{k}\cdot\bv{r}}
    \avrg{e^{j\Delta u_\parallel}X(\bv{x})Y(\bv{x}')},
\end{align}
where $\bv{r}=\bv{x}-\bv{x}'$, $j=-ik\mu$, 
$\Delta u_\parallel=u_\parallel(\bv{x})-u_\parallel(\bv{x}')$, and 
$X,Y \in \{\delg-\partial_\parallel u_\parallel, \Sij\}$
We suppress the tensor indices and use the real-space bias expansions in \eqs{eq:bias_exp_delg}{eq:bias_exp_Sij} for $\delg$ and $\Sij$, respectively. 
Writing $X=X(\bv{x})$ and $Y=Y(\bv{x}')$, the ensemble average can be expressed exactly in terms of cumulants \citep{Taruya+2010:TNS}:
\begin{align}
    \avrg{e^{j\Delta u_\parallel}XY}
    =
    \exp\!\left[\avrg{e^{j\Delta u_\parallel}}_{\mrm{c}}\right]
    \left[
    \avrg{e^{j\Delta u_\parallel}XY}_{\mrm{c}}
    +
    \avrg{e^{j\Delta u_\parallel}X}_{\mrm{c}}
    \avrg{e^{j\Delta u_\parallel}Y}_{\mrm{c}}
    \right],
    \label{eq:TNS_cumulant}
\end{align} 
where the subscript $\mrm{c}$ denotes a connected part. 
At leading order, the exponent in the prefactor includes 
$\avrg{(\Delta u_\parallel)^2}_{\mrm{c}} = 2\avrg{u_\parallel^2} - 2 \avrg{u_\parallel(\bv{x})u_\parallel(\bv{x}')}$,
which consists of a one-point part and a correlation between the two positions.
In our TNS-based model, we keep the one-point part of this factor unexpanded and evaluate the remaining terms up to one loop \citep{Taruya+2024:TNSofIA_1}. 
Since the one-point part is dominated by nonlinear velocities on small scales, we replace it with a phenomenological damping function,
\begin{align}
    \exp\!\left[j^2\avrg{u_\parallel^2}\right]
    \to
    D_{\mrm{FoG}}(kf\mu\sigma_v), 
    \label{eq:TNS_Gaussian_FoG}
\end{align} 
where the velocity dispersion $\sigma_v$ is a free parameter. 
In this work, we adopt a Gaussian damping form \citep{Okumura+2024:NLRSDofIA,Taruya+2024:TNSofIA_1}: 
$D_{\mrm{FoG}}(kf\mu\sigma_v) \equiv \exp\!\left[-(kf\mu\sigma_v)^2\right]$. 
The integrand then becomes
\begin{align}
    \avrg{e^{j\Delta u_\parallel}XY}
    &\simeq D_{\mrm{FoG}}(kf\mu\sigma_v)
    \Big[
    \avrg{XY}_{\mrm{tree+1\mathchar`-loop}}
    + j\avrg{\Delta u_\parallel XY}_{\mrm{c,tree}}
    \nonumber\\
    &\quad\quad\quad
    + j^2\avrg{\Delta u_\parallel X}_{\mrm{tree}}
         \avrg{\Delta u_\parallel Y}_{\mrm{tree}}
    - j^2\avrg{u_\parallel(\bv{x})u_\parallel(\bv{x}')}_{\mrm{tree}}
         \avrg{XY}_{\mrm{tree}}
    \Big].
    \label{eq:TNS_integrand}
\end{align}
Note that the last term retains the velocity correlation between the two positions, which is omitted in the original TNS model.
As shown in Appendix~\ref{app:relation_TNS}, the resulting deterministic spectra for $XY\in\{\mrm{gg},\mrm{g}E,EE,BB\}$ can be written as 
\begin{align}
    \tP_{XY}^{\mrm{TNS}}(k,\mu)
    = D_{\mrm{FoG}}(kf\mu\sigma_v)
    \left[
    \tP_{XY}^{\mrm{tree+1\mathchar`-loop}}(k,\mu)
    + (kf\mu\sigma_{v,\mrm{L}})^2\tP_{XY}^{(11)}(k,\mu)
    \right].
    \label{eq:TNS_EPT_relation}
\end{align}
The last term in the square brackets removes the expanded one-point velocity term already present in Eqs.~\eqref{eq:Pgg_13_rsd}, \eqref{eq:PgE_13_rsd}, and \eqref{eq:PEE_13_rsd}. 
Here, $\sigma_{v,\mrm{L}}^2$ is fixed by the linear power spectrum, as given in \eq{eq:13_zero_lag_moments}. 
We compute the terms in the square brackets from the same radial basis functions as in the EFT model. 
Note that we include the higher-derivative terms, which take the same form as the isotropic EFT counterterms $c_{\mrm{g},0}$ and $c_{S,0}$. 
We do not include the redshift-space counterterms $c_{\mrm{g},2}$, $c_{\mrm{g},4}$, and $c_{S,2}$, since the damping factor describes the corresponding small-scale velocity effects in this model. 
For the stochastic part, we adopt the same terms as the EFT model, Eqs.~\eqref{eq:Pgg_stoch_rsd}, \eqref{eq:PgE_stoch_rsd}, and \eqref{eq:PEE_stoch_real}--\eqref{eq:PBB_stoch_real}. 
We add them outside the damping factor, since contact terms contribute only at zero separation, where $\Delta u_\parallel=0$, and are thus not damped by the velocity dispersion. 
In total, this model has 22 free parameters:
12 bias parameters, 9 stochastic parameters, and the velocity dispersion $\sigma_v$. 
This model thus has the single parameter, $\sigma_v$, instead of the three redshift-space counterterms of the EFT model. 

\section{Comparison to \texorpdfstring{$N$}{N}-body simulations} \label{sec:nbody_sims} 
%
\begin{figure}
    \centering
    \includegraphics[width=1\columnwidth]{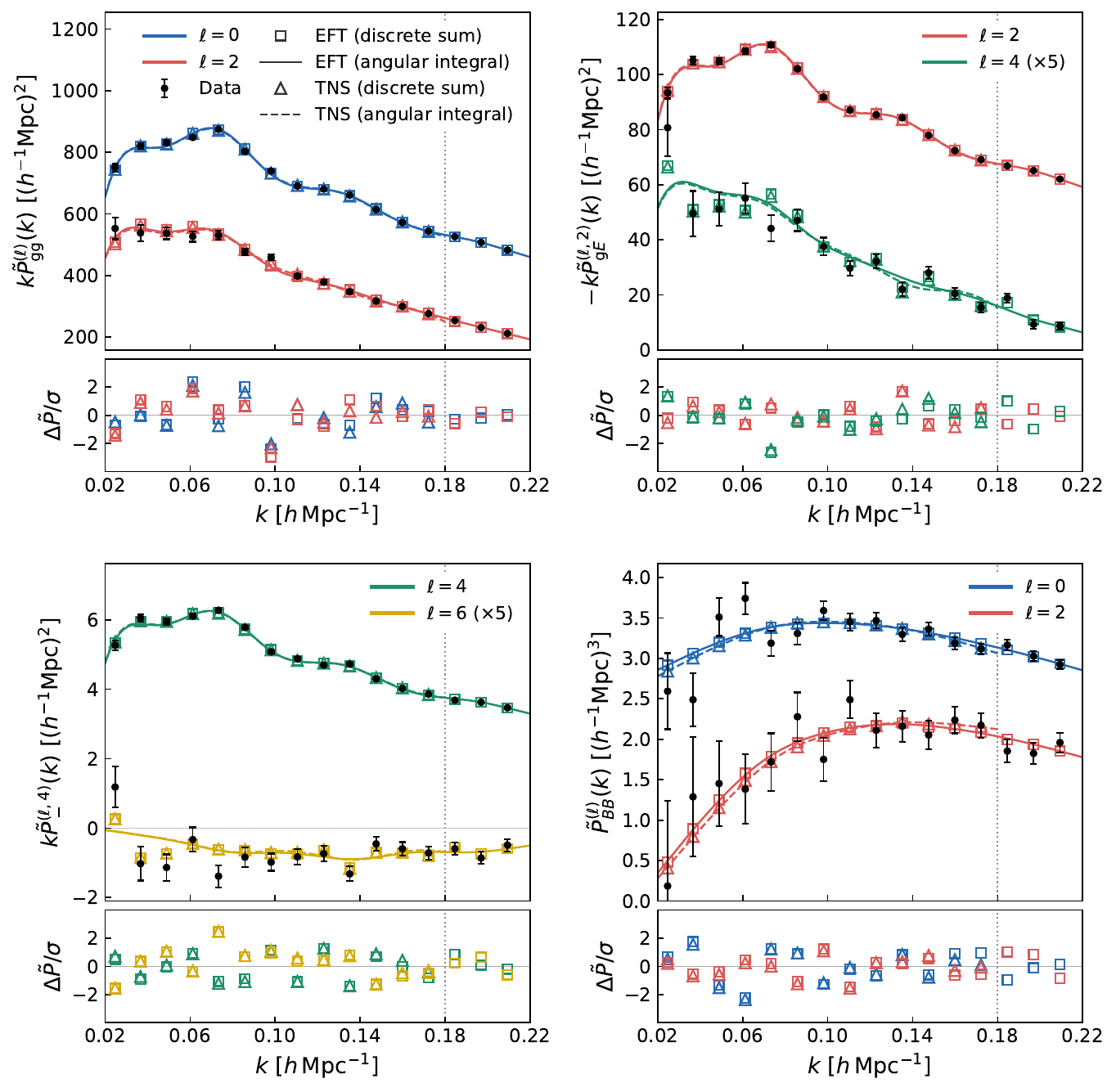}
    \caption{Comparison of the EFT and TNS-based models with the measured power spectra from low-mass halos, 
    $10^{12}\leq M_\mrm{vir}<10^{13}h^{-1}M_\odot$, at $z=0.484$. 
    In reading order, the panels show $kP_{\mrm{gg}}$, $-kP_{\mrm{g}E}$, $kP_-$, and $P_{BB}$. 
    Black points show the measurements. 
    Squares (EFT) and triangles (TNS) show the predictions averaged over the discrete Fourier modes in each shell. 
    Solid (EFT) and dashed (TNS) lines show predictions obtained by integrating over $\mu$, using the same MAP parameters. 
    We fit the EFT and TNS models up to $k_{\max}=0.22$ and $0.18\hMpci$, respectively. 
    The vertical dotted line corresponds to the $k_\mrm{max}$ for the TNS model. 
    The lower panels show
    $(\tilde{P}_{\mrm{model}}-\tilde{P}_{\mrm{data}})/\sigma$, where $\sigma$ is the measurement error. 
    }
    \label{fig:low_mass}
\end{figure}
\begin{figure}
    \centering
    \includegraphics[width=1\columnwidth]{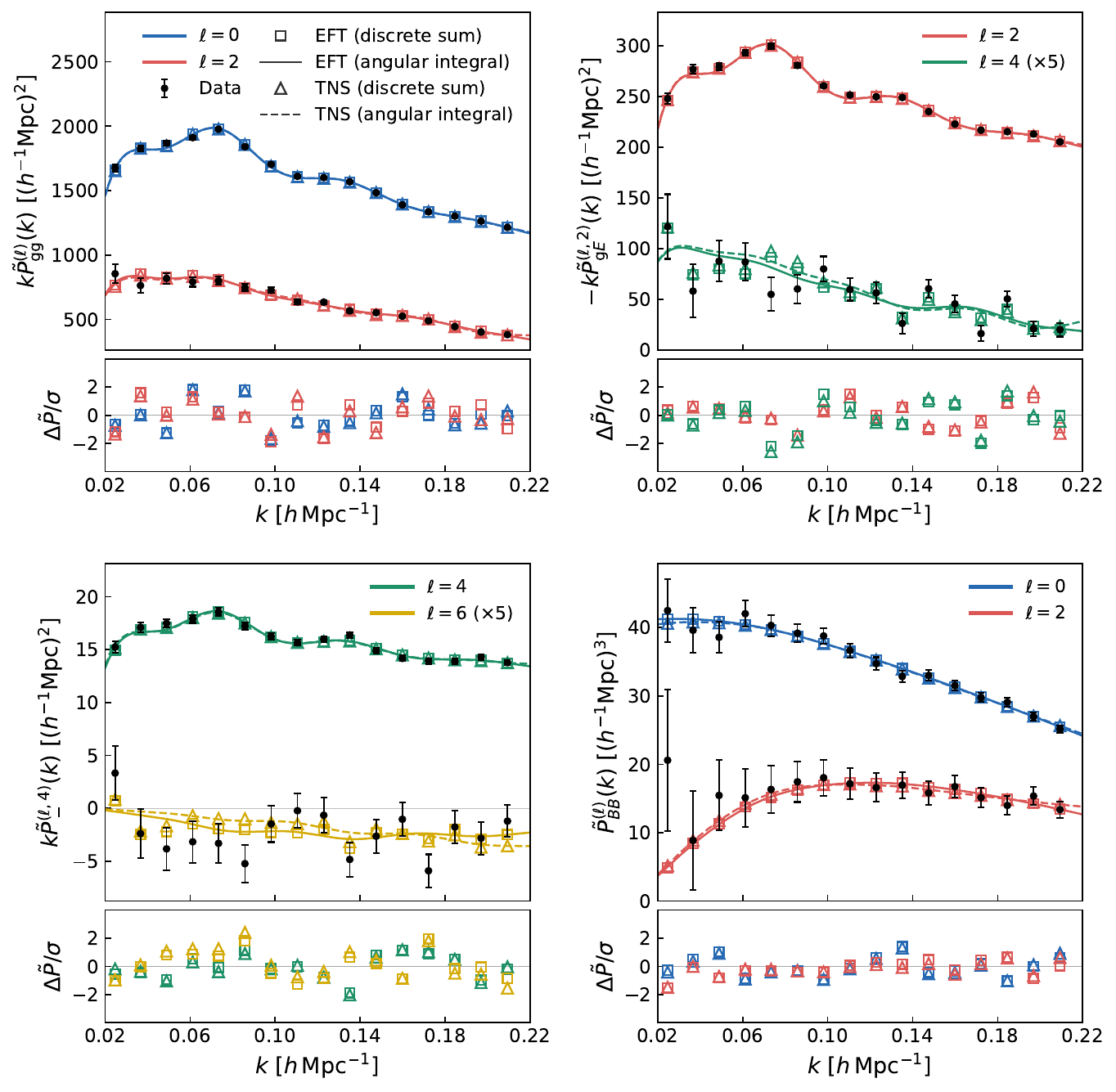}
    \caption{Same as Fig.~\ref{fig:low_mass}, but for the high-mass halo sample, $M_\mrm{vir}\geq10^{13}h^{-1}M_\odot$, at $z=0.484$.
    Both the EFT and TNS models are fitted up to
    $k_{\max}=0.22\hMpci$. 
    }
    \label{fig:high_mass}
\end{figure}
In this section, we compare our model predictions with power spectra measured in $N_r=20$ realizations of the high-resolution \texttt{Dark Quest} ($\texttt{HR}$) simulation suite \citep{Nishimichi+2019:DarkQuestI}. 
These simulations were run with a flat $\Lambda$CDM cosmology with \textit{Planck} TT,TE,EE+lowP in Table~3 of Ref.~\cite{Planck2015_cosmo}. 
Each simulation contains $N=2048^3$ particles in a cubic box of comoving side length $L=1h^{-1}\mrm{Gpc}$. 
The corresponding particle mass is $1.02\times10^{10}h^{-1}M_\odot$. 
We use halo positions and virial masses from the \texttt{Rockstar}
catalogs \citep{Behroozi+2013:Rockstar}. 
At $z=0.484$, we consider a low-mass sample with
$10^{12}\leq M_{\mrm{vir}}<10^{13}h^{-1}M_\odot$
and a high-mass sample with
$M_{\mrm{vir}}\geq10^{13}h^{-1}M_\odot$. 
The low-mass sample has a higher number density, allowing us to test the models with lower shot noise. 
The high-mass sample covers halo masses typical of luminous red galaxy hosts \citep{Guo+2014:BOSS_CMASS}. 
We use the shape catalogs of Ref.~\cite{Kurita+2021:IA_nbody}. 
The projected halo shapes were obtained from reduced inertia tensors, computed iteratively from the positions of member particles. (see Appendix~C of that work for details). 

We use the halo positions, velocities, and shapes in these catalogs to construct the halo density and shape fields in redshift space under the distant observer approximation. 
We measure the power spectrum multipoles by weighting each Fourier mode with an associated Legendre polynomial \citep{Kurita&Takada2022:AnalysisIAPS}.
For the $i$th spherical $k$-shell, we estimate the multipoles as \citep{Kurita+2021:IA_nbody} 
\begin{align}
    \hat{\tP}_{XY}^{(\ell,m)}(k_i)
    &=
    \frac{(2\ell+1)}{VN_i}
    \sum_{\bv{k}\in\mrm{shell}\,i}
    \tilde{X}(\bv{k})\tilde{Y}(-\bv{k})
    \bar{\mcal{L}}_\ell^m(\mu_{\bv{k}}), \quad 
    k_i=\frac{1}{N_i}
    \sum_{\bv{k}\in\mrm{shell}\,i}|\bv{k}|. 
\end{align}
Here, $V=L^3$ is the box volume, $N_i$ is the number of modes in the shell, and $k_i$ is their mean wavenumber. 
$\bar{\mcal{L}}_\ell^m\equiv\sqrt{(\ell-m)!/(\ell+m)!}\,\mcal{L}_\ell^m$ is the rescaled associated Legendre polynomial $\mcal{L}_\ell^m$. 
We use $m=0$ for $\tP_{\mrm{gg}}$ and $\tP_{BB}$, $m=2$ for $\tP_{\mrm{g}E}$, and $m=4$ for $\tP_-$. 
We set the $k$-shell width to $\Delta k \simeq 0.012\hMpci$. 

We compute the covariance matrix of the mean over the $N_r=20$ realizations in the Gaussian approximation following Ref.~\citep{Taruya+2024:TNSofIA_1}.%
\footnote{Ref.~\citep{Taruya+2024:TNSofIA_1} demonstrated that the Gaussian approximation provides a reasonable description of the diagonal covariance elements at $k\lesssim0.3\hMpci$, based on the same set of simulations. While the accuracy of the off-diagonal covariance elements was not explicitly tested, this result supports the use of the Gaussian approximation as a reasonable approximation in our analysis.}
The covariance between two multipoles is then
\begin{align}
    \mrm{Cov}
    \left[
    \tP_{XY}^{(\ell,m)}(k_i),
    \tP_{ZW}^{(\ell',m')}(k_j)
    \right]
    &=
    \frac{\delK_{ij}(2\ell+1)(2\ell'+1)}{N_rN_i}
    \int_{-1}^{1}\frac{\mrm{d}\mu}{2}
    \bar{\mcal{L}}_\ell^m(\mu) \bar{\mcal{L}}_{\ell'}^{m'}(\mu) \nonumber\\
    &\quad\quad\times
    \left[
    P_{XZ}^\mrm{tot}(k_i,\mu) P_{YW}^\mrm{tot}(k_i,\mu)
    + P_{XW}^\mrm{tot}(k_i,\mu) P_{YZ}^\mrm{tot}(k_i,\mu)
    \right]. \label{eq:gauss_cov} 
\end{align}
For $P^\mrm{tot}$, we use the linear redshift-space spectra and include
noise in the auto-power spectra. 
We adopt Poisson shot noise $1/\bar{n}_{\rm g}$ for the density. 
We estimate the shape noise from the $k\to0$ limit of the redshift-space $BB$ monopole and use this value for both $E$ and $B$ in the covariance. 
For the low- and high-mass samples, respectively,
$\bar{n}_{\mrm{g}}=\{4.18\times10^{-3},4.01\times10^{-4}\}(\hMpci)^3$
and the rms ellipticity per component is
$\sigma_e=\{0.329,0.314\}$.
The inferred constant shape noise is about $\{12,16\}\%$ higher than the Poisson value $\sigma_e^2/\bar{n}_{\mrm{g}}$.
We determine the fiducial biases in the covariance by fitting the real-space density auto and density--$E$ cross spectra over $0.02\leq k\leq0.08\hMpci$.
This gives $(b_1^{\mrm{fid}},b_K^{\mrm{fid}})=(1.20,-0.254)$ for the low-mass sample and $(1.95,-0.431)$ for the high-mass sample. 

We fit the two models to the same eight redshift-space multipoles: 
$\ell=0,2$ for $\tP_{\mrm{gg}}$ and $\tP_{BB}$ ($m=0$), $\ell=2,4$ for $\tP_{\mrm{g}E}$ ($m=2$), and $\ell=4,6$ for $\tP_-$ ($m=4$). 
For each spectrum, these are the two lowest multipoles in the associated Legendre basis. 
In each $k$-bin, we average the model prediction over the same Fourier modes and with the same multipole weights as in the measurements. 
This is particularly important for the higher multipoles of $\tP_{\mrm{g}E}$ and $\tP_-$, which are sensitive to the uneven sampling of directions in a finite shell. 

We vary the growth rate $f$ together with the nuisance parameters listed in Sections~\ref{subsec:ps_1loop_rsd_summary} and \ref{subsec:tns_model} (24 for the EFT model and 22 for the TNS model). 
We use dimensionless counterterms and stochastic coefficients, defined by
\begin{align}
    c_{X,n} = \frac{\hat c_{X,n}}{k_{\mrm{NL}}^2},
    \qquad
    N_{XY,0} = \bar{P}_{XY}\left(1+\alpha_{XY,0}\right),
    \qquad
    N_{XY,a} = \frac{\bar{P}_{XY}\,\alpha_{XY,a}}{k_{\mrm{NL}}^2},
    \label{eq:dimless_params}
\end{align}
with $k_{\mrm{NL}}=0.39\hMpci$.
Here, $X\in\{\mrm{g},S\}$, and the stochastic coefficients are normalized by the Poisson amplitudes
$\bar{P}_{\mrm{gg}}=1/\bar{n}_{\mrm{g}}$, $\bar{P}_{\mrm{g}S}=\sigma_e/\bar{n}_{\mrm{g}}$, and $\bar{P}_{SS}=\sigma_e^2/\bar{n}_{\mrm{g}}$.
The constant terms $N_{XY,0}$ appear only for $XY=\mrm{gg}$ and $SS$.
The label $a$ denotes the $k^2$ terms, $a=2,2\mu$ for all three spectra and additionally $a=2\Delta$ for $SS$.
Note that the TNS model includes only $c_{\mrm{g},0}$ and $c_{S,0}$ among the counterterms but uses all the stochastic coefficients. 
We adopt uniform priors $b_1\in[0,5]$, $b_K\in[-2,2]$, and $f\in[0,2]$, and, for the TNS model, $\sigma_v\in[0,10]\,h^{-1}\mrm{Mpc}$. 
All the other parameters, i.e., the higher-order biases and the dimensionless $\hat{c}$ and $\alpha$, have independent Gaussian priors $\mcal{N}(0,20)$. 
We use the minimum $\chi^2$ to check the fit and require $p>0.05$, assuming $\nu=N_{\mrm{data}}-N_{\mrm{par}}$ degrees of freedom. 
We also check whether the posterior means of $b_1$, $b_K$, and $f$ are within $2\sigma$ of their reference values.
For $b_1$ and $b_K$, the reference values come from the real-space measurements, and we use $\sigma^2=\sigma_{\mrm{post}}^2+\sigma_{\mrm{ref}}^2$ to include both errors \citep{Bakx+2023:EFTofIAvsSims}.

In Figures~\ref{fig:low_mass} and~\ref{fig:high_mass}, we compare the
measurements with the maximum a posteriori (MAP) predictions for the two halo samples. 
Among the multipoles, the $(\ell,m)=(4,2)$ component of $\tP_{\mrm{g}E}$ and the $(\ell,m)=(6,4)$ component of $\tP_-$ vanish in real space \citep{Kurita&Takada2022:AnalysisIAPS} and arise only from RSD. 
The former appears already at linear order through the Kaiser effect \citep{Kaiser1987:RSD} on the density field. 
The latter arises only from nonlinear terms, since RSD does not affect the shape field at linear order. 
These multipoles are therefore direct probes of RSD, and the agreement for them provides a useful check of the RSD treatment in the models. 
For the low-mass sample, the EFT model passes the $\chi^2$ and parameter-recovery tests up to $k_{\max}=0.22\hMpci$ ($\chi^2_{\mrm{red}}=1.03$, $p=0.40$), and the TNS-based model passes them up to $k_{\max}=0.18\hMpci$ ($\chi^2_{\mrm{red}}=1.11$, $p=0.23$).
At these cutoffs, the figure of merit of $f$, $\mathrm{FoM}\equiv1/\sigma_f$, is about 9\% higher for the EFT model than for the TNS-based model. 
For the high-mass sample, the EFT model ($\chi^2_{\mrm{red}}=0.96$, $p=0.61$) and the TNS-based model ($\chi^2_{\mrm{red}}=1.00$, $p=0.49$) both pass these tests at the common cutoff $k_{\max}=0.22\hMpci$.
At this cutoff, the FoM is about 27\% higher for the TNS-based model than for the EFT model, likely reflecting the smaller number of free parameters. 

Our tests are limited to two halo samples at a single redshift, and we varied only the growth rate $f$ among the cosmological parameters. 
A proper validation requires varying the cosmological parameters and including the AP effect \citep{Alcock&Paczynski1979,Matsubara&Suto1996:AP} to test whether a full-shape analysis recovers the input cosmology. 
Such an analysis would also show how much the shape field helps break parameter degeneracies \citep{Taruya&Okumura2020:IA_improvement,Shim+2025:IA_DDE_full,Shim+2025:IA_DDE_Letter}.
A more complete validation would also require testing the models at other redshifts, with different shape estimators, and on more realistic mock catalogs. We leave these tests to future work.

\section{Conclusions} \label{sec:conclusions} 
In this work, we have derived the one-loop power spectra of galaxy clustering and intrinsic alignments in redshift space. 
These include the cross spectrum between the galaxy density and the $E$-mode field of projected galaxy shapes, as well as the $E$- and $B$-mode auto spectra. 
Starting from the real-space bias expansions of the galaxy density and shape fields, we expanded the redshift-space mapping and computed the tree-level and one-loop contributions in Eulerian perturbation theory.
We showed that all the loop contributions can be written in terms of the loop functions already used for real-space shape spectra and redshift-space galaxy clustering. 
We evaluated them with the FFTLog method. 

We modeled the effects of small-scale LOS velocities on the large-scale signal with two models. 
One is the EFT model, which includes all counterterms required to absorb the UV sensitivity of the one-loop integrals. 
For the one-loop power spectra of the sky-projected shape field, we found that this model requires six renormalized bias parameters and two counterterms. 
The stochastic terms, determined by locality and the UV limits of the loop integrals, require two parameters for the cross spectrum and four for the auto spectra up to order $k^2$. 
The other is the TNS-based model, in which we identified the one-point velocity dispersion from the cumulant expansion and kept it in a phenomenological damping function. 
Our previous TNS-based model \citep{Taruya+2024:TNSofIA_1} treated the cross spectrum with linear bias. 
In this work, we included nonlinear bias and also modeled the shape auto spectra. 
We compared both models with the power spectra of halos measured in $N$-body simulations and found that they fit the measurements well on mildly nonlinear scales. 

The power spectra studied in this work are built from the positions of galaxies in redshift space and projected galaxy shapes.
The positions are obtained from spectroscopic surveys, and the shapes from imaging surveys, so these spectra can be measured only by combining the two types of surveys. 
Such data are becoming available from ongoing and upcoming surveys, including Subaru HSC/PFS \citep{Takada+2014:PFS}, DESI \citep{DESI_Collab2016:DESI_PartI}, Euclid \citep{Laureijs+2011:Euclid}, LSST \citep{Mandelbaum+2018:LSST_DESC_Science_Requirement}, the Roman Space Telescope \citep{Akeson+2019:Roman}, and the 4MOST survey \citep{deJong+2019:4MOST}. 
To extract cosmological information from these precise data, accurate models on nonlinear scales are essential.
With further validation, the models presented in this work can be used in a full-shape analysis that combines the density auto, density--shape cross, and shape auto spectra.
Such an analysis will show how much the IA power spectra can break parameter degeneracies and improve cosmological constraints beyond galaxy clustering alone.
We leave this to future work. 

\acknowledgments
We thank Yosuke Kobayashi for sharing his FFTLog code, which we extended for use in this work. 
We also thank Takahiro Nishimichi for providing the $N$-body simulation data. 
This work was supported in part by the Deutsche Forschungsgemeinschaft (DFG, German Research Foundation) under Germany's Excellence Strategy - EXC-2094/2 - 390783311.  
A.T. acknowledges support from the FY2025 MUSUBIME of Kyoto University and the Japan Society for the Promotion of Science (JSPS) KAKENHI Grant Numbers JP23K25868 and JP26H02044.
T.O. acknowledges support from the Taiwan National Science and Technology Council under grant No. NSTC 114-2112-M-001-004- and No. NSTC 115-2112-M-001-059-MY3, and the Academia Sinica Investigator Project grant No. AS-IV-114-M03 for the period of 2025–2029. 

\textit{AI usage statement}: We used GPT-6, Claude Opus 5.5, and earlier GPT and Claude models to improve the numerical implementation of the parameter inference code.  
We also used them to check the grammar. 
The authors take full responsibility for all results in this paper.

\clearpage
\appendix
\section{Details of the real-space power spectra}
\label{app:details_real} 

\subsection{Third-order contributions}
\label{subapp:3rd_kernel_real} 
The third-order kernel of the galaxy density field defined in \eq{eq:bias_exp_delg_real} is given by
\begin{align}
    K_3(\bv{q}_1,\bv{q}_2,\bv{q}_3) 
    &= 
    b_1 F_3(\bv{q}_1,\bv{q}_2,\bv{q}_3) 
    + 
    \frac{b_2}2 K_{\delta^2}^{(3)}(\bv{q}_1,\bv{q}_2,\bv{q}_3) \nonumber\\
    &\quad+ 
    b_{\mcal{G}_2} K_{\mcal{G}_2}^{(3)}(\bv{q}_1,\bv{q}_2,\bv{q}_3) 
    + 
    b_{\Gamma_3} K_{\Gamma_3}^{(3)}(\bv{q}_1,\bv{q}_2,\bv{q}_3), 
    \label{eq:density_kernel_3rd_real}
\end{align} 
where $F_3$ is the third-order matter kernel, and 
\begin{align}
    K_{\delta^2}^{(3)}(\bv{q}_1,\bv{q}_2,\bv{q}_3) 
    &= \frac23 \sum_\mrm{cyc.} F_2(\bv{q}_2,\bv{q}_3),\\
    K_{\mcal{G}_2}^{(3)}(\bv{q}_1,\bv{q}_2,\bv{q}_3) 
    &= \frac23 \sum_\mrm{cyc.} \sigma^2(\bv{q}_1,\bv{q}_{23}) F_2(\bv{q}_2,\bv{q}_3), \\
    K_{\Gamma_3}^{(3)}(\bv{q}_1,\bv{q}_2,\bv{q}_3) 
    &= \frac23 \sum_\mrm{cyc.} \sigma^2(\bv{q}_1,\bv{q}_{23}) \left[F_2(\bv{q}_2,\bv{q}_3) - G_2(\bv{q}_2,\bv{q}_3) \right].
\end{align}
Here, $G_2$ is the second-order velocity-divergence kernel \citep{Bernardeau+2002:PT_review}. 
The sums run over the three cyclic permutations of $(1,2,3)$. 
We compute the $(13)$ contribution by correlating $\delg^{(3)}$ with the linear density field $\delL$. 
After renormalization, we obtain 
\begin{align}
    \avrg{\delg^{(3)}(\bv{k}) \delL(\bv{k}')}_\mrm{ren.}' 
    &=\frac{b_1}{2}P_m^{(13)}(k) + \left(b_{\mcal G_2}+\frac25b_{\Gamma_3}\right) \mcal{F}_{\mcal{G}_2}(k),  \label{eq:delg3rd_delL_contraction} 
\end{align}
where the loop functions are defined by 
\begin{align}
    P_m^{(13)}(k) 
    &\equiv 6\PL(k)\int_\bv{q} F_3(\bv{k},\bv{q},-\bv{q})\PL(q), 
    \label{eq:def_Pm13}\\
    \mcal{F}_{\mcal{G}_2}(k)
    &\equiv
    3\PL(k)\int_{\bv{q}} 
    K_{\mcal{G}_2}^{(3)}(\bv{k},\bv{q},-\bv{q})\PL(q). 
    \label{eq:def_FG2}
\end{align} 
The contribution proportional to $b_2$ is absorbed into $b_1$. 
For the configuration $(\bv{k},\bv{q},-\bv{q})$, the kernels are explicitly given by the following expressions, with $x\equiv\hbv{k}\cdot\hbv{q}$:
\begin{align}
    F_3(\bv{k},\bv{q},-\bv{q}) 
    &= 
    \frac16 \sum_{\sigma=\pm1}
    \left(\frac{5-14x^2}{9} - \sigma\frac{k}{q}x\right)
    \frac{k^2G_2(\bv{k},\sigma\bv{q})}{|\bv{k}+\sigma\bv{q}|^2} \nonumber\\
    &= \frac{k^2(1-x^2)}{63|\bv{k}-\bv{q}|^2|\bv{k}+\bv{q}|^2}\left[5(k^2+q^2)+(4k^2-14q^2)x^2\right] - \frac{x^2}{6}\frac{k^2}{q^2}, 
    \label{eq:def_F3_13config} \\
    K_{\mcal{G}_2}^{(3)}(\bv{k},\bv{q},-\bv{q}) 
    &= 
    \frac{2}{3}\left[\sigma^2(\bv{q},\bv{k}-\bv{q})F_2(\bv{k},-\bv{q}) + \sigma^2(-\bv{q},\bv{k}+\bv{q})F_2(\bv{k},\bv{q})\right]\nonumber\\
    &=
    \frac{2}{3}(x^2-1) \sum_{\sigma=\pm1}\frac{k^2F_2(\bv{k},\sigma\bv{q})}{|\bv{k}+\sigma\bv{q}|^2} 
    = 
    -\frac{20}{21}
    \frac{k^2(1-x^2)^2(k^2+q^2)}{|\bv{k}-\bv{q}|^2|\bv{k}+\bv{q}|^2}. 
    \label{eq:def_K_G2_3rd_13config}
\end{align}

For the shape field, the third-order kernel in \eq{eq:bias_exp_Sij_real} is 
\begin{align}
    K_{3,ij}(\bv{q}_1,\bv{q}_2,\bv{q}_3)
    &= b_K \mcal{D}_{ij}(\bv{q}_{123})F_3(\bv{q}_1,\bv{q}_2,\bv{q}_3) \nonumber\\
    &\quad+
    \sum_{a=1}^3b_{2,a}K_{Q_a,ij}^{(3)}(\bv{q}_1,\bv{q}_2,\bv{q}_3)
    +
    \sum_{a=1}^2b_{3,a}K_{C_a,ij}^{(3)}(\bv{q}_1,\bv{q}_2,\bv{q}_3).
    \label{eq:shape_kernel_3rd_real}
\end{align}
To write down these kernels, we denote the second- and third-order kernels of $\Pi_{ij}^{[2]}$ by 
$\pi_{ij}^{(2,2)}$ and $\pi_{ij}^{(2,3)}$, respectively \citep{Vlah+2020:IA_EFT}:
\begin{align}
    \pi_{ij}^{(2,2)}(\bv{q}_1,\bv{q}_2)
    &=
    \h{q}_{12,i}\h{q}_{12,j}F_2(\bv{q}_1,\bv{q}_2)
    - 
    \frac{\mu_{12}}{2}
    \left(\frac{q_1}{q_2}\h{q}_{1i}\h{q}_{1j}+\frac{q_2}{q_1}\h{q}_{2i}\h{q}_{2j}\right), \label{eq:pi22_kernel}\\
    \pi_{ij}^{(2,3)}(\bv{q}_1,\bv{q}_2,\bv{q}_3)
    &=2\h{q}_{123,i}\h{q}_{123,j}F_3(\bv{q}_1,\bv{q}_2,\bv{q}_3) \nonumber\\
    &\quad- 
    \frac13 \sum_\mrm{cyc.} \mu_{1,23}
    \left[
    \h{q}_{23,i}\h{q}_{23,j}\frac{q_{23}}{q_1}F_2(\bv{q}_2,\bv{q}_3)
    +
    \h{q}_{1,i}\h{q}_{1,j}\frac{q_1}{q_{23}}G_2(\bv{q}_2,\bv{q}_3)
    \right], \label{eq:pi23_kernel}
\end{align}
where $\mu_{1,23} \equiv \hbv{q}_1\cdot\hbv{q}_{23}$. 
The third-order kernels can then be obtained from the definitions of the operators: 
\begin{align}
    K_{Q_1,ij}^{(3)}(\bv{q}_1,\bv{q}_2,\bv{q}_3)
    &= \frac23 \sum_\mrm{cyc.} K_{Q_1,ij}(\bv{q}_1,\bv{q}_{23}) F_2(\bv{q}_2,\bv{q}_3), \label{eq:cubic_Q1_kernel}\\
    K_{Q_2,ij}^{(3)}(\bv{q}_1,\bv{q}_2,\bv{q}_3) 
    &= \frac7{20}\left[\TF{\pi_{ij}^{(2,3)}(\bv{q}_1,\bv{q}_2,\bv{q}_3)}
    -K_{Q_1,ij}^{(3)}(\bv{q}_1,\bv{q}_2,\bv{q}_3)\right], \label{eq:cubic_Q2_kernel}\\
    K_{Q_3,ij}^{(3)}(\bv{q}_1,\bv{q}_2,\bv{q}_3) 
    &= \frac13\sum_\mrm{cyc.} 
    \left[\mcal D_{ij}(\bv{q}_1)+\mcal D_{ij}(\bv{q}_{23})\right] 
    F_2(\bv{q}_2,\bv{q}_3) \nonumber\\
    &\quad-
    \frac7{20} \TF{\pi_{ij}^{(2,3)}(\bv{q}_1,\bv{q}_2,\bv{q}_3)}
    - 
    \frac{13}{20}K_{Q_1,ij}^{(3)}(\bv{q}_1,\bv{q}_2,\bv{q}_3), \label{eq:cubic_Q3_kernel}\\
    K_{C_2,ij}^{(3)}(\bv{q}_1,\bv{q}_2,\bv{q}_3) 
    &= \frac13 \sum_\mrm{cyc.} 
    \TF{
    \h{q}_{1i}\h{q}_{1k}\pi_{kj}^{(2,2)}(\bv{q}_2,\bv{q}_3)
    +\pi_{ik}^{(2,2)}(\bv{q}_2,\bv{q}_3)\h{q}_{1k}\h{q}_{1j}
    }, \label{eq:cubic_C2_kernel}\\
    K_{C_1,ij}^{(3)}(\bv{q}_1,\bv{q}_2,\bv{q}_3)
    &= 4 K_{C_2,ij}^{(3)}(\bv{q}_1,\bv{q}_2,\bv{q}_3)
    - 
    \frac43 \sum_\mrm{cyc.} \TF{\pi_{ij}^{(2,2)}(\bv{q}_2,\bv{q}_3)}. \label{eq:cubic_C1_kernel}
\end{align}
The second-order kernel $K_{Q_1,ij}$ is given in \eq{eq:def_KQ1ij}. 
We first consider the $(13)$ contribution from $Q_{1,ij}$.
The renormalized contraction with the linear density field is 
\begin{align}
    \avrg{Q_{1,ij}^{(3)}(\bv{k}) \delL(\bv{k}')}_\mrm{ren.}'
    = 
    3\PL(k) 
    \int_{\bv{q}}
    \left[
    K_{Q_{1},ij}^{(3)}(\bv{k},\bv{q},-\bv{q}) 
    - 
    K_{Q_{1},ij}^{(3),\mrm{UV}}(\bv{k},\bv{q},-\bv{q}) 
    \right]
    \PL(q), 
    \label{eq:Q13rd_delL_contraction}
\end{align} 
where $K_{Q_1,ij}^{(3),\mrm{UV}}$ is the leading term of the kernel in the UV limit $k\ll q$.
Its contribution is absorbed into the linear shape bias $b_K$. 
Since $\PL(q)$ is independent of $\varphi$, we can replace the kernels in
\eq{eq:Q13rd_delL_contraction} by their azimuthal averages. 
Using the azimuthal averaging formulae in Appendix~\ref{subapp:azm_avrg_13}, we obtain 
\begin{align}
    \avrg{K_{Q_{1},ij}^{(3)}}_\varphi
    &= 
    \frac{2}{3}\sum_{\sigma=\pm1}
    \avrg{K_{Q_1,ij}(-\sigma\bv{q},\bv{k}+\sigma\bv{q})}_\varphi F_2(\bv{k},\sigma\bv{q}) \nonumber\\
    &= 
    \frac{2}{3} \mcal{D}_{ij}(\bv{k}) \sum_{\sigma=\pm1} 
    \left[x^2 + \frac{q^2}{k^2}\mcal{L}_2(x) + \sigma\frac{q}{k}x\left(1+\mcal{L}_2(x)\right)\right]
    \frac{k^2F_2(\bv{k},\sigma\bv{q})}{|\bv{k}+\sigma\bv{q}|^2},  
    \label{eq:K_Q1ij_3rd_iso} \\
    \avrg{K_{Q_{1},ij}^{(3),\mrm{UV}}}_\varphi
    &= 
    \avrg{\frac{2}{3} K_{Q_1,ij}(\bv{k},\bv{q}) +  \frac{-6x^2+20}{21}\mcal{D}_{ij}(\hbv{q})}_\varphi
    =
    \frac{-9x^4+47x^2-10}{21}\mcal{D}_{ij}(\bv{k}).  
    \label{eq:K_Q1ij_3rd_UV_iso}
\end{align}
Note that statistical isotropy requires the contraction to be proportional to
$\mcal{D}_{ij}(\bv{k})$. 
We define $\mcal{F}_{Q_1}$ as its scalar coefficient: 
\begin{align}
    \avrg{Q_{1,ij}^{(3)}(\bv{k}) \delL(\bv{k}')}_\mrm{ren.}' 
    \equiv 
    \mcal{D}_{ij}(\bv{k}) \mcal{F}_{Q_1}(k) 
    = 
    \mcal{D}_{ij}(\bv{k})
    \left[
    3\PL(k) 
    \int_{\bv{q}} 
    K_{Q_{1}}^{(3)}(\bv{k},\bv{q},-\bv{q})
    \PL(q)
    \right], 
\end{align} 
where the renormalized scalar kernel is 
\begin{align}
    K_{Q_{1}}^{(3)}(\bv{k},\bv{q},-\bv{q}) 
    \equiv 
    \frac{10}{21}
    \frac{k^2(1-x^2)^2(k^2+q^2-6q^2x^2)}{|\bv{k}-\bv{q}|^2|\bv{k}+\bv{q}|^2}. 
    \label{eq:def_K_Q1_3rd}
\end{align}
Our $\mcal{F}_{Q_1}$ equals $J_3$ with the corrected kernel given in Ref.~\cite{Bakx+2023:EFTofIAvsSims}. 
The corresponding integrand kernels differ by a polynomial whose angular average vanishes.  

For the other quadratic and cubic operators, we obtain 
\begin{align}
    \avrg{Q_{2,ij}^{(3)}(\bv{k}) \delL(\bv{k}')}_{\rm ren}' 
    &= 
    \mcal{D}_{ij}(\bv{k})
    \left[
    - \frac7{30}\mcal{F}_{\mcal{G}_2}(k) 
    - \frac7{100}\mcal{F}_{Q_1}(k)
    \right], \nonumber\\
    \avrg{Q_{3,ij}^{(3)}(\bv{k}) \delL(\bv{k}')}_{\rm ren}' 
    &= 
    \mcal{D}_{ij}(\bv{k}) 
    \left[
    - \frac1{60}\mcal{F}_{\mcal{G}_2}(k)
    + \frac7{100}\mcal{F}_{Q_1}(k)
    \right], \nonumber\\
    \avrg{C_{1,ij}^{(3)}(\bv{k}) \delL(\bv{k}')}_{\rm ren}' 
    &= 
    \mcal{D}_{ij}(\bv{k}) \mcal{F}_{\mcal{G}_2}(k), \nonumber\\
    \avrg{C_{2,ij}^{(3)}(\bv{k}) \delL(\bv{k}')}_{\rm ren}' 
    &= 
    \mcal{D}_{ij}(\bv{k}) \mcal{F}_{Q_1}(k). 
    \label{eq:quadratic_descendant_responses}
\end{align}
Using these results, we can write the effective third-order bias parameters
in \eq{eq:PgE_13} and \eq{eq:PEE_13} as 
\begin{align}
    \bar{b}_{3,1} 
    \equiv b_{3,1} - \frac7{30} b_{2,2} - \frac1{60} b_{2,3}, \quad
    \bar{b}_{3,2} 
    \equiv b_{3,2} + b_{2,1} - \frac7{100} b_{2,2} + \frac7{100} b_{2,3}.
    \label{eq:bar_b3_map}
\end{align}
Therefore, we obtain 
\begin{align}
    \avrg{\Sij^{(3)}(\bv{k}) \delL(\bv{k}')}_\mrm{ren.}'  
    &=
    \mcal{D}_{ij}(\bv{k}) \left[
    \frac{b_K}{2} P_m^{(13)}(k) 
    + \bar{b}_{3,1} \mcal{F}_{\mcal{G}_2}(k)
    + \bar{b}_{3,2} \mcal{F}_{Q_1}(k)
    \right]. \label{eq:Sij3rd_delL_contraction} 
\end{align}

\subsection{Helicity decomposition and the (22)-type loop integrals}
\label{subapp:helicity_22} 
The loop integrals defined in \eqs{eq:def_mcalI_OQaE}{eq:def_mcalI_QaQbAA} can be evaluated directly. 
However, using a helicity basis makes the expressions more compact and easier to compare with previous results in Refs.~\cite{Vlah+2020:IA_EFT,Bakx+2023:EFTofIAvsSims}. 
Here, we briefly derive the relations between the power spectra of projected galaxy shapes and the helicity power spectra of the three-dimensional shape field. 

Assuming statistical homogeneity, we define the Cartesian galaxy density--shape cross-power spectrum and the galaxy-shape auto-power spectrum as
\begin{align}
    \avrg{\delg(\bv{k}) \Sij(\bv{k}')}
    &\equiv (2\pi)^3\delD_{\bv{k}+\bv{k}'}
    P_{\mrm{g} S,ij}(\bv{k}), \label{eq:def_PgS_Cartesian}\\
    \avrg{\Sij(\bv{k}) \Skl(\bv{k}')}
    &\equiv (2\pi)^3\delD_{\bv{k}+\bv{k}'}
    P_{SS,ijkl}(\bv{k}).
    \label{eq:def_PSS_Cartesian}
\end{align} 
To identify their independent components under statistical isotropy, we employ the helicity decomposition following Ref.~\cite{Vlah+2020:IA_EFT}. 
A symmetric trace-free rank-two field can be decomposed into five helicity components, 
\begin{align}
    \Sij(\bv{k}) 
    \equiv \sum_{m=-2}^{2} S^{(m)}(\bv{k}) Y_{ij}^{(m)}(\bv{k}), 
    \label{eq:Sij_helicity_decomp}
\end{align}
where $Y_{ij}^{(m)}$, with $m=0,\pm 1,\pm 2$, are orthonormal helicity tensors. 
We adopt the phase convention 
\begin{align}
    Y_{ij}^{(m)*}(\bv{k}) Y_{ij}^{(m')}(\bv{k}) = \delK_{mm'}, \quad
    Y_{ij}^{(m)*}(\bv{k})
    = (-1)^m Y_{ij}^{(-m)}(\bv{k}) 
    = Y_{ij}^{(m)}(-\bv{k}).
\end{align} 
In particular, $Y_{ij}^{(0)}(\bv{k}) = \sqrt{3/2} \mcal{D}_{ij}(\bv{k})$. 
In real space, statistical isotropy implies that different helicity components are uncorrelated and that the scalar density field correlates only with the helicity-0 component. 
Parity invariance further requires the power spectra of opposite helicities, $+m$ and $-m$, to be equal. 
The density--shape and shape-shape spectra are therefore specified by one and three scalar functions, respectively: 
\begin{align}
    P_{\mrm{g}S}^{(0)}(k)
    \equiv
    Y_{ij}^{(0)}
    P_{\mrm{g}S,ij}, \quad
    P_{SS}^{(\lambda)}(k)
    \equiv
    \operatorname{Re}\left[
    Y_{ij}^{(\lambda)*}
    Y_{kl}^{(\lambda)}
    \right]
    P_{SS,ijkl},
    \label{eq:def_PgS_PSS_Helicity}
\end{align}
where $\lambda=|m|=0,1,2$. 
Using the completeness of the helicity tensors, we have the inverse relation: 
\begin{align}
    P_{\mrm{g}S,ij}
    =Y_{ij}^{(0)} P_{\mrm{g}S}^{(0)}(k), \quad
    P_{SS,ijkl}
    = Y_{ij}^{(0)} Y_{kl}^{(0)}P_{SS}^{(0)}(k) 
    + 2\sum_{\lambda=1}^2 
    \mrm{Re}\left[
    Y_{ij}^{(\lambda)*}
    Y_{kl}^{(\lambda)}
    \right] P_{SS}^{(\lambda)}(k). 
    \label{eq:def_PgS_PSS_Helicity_Inv}
\end{align} 
By contracting these expressions with the projection tensors $M_{ij}^E$ and $M_{ij}^B$, we obtain \citep{Kurita&Takada2022:AnalysisIAPS,Bakx+2023:EFTofIAvsSims} 
\begin{align}
    P_{\mrm{g}E}(k,\mu)
    &=
    \sqrt\frac32\frac{1-\mu^2}{2}P_{\mrm{g}S}^{(0)}(k), \label{eq:PgE_real_PgS}\\
    P_{EE}(k,\mu)
    &=
    \frac{3(1-\mu^2)^2}{8}P_{SS}^{(0)}(k)
    +
    \frac{\mu^2(1-\mu^2)}{2}P_{SS}^{(1)}(k)
    +
    \frac{(1+\mu^2)^2}{8}P_{SS}^{(2)}(k), \label{eq:PEE_real_PSS}\\
    P_{BB}(k,\mu)
    &=
    \frac{1-\mu^2}{2}P_{SS}^{(1)}(k)
    +
    \frac{\mu^2}{2}P_{SS}^{(2)}(k). \label{eq:PBB_real_PSS}
\end{align}
Thus, the angular dependence is entirely determined by the projection, while the helicity spectra depend only on $k$. 

To express the (22)-type loop contributions, we define the helicity components of the quadratic kernels
\begin{align}
    K_{Q_a}^{(m)}(\bv{q}_1,\bv{q}_2)
    \equiv
    Y_{ij}^{(m)*}(\bv{k})
    K_{Q_a,ij}(\bv{q}_1,\bv{q}_2),  
    \label{eq:def_Q_helicity}
\end{align}
with $\bv{k}=\bv{q}_{12}$. 
At this order, the definitions in \eqs{eq:def_Q2ij}{eq:def_Q3ij} imply
\begin{align}
    K_{Q_2}^{(\pm1)}
    =
    K_{Q_2}^{(\pm2)}
    =
    K_{Q_3}^{(0)}
    =
    K_{Q_3}^{(\pm1)}
    =
    0.
\end{align} 
We then define the $(22)$-type loop integrals for helicity components as 
\begin{align}
    \mcal{I}_{X,Q_a}^{(0)}(k)
    &\equiv 
    2\int_{\bv{q}}
    K_X(\bv{q},\bv{k}-\bv{q})K_{Q_a}^{(0)}(\bv{q},\bv{k}-\bv{q})
    \PL(q)\PL(|\bv{k}-\bv{q}|), \label{eq:def_mcalI_XQ}\\
    \mcal{I}_{Q_a,Q_b}^{(\lambda)}(k)
    &\equiv 
    2\int_{\bv{q}}
    \mrm{Re}\left[
    K_{Q_a}^{(\lambda)}(\bv{q},\bv{k}-\bv{q})
    K_{Q_b}^{(\lambda)*}(\bv{q},\bv{k}-\bv{q})
    \right]
    \PL(q)\PL(|\bv{k}-\bv{q}|), \label{eq:def_mcalI_QQ}
\end{align}
where $X\in\{\delta,\delta^2,\mcal{G}_2\}$ and $\lambda=0,1,2$. 
The azimuthal integration about $\bv{k}$ eliminates products of different helicities. 
Therefore, the same projection factors relate these functions to the projected loop functions: 
\begin{align}
    \mcal{I}_{X,Q_a}^{E}(k,\mu)
    &=
    \sqrt{\frac32}\,\frac{1-\mu^2}{2}
    \mcal{I}_{X,Q_a}^{(0)}(k), \label{eq:relation_mcalI_gE_helicity}\\
    \mcal{I}_{Q_a,Q_b}^{EE}(k,\mu)
    &=
    \frac{3(1-\mu^2)^2}{8}\mcal{I}_{Q_a,Q_b}^{(0)}(k)
    +
    \frac{\mu^2(1-\mu^2)}{2}\mcal{I}_{Q_a,Q_b}^{(1)}(k)
    +
    \frac{(1+\mu^2)^2}{8}\mcal{I}_{Q_a,Q_b}^{(2)}(k), \label{eq:relation_mcalI_EE_helicity}\\
    \mcal{I}_{Q_a,Q_b}^{BB}(k,\mu)
    &=
    \frac{1-\mu^2}{2}\mcal{I}_{Q_a,Q_b}^{(1)}(k)
    +
    \frac{\mu^2}{2}\mcal{I}_{Q_a,Q_b}^{(2)}(k). 
    \label{eq:relation_mcalI_BB_helicity}
\end{align}
These expressions separate the geometrical $\mu$ dependence from the scalar loop functions of $k$. 

\subsection{Degeneracies among the (22)-type loop integrals}
\label{subapp:degeneracy_22} 
The loop functions of helicity components are not all independent. 
The four relations given in Eqs.~(B.45) and (B.46) of Ref.~\cite{Vlah+2020:IA_EFT} correspond to the following four relations in our notation: 
\begin{align}
    \mcal{I}_{\delta,Q_1}^{(0)} 
    &= \sqrt\frac23 \left(\mcal{I}_{\delta,\delta^2} + \mcal{I}_{\delta,\mcal{G}_2}\right) 
    - \frac57 \mcal{I}_{\mcal{G}_2,Q_1}^{(0)} 
    -\frac1{28}\sqrt\frac23 \left(\mcal{I}_{\delta^2,\mcal{G}_2} + \mcal{I}_{\mcal{G}_2,\mcal{G}_2}\right), \\
    \mcal{I}_{Q_1,Q_1}^{(2)} 
    &= \frac16\mcal{I}_{Q_1,Q_1}^{(0)} 
    - \frac13\sqrt\frac23 \left(\mcal{I}_{\delta^2,Q_1}^{(0)} + \mcal{I}_{\mcal{G}_2,Q_1}^{(0)}\right) 
    + \frac19\left(\mcal{I}_{\delta^2,\delta^2}+2\mcal{I}_{\delta^2,\mcal{G}_2}+\mcal{I}_{\mcal{G}_2,\mcal{G}_2}\right), \\
    \mcal{I}_{Q_1,Q_3}^{(2)} 
    &= \frac18\sqrt\frac23\mcal{I}_{\mcal{G}_2,Q_1}^{(0)} 
    - \frac1{12}\left(\mcal{I}_{\delta^2,\mcal{G}_2} + \mcal{I}_{\mcal{G}_2,\mcal{G}_2}\right)
    , \\
    \mcal{I}_{Q_3,Q_3}^{(2)} 
    &= \frac1{16}\mcal{I}_{\mcal{G}_2,\mcal{G}_2}.
\end{align} 
Furthermore, from $Q_2^{(0)} = -\sqrt{2/3}\sigma^2/4$, we have 
\begin{align}
    \mcal{I}_{X,Q_2}^{(0)} = -\frac14\sqrt\frac23 \mcal{I}_{X,\mcal{G}_2},\quad 
    \mcal{I}_{Q_1,Q_2}^{(0)} = -\frac14\sqrt\frac23 \mcal{I}_{\mcal{G}_2,Q_1}^{(0)},\quad 
    \mcal{I}_{Q_2,Q_2}^{(0)} = \frac1{24}\mcal{I}_{\mcal{G}_2,\mcal{G}_2}, 
\end{align}
with $X\in\{\delta,\delta^2,\mcal{G}_2\}$. 
In this work, we find an additional relation for the helicity-1 integral:\footnote{In the notation of Refs.~\cite{Vlah+2020:IA_EFT,Bakx+2023:EFTofIAvsSims}, this corresponds to 
\begin{align*}
    I_{22} - 9I_{23} - 4\sqrt6I_{24} + 24I_{44} + 36I_{55} = 0. 
\end{align*}
The complete correspondence with the notation of these references is 
\begin{align*}
    \mcal{I}_{\delta,\delta}&=2I_{11}, &
    \mcal{I}_{\delta,\delta^2}&=2I_{13}, &
    \mcal{I}_{\delta,\mcal{G}_2}&=2(I_{12}-I_{13}), \nonumber\\
    \mcal{I}_{\delta^2,\delta^2}&=2I_{33}, &
    \mcal{I}_{\delta^2,\mcal{G}_2}&=2(I_{23}-I_{33}), &
    \mcal{I}_{\mcal{G}_2,\mcal{G}_2}&=2(I_{22}-2I_{23}+I_{33}), \nonumber\\
    \mcal{I}_{\delta,Q_1}&=2I_{14}, &
    \mcal{I}_{\delta^2,Q_1}&=2I_{34}, &
    \mcal{I}_{\mcal{G}_2,Q_1}&=2(I_{24}-I_{34}),\nonumber\\
    \mcal{I}_{Q_1,Q_1}^{(0)}&=2I_{44}, &
    \mcal{I}_{Q_1,Q_1}^{(1)}&=2I_{55}, &&&\nonumber\\
    \mcal{I}_{Q_1,Q_1}^{(2)}&=2I_{66}, &
    \mcal{I}_{Q_1,Q_3}^{(2)}&=2(I_{67}-I_{66}), &
    \mcal{I}_{Q_3,Q_3}^{(2)}&=2(I_{66}-2I_{67}+I_{77}).
\end{align*}
}
\begin{align}
    \mcal{I}_{Q_1,Q_1}^{(1)} 
    &= -\frac23\mcal{I}_{Q_1,Q_1}^{(0)} 
    + \frac13\sqrt\frac23 \left(\mcal{I}_{\delta^2,Q_1}^{(0)} + \mcal{I}_{\mcal{G}_2,Q_1}^{(0)}\right) 
    + \frac29\mcal{I}_{\delta^2,\delta^2} 
    + \frac7{36}\mcal{I}_{\delta^2,\mcal{G}_2} 
    - \frac1{36}\mcal{I}_{\mcal{G}_2,\mcal{G}_2}. 
\end{align}
From these reductions, the only integrals needed to be evaluated in addition to those of density auto-spectrum are
\begin{align}
    \left\{ \mcal{I}_{\delta^2,Q_1}^{(0)}, \mcal{I}_{\mcal{G}_2,Q_1}^{(0)}, \mcal{I}_{Q_1,Q_1}^{(0)} \right\}. 
\end{align}
We here write down the analytic expressions of these integrals explicitly: 
\begin{align}
    \mcal{I}_{\delta^2,Q_1}^{(0)}(k) &\equiv 2\int_\bv{q} K_{Q_1}^{(0)}(\bv{q},\bv{k}-\bv{q}) \PL(q) \PL(|\bv{k}-\bv{q}|), \label{eq:def_mcalI_delta2Q1}\\
    \mcal{I}_{\mcal{G}_2,Q_1}^{(0)}(k) &\equiv 2\int_\bv{q} \sigma^2(\bv{q},\bv{k}-\bv{q}) K_{Q_1}^{(0)}(\bv{q},\bv{k}-\bv{q}) \PL(q) \PL(|\bv{k}-\bv{q}|), \label{eq:def_mcalI_G2Q1}\\
    \mcal{I}_{Q_1,Q_1}^{(0)}(k) &\equiv 2\int_\bv{q} \left[K_{Q_1}^{(0)}(\bv{q},\bv{k}-\bv{q})\right]^2 \PL(q) \PL(|\bv{k}-\bv{q}|), \label{eq:def_mcalI_Q1Q1_lam0}
\end{align}
where the helicity-0 component of the kernel $K_{Q_1,ij}$ is given by 
\begin{align}
    K_{Q_1}^{(0)}(\bv{q},\bv{p}) &\equiv 
    \frac{(\bv{q}\cdot\bv{p})\left[2(\bv{q}\cdot\bv{p})(q^2+p^2) + 3q^2p^2 + (\bv{q}\cdot\bv{p})^2\right]}{\sqrt6 k^2q^2p^2}. 
\end{align} 
Note that $\mcal{I}_{Q_1,Q_1}^{(0)}$ approaches a nonzero constant as $k\to0$: 
$\mcal{I}_{Q_1,Q_1}^{(0)}(0) = 4\int_\bv{q}\PL^2(q)/15$. 
We subtract this in our numerical implementation.

\subsection{Stochasticity in real space}
\label{subapp:stochasticity} 
The stochastic contributions to the galaxy-shape power spectra have been discussed in Refs.~\cite{Vlah+2020:IA_EFT,Chen&Kokron2024:LEFTofIA}. 
We first consider a local derivative expansion of the stochastic shape field: 
\begin{align}
    S_{ij}^\mrm{stoch.} = 
    \epsilon_{ij} 
    + c_{\epsilon,1} \nabla^2\epsilon_{ij}
    + c_{\epsilon,2} \TF{\partial_{(i}\partial_k\epsilon_{kj)}}
    + \cdots, \label{eq:Sij_stoch_local_exp}
\end{align}
where $\epsilon_{ij}$ is a symmetric trace-free white-noise field, and parentheses denote symmetrization with a factor of $1/2$. 
We normalize its power spectrum as 
\begin{align}
    \avrg{\epsilon_{ij}(\bv{k})\epsilon_{kl}(\bv{k}')}' 
    = 
    N_{SS,0} \left(\delK_{ik}\delK_{jl} + \delK_{il}\delK_{jk} -\frac23\delK_{ij}\delK_{kl}\right). 
\end{align} 
We write the stochastic helicity spectra up to order $k^2$ as 
\begin{align}
    N_{\mrm{g}S}(k)
    = \sqrt\frac23N_{\mrm{g}S,2}k^2,\quad
    N_{SS}^{(\lambda)}(k)
    = 2\left(N_{SS,0}+N_{SS,2}^{(\lambda)}k^2\right),
    \quad \lambda=0,1,2. 
    \label{eq:PgS_PSS_helicity_stoch_real}
\end{align}
Statistical isotropy and locality rule out a constant stochastic term in the cross spectrum.
They also require the same constant term for all helicities in the auto spectra. 
The cross terms between $\epsilon_{ij}$ and the derivative terms in \eq{eq:Sij_stoch_local_exp} give 
\begin{align}
    N_{SS,2}^{(0)}
    &= 
    -2N_{SS,0} \left(c_{\epsilon,1} + \frac23 c_{\epsilon,2}\right), \\
    N_{SS,2}^{(1)}
    &= 
    -2N_{SS,0} \left(c_{\epsilon,1} + \frac12 c_{\epsilon,2}\right), \\
    N_{SS,2}^{(2)}
    &= 
    -2N_{SS,0} c_{\epsilon,1}. 
\end{align}
Thus, the local expansion has only two independent $k^2$ coefficients, which satisfy \citep{Chen&Kokron2024:LEFTofIA} 
\begin{align}
    3N_{SS,2}^{(0)}-4N_{SS,2}^{(1)}+N_{SS,2}^{(2)}=0. 
    \label{eq:relation_stoch_k2_helicity} 
\end{align} 

In the renormalized power spectra, however, the stochastic coefficients also absorb the UV dependence of the (22)-type loop integrals. 
These contributions need not satisfy \eq{eq:relation_stoch_k2_helicity}. 
For example, for $k\ll q$, the second-order kernels of $Q_{1,ij}$ and $Q_{2,ij}$ are  
\begin{align}
    K_{Q_1,ij}(\bv{q},\bv{k}-\bv{q}) 
    &= 
    \mcal{D}_{ij}(\bv{q}) + \mcal{O}\left(k/q\right), \\
    K_{Q_2,ij}(\bv{q},\bv{k}-\bv{q}) 
    &= \frac{1-x^2}{4} \frac{k^2}{q^2} \mcal{D}_{ij}(\bv{k})  + \mcal{O}\left((k/q)^3\right). 
\end{align} 
Therefore, the UV contribution to their (22)-type loop integral between $Q_{1,ij}$ and $Q_{2,ij}$ is 
\begin{align}
    \left. \mcal{I}_{Q_1,Q_2}^{(0)}(k)\right|_\mrm{UV} = -\frac{2}{45}k^2\mcal{J}_\mrm{UV}(\Lambda), \qquad 
    \mcal{I}_{Q_1,Q_2}^{(1)}(k) = \mcal{I}_{Q_1,Q_2}^{(2)}(k) = 0, 
\end{align}
where $\Lambda$ is an arbitrary separation wavenumber satisfying $k\ll\Lambda$, and 
\begin{align}
    \mcal{J}_\mrm{UV}(\Lambda) \equiv \int_{\bv{q},\,q>\Lambda} \frac{\PL^2(q)}{q^2}.
\end{align}
This contribution affects only helicity 0 and does not satisfy \eq{eq:relation_stoch_k2_helicity}. 
We therefore treat the three $N_{SS,2}^{(\lambda)}$ as independent. 
Substituting \eq{eq:PgS_PSS_helicity_stoch_real} into Eqs.~\eqref{eq:PgE_real_PgS}--\eqref{eq:PBB_real_PSS}, we obtain 
\begin{align}
    P_{\mrm{g}E}^\mrm{stoch}(k,\mu)
    &=
    \frac{1-\mu^2}{2}N_{\mrm{g}S,2}k^2, \label{eq:PgE_stoch_real_3Dnotation}\\
    P_{EE}^\mrm{stoch}(k,\mu)
    &=
    N_{SS,0} + 
    \left[
    \frac{3(1-\mu^2)^2}{4}N_{SS,2}^{(0)} 
    + 
    \mu^2(1-\mu^2)N_{SS,2}^{(1)} 
    +
    \frac{(1+\mu^2)^2}{4}N_{SS,2}^{(2)} 
    \right]k^2, \label{eq:PEE_stoch_real_3Dnotation}\\
    P_{BB}^\mrm{stoch}(k,\mu)
    &=
    N_{SS,0} + 
    \left[
    (1-\mu^2)N_{SS,2}^{(1)} 
    + 
    \mu^2N_{SS,2}^{(2)} 
    \right]k^2. \label{eq:PBB_stoch_real_3Dnotation} 
\end{align} 
By redefining the coefficients as 
\begin{align}
    N_{SS,2} \equiv N_{SS,2}^{(1)}, \quad 
    N_{SS,2\mu} \equiv N_{SS,2}^{(2)} - N_{SS,2}^{(1)}, \quad
    N_{SS,2\Delta} \equiv 3N_{SS,2}^{(0)}-4N_{SS,2}^{(1)}+N_{SS,2}^{(2)}, 
\end{align}
we obtain Eqs.~\eqref{eq:PgE_stoch_real}--\eqref{eq:PBB_stoch_real}. 
The locality relation \eq{eq:relation_stoch_k2_helicity} is equivalent to $N_{SS,2\Delta}=0$. 

In Cartesian form, the UV contribution to the $Q_1$-$Q_2$ cross spectrum is 
\begin{align}
    \left. \avrg{Q_{1,ij}(\bv{k}) Q_{2,kl}(\bv{k}')}' \right|_\mrm{UV}
    =
    -\frac{2}{45}k^2 Y_{ij}^{(0)}(\bv{k}) Y_{kl}^{(0)}(\bv{k}) 
    \mcal{J}_\mrm{UV}(\Lambda) + \mcal{O}\left(k^4\right). 
    \label{eq:UV_Q1Q2}
\end{align}
The inverse Laplacian in $Q_2$ leaves a tensor structure proportional to $k^2 \mcal{D}_{ij}(\bv{k}) \mcal{D}_{kl}(\bv{k})$. 
This structure can be represented by adding a spatially nonlocal operator to  \eq{eq:Sij_stoch_local_exp}: 
\begin{align}
    S_{ij} \supset c_{\epsilon,3}\mcal{D}_{ij}(\partial) \partial_k\partial_l\epsilon_{kl}, \quad 
    \mcal{D}_{ij}(\partial) \equiv \left(\frac{\partial_i\partial_j}{\nabla^2} - \frac{\delK_{ij}}{3}\right). 
\end{align}
The two cross terms between this operator and $\epsilon_{ij}$ contribute 
\begin{align}
    P_{SS,ijkl}^\mrm{stoch.}(\bv{k}) \supset -4N_{SS,0} c_{\epsilon,3} k^2 \mcal{D}_{ij}(\bv{k}) \mcal{D}_{kl}(\bv{k}). 
\end{align}
This reproduces the tensor structure of the UV contribution in \eq{eq:UV_Q1Q2}. 


\section{Details of the redshift-space power spectra}
\label{app:details_rsd} 

\subsection{Perturbation kernels in redshift space}
\label{subapp:PT_kernel_rsd} 
Expanding \eqs{eq:tdelg_exp_cubic}{eq:tSij_exp_cubic} to third order in $\delL$, we obtain 
\begin{align}
    \tdelg^{(1)} &= \delg^{(1)} - \partial_\parallel u_\parallel^{(1)}, \label{eq:delg_rsd_1st}\\
    \tdelg^{(2)} &= \delg^{(2)} - \partial_\parallel u_\parallel^{(2)} 
    - \partial_\parallel \left[u_\parallel^{(1)}\delg^{(1)}\right] 
    + \frac{1}{2}\partial_\parallel^2\left[\left(u_\parallel^{(1)}\right)^2\right], \label{eq:delg_rsd_2nd}\\
    \tdelg^{(3)} &= \delg^{(3)} - \partial_\parallel u_\parallel^{(3)} 
    - \partial_\parallel \left[u_\parallel^{(1)}\delg^{(2)}+u_\parallel^{(2)}\delg^{(1)}\right] \nonumber\\
    &\quad+ \frac{1}{2}\partial_\parallel^2\left[2u_\parallel^{(1)}u_\parallel^{(2)} + \left(u_\parallel^{(1)}\right)^2\delg^{(1)}\right] - \frac{1}{6}\partial_\parallel^3\left[\left(u_\parallel^{(1)}\right)^3\right], \label{eq:delg_rsd_3rd}
\end{align} 
for the density field, and 
\begin{align}
    \tSij^{(1)} &= \Sij^{(1)}, \label{eq:Sij_rsd_1st}\\
    \tSij^{(2)} &= \Sij^{(2)} - \partial_\parallel \left[u_\parallel^{(1)}\Sij^{(1)}\right] , \label{eq:Sij_rsd_2nd}\\
    \tSij^{(3)} &= \Sij^{(3)} - \partial_\parallel \left[u_\parallel^{(1)}\Sij^{(2)}+u_\parallel^{(2)}\Sij^{(1)}\right] 
    + \frac{1}{2}\partial_\parallel^2\left[\left(u_\parallel^{(1)}\right)^2\Sij^{(1)}\right], \label{eq:Sij_rsd_3rd}
\end{align}
for the shape field. 

To obtain the Fourier-space kernels, we use the real-space kernels defined in \eqs{eq:bias_exp_delg_real}{eq:bias_exp_Sij_real} with the expansion of the velocity field
\begin{align}
    u_\parallel^{(n)}(\bv{k})
    =
    if\frac{k_\parallel}{k^2} \int_{\bv{q}_1,\cdots,\bv{q}_n} (2\pi)^3\delD_{\bv{k}-\bv{q}_{1\cdots n}} G_n(\bv{q}_1,\cdots,\bv{q}_n) \prod_{m=1}^n \delL({\bv{q}_m}), 
\end{align}
where $G_n$ is the $n$th-order velocity-divergence kernel \citep{Bernardeau+2002:PT_review}. 
We isolate the mapping contributions to the redshift-space kernels defined in \eqs{eq:bias_exp_delg_redshift}{eq:bias_exp_Sij_redshift} by writing 
\begin{align}
    Z_n = K_n + \Delta Z_n,\quad Z_{n,ij} = K_{n,ij} + \Delta Z_{n,ij}. 
\end{align}
For the density field, these contributions take the following symmetrized form: 
\begin{align}
    \Delta Z_1(\bv{k}) &= f\frac{k_\parallel^2}{k^2}, \label{eq:DelZ_1st}\\
    \Delta Z_2(\bv{q}_1,\bv{q}_2) &= f\frac{k_\parallel^2}{k^2} G_2(\bv{q}_1,\bv{q}_2)
    +
    \frac{f b_1k_\parallel}{2}
    \left[
    \frac{q_{1\parallel}}{q_1^2}
    +
    \frac{q_{2\parallel}}{q_2^2}
    \right] 
    +
    \frac{f^2 k_\parallel^2}{2}
    \frac{q_{1\parallel}q_{2\parallel}}{q_1^2q_2^2}, \label{eq:def_DelZ2}\\
    \Delta Z_3(\bv{q}_1,\bv{q}_2,\bv{q}_3) &= f\frac{k_\parallel^2}{k^2} G_3(\bv{q}_1,\bv{q}_2,\bv{q}_3)
    +
    \frac{f k_\parallel}{3}
    \sum_\mrm{cyc.}
    \left[
    \frac{q_{1\parallel}}{q_1^2}
    K_2(\bv{q}_2,\bv{q}_3)
    +
    b_1 
    \frac{q_{12,\parallel}}{q_{12}^2}
    G_2(\bv{q}_1,\bv{q}_2)
    \right] \nonumber\\
    &\quad+
    \frac{f^2 k_\parallel^2}{3}
    \sum_\mrm{cyc.}
    \left[
    \frac{q_{1\parallel}}{q_1^2} \frac{q_{23,\parallel}}{q_{23}^2} G_2(\bv{q}_2,\bv{q}_3) 
    + \frac{b_1}{2} \frac{q_{1\parallel}q_{2\parallel}}{q_1^2q_2^2} 
    \right]
    +
    \frac{f^3 k_\parallel^3}{6}
    \frac{q_{1\parallel}q_{2\parallel}q_{3\parallel}}{q_1^2q_2^2q_3^2}, \label{eq:def_DelZ3}
\end{align} 
where $\bv{k}=\bv{q}_{1\cdots n}$ at each order. 
For the shape field, $\Delta Z_{1,ij}=0$, and the second- and third-order corrections are 
\begin{align}
    \Delta Z_{2,ij}(\bv{q}_1,\bv{q}_2) 
    &= \frac{f k_\parallel}{2} 
    \left[
    \frac{q_{1\parallel}}{q_1^2}K_{1,ij}(\bv{q}_2) 
    + \frac{q_{2\parallel}}{q_2^2}K_{1,ij}(\bv{q}_1)
    \right], \label{eq:def_DelZ2ij}\\
    \Delta Z_{3,ij}(\bv{q}_1,\bv{q}_2,\bv{q}_3) 
    &= 
    \frac{f k_\parallel}{3}
    \sum_\mrm{cyc.}
    \left[
    \frac{q_{1\parallel}}{q_1^2} K_{2,ij}(\bv{q}_2,\bv{q}_3) 
    \right]
    +
    \frac{f k_\parallel}{3}
    \sum_\mrm{cyc.}
    \left[
    \frac{q_{12,\parallel}}{q_{12}^2}G_2(\bv{q}_1,\bv{q}_2) K_{1,ij}(\bv{q}_3) 
    \right]\nonumber\\
    &\quad+
    \frac{f^2 k_\parallel^2}{6}
    \sum_\mrm{cyc.}
    \left[
    \frac{q_{1\parallel}q_{2\parallel}}{q_1^2q_2^2} K_{1,ij}(\bv{q}_3)
    \right]. \label{eq:def_DelZ3ij}
\end{align}

\subsection{The (22)-type loop integrals}
\label{subapp:22_integral} 
Here we derive the reductions of the additional (22) integrals in redshift space in the galaxy density power spectrum, \eq{eq:Pgg_22_rsd}, and the galaxy shape power spectra, Eqs.~\eqref{eq:PgE_22_rsd}--\eqref{eq:PBB_22_rsd}. 
We set $\bv{p}=\bv{k}-\bv{q}$ and use the transverse decomposition and azimuthal averages summarized in Appendix~\ref{subapp:azm_avrg_22}.

\subsubsection{Density auto-power spectrum} \label{subsubapp:22_integral_density_auto} 
The LOS-independent kernels are given by 
\begin{align}
    K_\delta = F_2,\quad
    K_{\delta^2} = 1,\quad
    K_{\mcal{G}_2} = \sigma^2,\quad 
    K_\theta = G_2 = K_\delta + \frac{2}{7}K_{\mcal{G}_2}. 
    \label{eq:def_KX_iso}
\end{align}
These kernels are independent of the azimuthal angle $\varphi$ of $\bv{q}$ about $\bv{k}$.
The LOS-dependent kernels can be written as 
\begin{align}
    K_{u\delta}
    &\equiv \frac{k\mu}{2}\left(\frac{q_\parallel}{q^2} + \frac{p_\parallel}{p^2}\right)
    = \mu^2\left(K_\delta -\frac{2}{7}K_{\mcal{G}_2}\right) 
    + \frac{k\mu(p^2-q^2)}{2q^2p^2}(\hbv{n}\cdot\bv{q}_\perp), \label{eq:def_Kud}\\
    K_{uu}
    &\equiv \frac{k^2q_\parallel p_\parallel}{2q^2p^2}
    =K_{u\delta} - \frac{k^2}{2q^2p^2}\left[(\hbv{n}\cdot\bv{q}_\perp)^2 + \mu^2q_\perp^2\right]. \label{eq:def_Kuu}
\end{align} 
$K_{u\delta}$ contains the first moment in $\bv{q}_\perp$ and $K_{uu}$ contains the moments up to second moment with the same first moment as $K_{u\delta}$. 
Their azimuthal averages are 
\begin{align}
    \avrg{K_{u\delta}}_\varphi 
    = \mu^2\left(K_\delta - \frac{2}{7}K_{\mcal{G}_2}\right),\quad 
    \avrg{K_{uu}}_\varphi 
    = \mu^2K_\delta + \left(\frac{1}{4}-\frac{\mu^2}{28}\right)K_{\mcal{G}_2}. 
    \label{eq:azm_avrg_Kud_and_Kuu}
\end{align}
Since $P(q)P(p)$ is independent of $\varphi$, these averages can be used directly when the other kernel is LOS independent. 
For $X \in \{\delta,\delta^2,\mcal{G}_2\}$, we obtain 
\begin{align}
    \mcal{I}_{X,\theta} &= \mcal{I}_{X,\delta} + \frac{2}{7}\mcal{I}_{X,\mcal{G}_2}, \quad
    \mcal{I}_{\theta,\theta} = \mcal{I}_{\delta,\delta} + \frac{4}{7}\mcal{I}_{\delta,\mcal{G}_2} + \frac{4}{49}\mcal{I}_{\mcal{G}_2,\mcal{G}_2}, \label{eq:calI_theta_iso}\\
    \mcal{I}_{X,u\delta} &= \mu^2 \left(\mcal{I}_{X,\delta} - \frac{2}{7}\mcal{I}_{X,\mcal{G}_2}\right), \quad
    \mcal{I}_{\theta,u\delta} = \mu^2 \left(\mcal{I}_{\delta,\delta} - \frac{4}{49}\mcal{I}_{\mcal{G}_2,\mcal{G}_2}\right), \label{eq:calI_ud_iso}\\
    \mcal{I}_{X,uu} &= \mu^2\mcal{I}_{X,\delta} + \left(\frac{1}{4}-\frac{\mu^2}{28}\right)\mcal{I}_{X,\mcal{G}_2}, \quad
    \mcal{I}_{\theta,uu} = \mu^2 \mcal{I}_{\delta,\delta} + \frac{1+\mu^2}{4}\mcal{I}_{\delta,\mcal{G}_2} + \left(\frac{1}{14}-\frac{\mu^2}{98}\right)\mcal{I}_{\mcal{G}_2,\mcal{G}_2}. \label{eq:calI_uu_iso}
\end{align}
Next, we consider the loop integrals with products of the LOS-dependent kernels, $K_{u\delta}$ and $K_{uu}$. 
Using \eqs{eq:def_Kud}{eq:azm_avrg_Kud_and_Kuu}, we find 
\begin{align}
    \avrg{K_{u\delta}^2}_\varphi
    &= \avrg{K_{u\delta}}_\varphi^2 + \left[\frac{k\mu(p^2-q^2)}{2q^2p^2}\right]^2\h{n}_i \h{n}_j \avrg{q_{\perp i}q_{\perp j}}_\varphi\nonumber\\
    &= \avrg{K_{u\delta}}_\varphi^2 - \mu^2(1-\mu^2)\frac{\sigma^2(p^2-q^2)^2}{8q^2p^2}, 
    \label{eq:azm_avrg_Kud2}
\end{align}
where we have used $\sigma^2 \equiv (\hbv{q}\cdot\hbv{p})^2-1=-k^2q_\perp^2/(q^2p^2)$ in the second line. 
We define the RSD-induced distinct integral as 
\begin{align}
    \mcal{I}_\mrm{RSD}(k) \equiv 2\int_\bv{q} \frac{-\sigma^2(\bv{q},\bv{p})(p^2-q^2)^2} {8q^2p^2} \PL(q) \PL(p).
    \label{eq:def_I_rsd}
\end{align}
Substituting \eq{eq:azm_avrg_Kud_and_Kuu} into \eq{eq:azm_avrg_Kud2}, we obtain 
\begin{align}
    \mcal{I}_{u\delta,u\delta}
    = 
    \mu^4\left(\mcal{I}_{\delta,\delta} - \frac{4}{7}\mcal{I}_{\delta,\mcal{G}_2} + \frac{4}{49}\mcal{I}_{\mcal{G}_2,\mcal{G}_2}\right) 
    + 
    \mu^2(1-\mu^2)\mcal{I}_\mrm{RSD}. \label{eq:def_I_ud_ud}
\end{align}
Similarly, using \eqs{eq:nn_qperp_rank2}{eq:nnnn_qperp_rank4}, we find 
\begin{align}
    \avrg{K_{u\delta}K_{uu}}_\varphi 
    &= \avrg{K_{u\delta}}_\varphi\avrg{K_{uu}}_\varphi 
    - \mu^2(1-\mu^2)\frac{\sigma^2(p^2-q^2)^2}{8q^2p^2}, \label{eq:azm_avrg_Kud_Kuu}\\
    \avrg{K_{uu}^2}_\varphi 
    &= \avrg{K_{uu}}_\varphi^2 
    + \frac{(1-\mu^2)^2}{32}\left(\sigma^2\right)^2
    - \mu^2(1-\mu^2)\frac{\sigma^2(p^2-q^2)^2}{8q^2p^2}. \label{eq:azm_avrg_Kuu2} 
\end{align}
Therefore, the remaining integrals are 
\begin{align}
    \mcal{I}_{u\delta,uu}
    &= 
    \mu^4 \mcal{I}_{\delta,\delta} 
    + \mu^2\left(\frac{1}{4}-\frac{9\mu^2}{28}\right)\mcal{I}_{\delta,\mcal{G}_2} 
    - \mu^2\left(\frac{1}{14}-\frac{\mu^2}{98}\right)\mcal{I}_{\mcal{G}_2,\mcal{G}_2}
    + \mu^2(1-\mu^2)\mcal{I}_\mrm{RSD}, \label{eq:def_I_ud_uu}\\
    \mcal{I}_{uu,uu}
    &=
    \mu^4 \mcal{I}_{\delta,\delta} 
    + \mu^2\left(\frac{1}{2}-\frac{\mu^2}{14}\right)\mcal{I}_{\delta,\mcal{G}_2} 
    + \left(\frac{3}{32} - \frac{9\mu^2}{112} + \frac{51\mu^4}{1568}\right) \mcal{I}_{\mcal{G}_2,\mcal{G}_2}
    + \mu^2(1-\mu^2)\mcal{I}_\mrm{RSD}. \label{eq:def_I_uu_uu}
\end{align}
These relations specify all the additional integrals in \eq{eq:Pgg_22_rsd}. 

\subsubsection{density--shape cross- and shape auto-power spectra}
\label{subsubapp:22_integral_shape} 
In real space, the projected second-order shape kernels defined in \eq{eq:def_EB_kernel} are explicitly given by 
\begin{align}
    K_{Q_1}^E
    &=\frac{\bv{q}\cdot\bv{p}}{q^2p^2}
    M_{ij}^E q_ip_j,
    \label{eq:def_KQ1_E}\\
    K_{Q_2}^E
    &=-\frac{1-\mu^2}{8}K_{\mcal{G}_2}
    = \frac{k^2}{q^2p^2} \frac{1-\mu^2}{8} q_\perp^2,
    \label{eq:def_KQ2_E}\\
    K_{Q_3}^E
    &=\frac{k^2}{2q^2p^2}
    \left[
    M_{ij}^Eq_{\perp i}q_{\perp j}
    +\frac{1-\mu^2}{4}q_\perp^2
    \right].
    \label{eq:def_KQ3_E}
\end{align}
From \eq{eq:Sij_rsd_2nd}, the redshift-space mapping introduces the additional kernel
\begin{align}
    K_{uK}^A
    &\equiv
    \frac{k\mu}{2}M_{ij}^A
    \left[
    \frac{q_\parallel}{q^2}\mcal{D}_{ij}(\bv{p})
    +\frac{p_\parallel}{p^2}\mcal{D}_{ij}(\bv{q})
    \right],
    \quad A=E,B.
    \label{eq:def_KuK_A}
\end{align} 
Using $\bv{k} = \bv{q} + \bv{p}$ and \eqs{eq:def_KQ2_E}{eq:def_KQ3_E}, we obtain 
\begin{align}
    K_{uK}^A - K_{uu}M_{ij}^A\h{k}_i\h{k}_j
    &=
    \frac{M_{ij}^A}{2q^2p^2}
    (q_\parallel p_i-p_\parallel q_i)
    (q_\parallel p_j-p_\parallel q_j)
    =
    \frac{k^2}{2q^2p^2}M_{ij}^A
    \left[
    q_{\perp i}q_{\perp j}
    +q_\perp^2\h{k}_i\h{k}_j
    \right] \nonumber\\
    &= 
    K_{Q_2}^A + K_{Q_3}^A, 
\end{align}
where we used 
$q_\parallel\bv{p} - p_\parallel\bv{q} = (\hbv{n}\cdot\bv{q}_\perp)\bv{k} - k\mu\bv{q}_\perp$, and the projection tensors defined in \eqs{eq:def_M_E}{eq:def_M_B}. 
Thus, we find 
\begin{align}
    K_{uK}^E
    = K_{Q_2}^E + K_{Q_3}^E
    + \frac{1-\mu^2}{2}K_{uu}, \quad
    K_{uK}^B
    = K_{Q_3}^B.
    \label{eq:KuK_EB_reduction} 
\end{align}
These relations lead to the reduced kernels in \eqs{eq:def_Z2_E_red}{eq:def_Z2_B_red}. 

The scalar integrals in \eq{eq:PgE_22_rsd} were reduced in Appendix~\ref{subsubapp:22_integral_density_auto}. 
We therefore consider the integrals involving the shape kernels. 
From \eqs{eq:def_KX_iso}{eq:relation_mcalI_gE_helicity}, 
\begin{align}
    \mcal{I}_{\theta,Q_a}^E 
    = \mcal{I}_{\delta,Q_a}^E + \frac{2}{7}\mcal{I}_{\mcal{G}_2,Q_a}^E 
    = \sqrt{\frac32} \frac{1-\mu^2}{2}
    \left[
    \mcal{I}_{\delta,Q_a}^{(0)} + \frac{2}{7}\mcal{I}_{\mcal{G}_2,Q_a}^{(0)}
    \right], \quad a=1,2,3. 
\end{align}
Also, since $K_{Q_2}^E$ is proportional to $K_{\mcal{G}_2}^E$, 
\begin{align}
    \mcal{I}_{u\delta,Q_2}^E = -\frac{1-\mu^2}{8} \mcal{I}_{\mcal{G}_2,u\delta},\quad 
    \mcal{I}_{uu,Q_2}^E = -\frac{1-\mu^2}{8} \mcal{I}_{\mcal{G}_2,uu}. 
\end{align}
Thus, the remaining non-trivial integrals involve $K_{u\delta}$ or $K_{uu}$ paired with $K_{Q_1}^E$ or $K_{Q_3}^E$. 

We start with $K_{Q_3}^E$. 
Since $Q_{3,ij}$ is a transverse traceless tensor as shown in \eq{eq:def_KQ3ij}, i.e., including helicity$\pm2$ modes only, the azimuthal average of $K_{Q_3}^E$ vanishes:
\begin{align}
    \avrg{K_{Q_3}^E}_\varphi = 0. 
\end{align}
Furthermore, since $K_{u\delta}$ in \eq{eq:def_Kud} contains terms up to first-order moment in $\bv{q}_\perp$, 
\begin{align}
    \avrg{K_{u\delta}K_{Q_3}^E}_\varphi = 0. 
\end{align} 
The cross term with $K_{uu}$ in \eq{eq:def_Kuu} which has a second-order moment in $\bv{q}_\perp$ does not vanish instead as 
\begin{align}
    \avrg{K_{uu}K_{Q_3}^E}_\varphi 
    &= -\left[\frac{k^2}{2q^2p^2}\right]^2 \avrg{(\hbv{n}\cdot\bv{q}_\perp)^2\left(M_{ij}^Eq_{\perp i}q_{\perp j}+\frac{1-\mu^2}{4}q_\perp^2 \right)}_\varphi \nonumber\\
    &= -\left[\frac{k^2}{2q^2p^2}\right]^2 
    \frac{1-\mu^4}{16}q_\perp^4 
    = -\frac{1-\mu^4}{64} \left(\sigma^2\right)^2. 
\end{align}
Therefore, we obtain 
\begin{align}
    \mcal{I}_{u\delta,Q_3}^E = 0, \quad 
    \mcal{I}_{uu,Q_3}^E = -\frac{1-\mu^4}{64} \mcal{I}_{\mcal{G}_2,\mcal{G}_2}. 
\end{align}

For $K_{Q_1}^E$ in \eq{eq:def_KQ1_E}, we first perform the helicity decomposition as
\begin{align}
    K_{Q_1}^E 
    = \frac{1-\mu^2}{2}\sqrt\frac32 K_{Q_1}^{(0)} 
    - \frac{\bv{q}\cdot\bv{p}}{k^2}\left[K_{u\delta} - \avrg{K_{u\delta}}_\varphi\right]
    -\frac{2(\bv{q}\cdot\bv{p})}{k^2} K_{Q_3}^E, 
\end{align}
where we used 
$q_i = (\hbv{k}\cdot\bv{q})\h{k}_i + q_{\perp i}$, 
$p_i = (\hbv{k}\cdot\bv{p})\h{k}_i - q_{\perp i}$, and 
\begin{align}
    \sqrt\frac32 K_{Q_1}^{(0)} = \frac{\bv{q}\cdot\bv{p}}{q^2p^2}
    \left[
    (\hbv{k}\cdot\bv{q})(\hbv{k}\cdot\bv{p}) + \frac{q_\perp^2}{2}
    \right]. 
\end{align}
The azimuthal average of the product of $K_{u\delta}$ and $K_{Q_1}^E$ is then 
\begin{align}
    \avrg{K_{u\delta}K_{Q_1}^E}_\varphi 
    &= \avrg{K_{u\delta}}_\varphi \avrg{K_{Q_1}^E}_\varphi 
    - \frac{\bv{q}\cdot\bv{p}}{k^2}\left[\avrg{K_{u\delta}^2}_\varphi - \avrg{K_{u\delta}}_\varphi^2\right]\nonumber\\
    &= \frac{(1-\mu^2)\mu^2}{2} 
    \sqrt\frac32 \left(K_\delta - \frac27K_{\mcal{G}_2} \right)K_{Q_1}^{(0)} 
    + \frac{(1-\mu^2)\mu^2}{2} 
    \frac{(\bv{q}\cdot\bv{p})(p^2-q^2)^2}{4k^2q^2p^2}\sigma^2, 
\end{align}
where we used \eqs{eq:azm_avrg_Kud_and_Kuu}{eq:azm_avrg_Kud2} in the second line. 
The function in the second term can be reduced further: 
\begin{align}
    \frac{(\bv{q}\cdot\bv{p})(p^2-q^2)^2}{4k^2q^2p^2} 
    = 
    \frac12K_\delta - \frac13K_{\delta^2} + \frac1{42}K_{\mcal{G}_2} - \frac23\sqrt\frac32K_{Q_1}^{(0)}. 
\end{align}
Therefore, we finally obtain
\begin{align}
    \mcal{I}_{u\delta,Q_1}^E = \frac{(1-\mu^2)\mu^2}{2} 
    \left[
    \frac12 \mcal{I}_{\delta,\mcal{G}_2}
    - \frac13 \mcal{I}_{\delta^2,\mcal{G}_2}
    + \frac1{42} \mcal{I}_{\mcal{G}_2,\mcal{G}_2} 
    + \sqrt\frac32 \mcal{I}_{\delta,Q_1}^{(0)} 
    - \frac{20}{21}\sqrt\frac32 \mcal{I}_{\mcal{G}_2,Q_1}^{(0)}
    \right]. 
\end{align}
Similarly, using \eq{eq:def_Kuu}, the cross term $K_{uu}$ and $K_{Q_1}^E$ becomes 
\begin{align}
    \avrg{K_{uu}K_{Q_1}^E}_\varphi 
    &= \avrg{K_{u\delta}K_{Q_1}^E}_\varphi 
    - \frac{k^2\mu^2q_\perp^2}{2q^2p^2} \avrg{K_{Q_1}^E}_\varphi 
    - \frac{k^2}{2q^2p^2} \avrg{(\hbv{n}\cdot\bv{q}_\perp)^2K_{Q_1}^E}\nonumber\\
    &= \avrg{K_{u\delta}K_{Q_1}^E}_\varphi 
    + \frac{(1-\mu^2)(1+\mu^2)}{48} 
    \left[5\sqrt\frac32\sigma^2K_{Q_1}^{(0)} + \sigma^2 + \left(\sigma^2\right)^2\right]. 
\end{align}
Hence, 
\begin{align}
    \mcal{I}_{uu,Q_1}^E = \mcal{I}_{u\delta,Q_1}^E 
    + 
    \frac{(1-\mu^2)(1+\mu^2)}{48} 
    \left[
    5\sqrt\frac32 \mcal{I}_{\mcal{G}_2,Q_1}^{(0)}
    + \mcal{I}_{\delta^2,\mcal{G}_2}
    + \mcal{I}_{\mcal{G}_2,\mcal{G}_2}
    \right]. 
\end{align} 
We have now obtained the reduced expressions for all the integrals required in \eqs{eq:PgE_22_rsd}{eq:PEE_22_rsd}. 
The redshift-space $B$-mode spectrum in \eq{eq:PBB_22_rsd} can be obtained just by shifting the bias parameters in the real-space $B$-mode spectrum in \eq{eq:PBB_22}. 

\subsubsection{UV limits and redshift-space stochasticity}
\label{subsubapp:UVlim_2nd_kernels}
We identify the (22)-type contributions from the UV modes, which are absorbed into the stochastic parameters. 
For later use, we first summarize the expressions of the second-order kernels in the UV limit, $k\ll q$, up to $k^2/q^2$ at the (22) configuration, $(\bv{q},\bv{k}-\bv{q})$: 
\begin{align}
    &K_\delta^\mrm{UV} = \frac{3-10x^2}{14} \frac{k^2}{q^2}, \quad 
    K_{\delta^2} = 1, \quad 
    K_{\mcal{G}_2} = (x^2-1) \frac{k^2}{q^2}, \\
    &K_{Q_1,ij}^\mrm{UV} = 
    \mcal{D}_{ij}(\bv{q}) 
    + \left[x\h{q}_i\h{q}_j - \h{k}_{(i}\h{q}_{j)}\right]\frac{k}{q}
    + \left[(2x^2-1)\h{q}_i\h{q}_j - x\h{k}_{(i}\h{q}_{j)} + \frac{1-x^2}3 \delK_{ij}\right]\frac{k^2}{q^2}, \\
    &K_{Q_2,ij}^\mrm{UV} = 
    \frac{1-x^2}{4} \mcal{D}_{ij}(\bv{k}) \frac{k^2}{q^2}, \quad 
    K_{Q_3,ij}^\mrm{UV} = 
    \frac12 \left[(\h{q}_i - x\h{k}_i)(\h{q}_j - x\h{k}_j) - \frac{1-x^2}2 \mcal{P}_{ij}(\bv{k})\right]\frac{k^2}{q^2}, \\
    &K_\theta^\mrm{UV} = -\frac{1+6x^2}{14} \frac{k^2}{q^2}, \quad 
    K_{u\delta}^\mrm{UV} = 
    \frac{\mu}{2} \left(\mu - 2x\h{q}_\parallel\right) \frac{k^2}{q^2}, \quad 
    K_{uu}^\mrm{UV} = 
    -\frac12 \h{q}_\parallel^2  \frac{k^2}{q^2}. 
\end{align}
Among these, only $K_{\delta^2}$ and $K_{Q_1,ij}^\mrm{UV}$ have zeroth order contributions and otherwise suppressed by $k^2/q^2$. 

The UV contributions to the real-space power spectra are already given in Sections~\ref{subsubsec:pgE_pEE_pBB_real} and Appendix~\ref{subapp:stochasticity}. 
Therefore, here we consider the residual terms in the redshift-space power spectra: 
\begin{align}
    \left.\Delta \tP_{XY}^{(22)}\right|_\mrm{UV} 
    \equiv 
    \left.\tP_{XY}^{(22)}\right|_\mrm{UV} 
    - 
    \left.P_{XY}^{(22)}\right|_\mrm{UV}. \label{eq:Delta_P22_UV}
\end{align}
where $XY\in\{\mrm{g}E,EE,BB\}$. 
The required UV limits of the integrals are 
\begin{align}
    \left\{
    \left.\mcal{I}_{\delta^2,Q_2}^E \right|_\mrm{UV}, 
    \left.\mcal{I}_{\delta^2,uu} \right|_\mrm{UV}, 
    \left.\mcal{I}_{\theta,Q_1} \right|_\mrm{UV} 
    \right\}
    &= 
    \left\{
    \frac{1-\mu^2}{6}, 
    -\frac13 , 
    -\frac{2(1-\mu^2)}{35} 
    \right\}
    k^2\mcal{J}_\mrm{UV}(\Lambda), 
\end{align}
for the cross power spectrum, and 
\begin{align}
    \left\{
    \left.\mcal{I}_{Q_1,Q_2}^{EE} \right|_\mrm{UV}, 
    \left.\mcal{I}_{Q_1,Q_3}^{EE} \right|_\mrm{UV}, 
    \left.\mcal{I}_{Q_1,Q_3}^{BB} \right|_\mrm{UV} 
    \right\}
    &= 
    \left\{
    -\frac{(1-\mu^2)^2}{60}, 
    \frac{(1+\mu^2)^2}{60}, 
    \frac{\mu^2}{15} 
    \right\}
    k^2\mcal{J}_\mrm{UV}(\Lambda), 
\end{align}
for the auto power spectra. 
Substituting these to \eq{eq:Delta_P22_UV}, we obtain 
\begin{align}
    \left.\Delta \tP_{\mrm{g}E}^{(22)}\right|_\mrm{UV} 
    &= 
    -\frac{2}{35}fb_{2,1}(1-\mu^2)\mu^2 k^2\mcal{J}_\mrm{UV}(\Lambda), \nonumber\\
    \left.\Delta \tP_{EE}^{(22)}\right|_\mrm{UV} 
    &= 
    \left.\Delta \tP_{BB}^{(22)}\right|_\mrm{UV} 
    = 
    \frac{2}{15}fb_Kb_{2,1}\mu^2 k^2\mcal{J}_\mrm{UV}(\Lambda). 
\end{align}
The contributions to $\tP_{EE}$ and $\tP_{BB}$ are absorbed into $N_{SS,2\mu}$ in \eqs{eq:PEE_stoch_real}{eq:PBB_stoch_real}. 
Therefore, the required free parameters are the same as in real space. 
On the other hand, the contribution to $\tP_{\mrm{g}E}$ has an additional factor of $\mu^2$ compared with \eq{eq:PgE_stoch_real}. 
To absorb this contribution, we introduce a new parameter, $N_{\mrm{g}S,2\mu}$, in \eq{eq:PgE_stoch_rsd}.

\subsection{The (13)-type loop integrals}
\label{subapp:13_integral} 
Here, we derive the (13)-type contributions to the redshift-space power spectra. 
The real-space third-order contributions are computed in \eqs{eq:delg3rd_delL_contraction}{eq:Sij3rd_delL_contraction}. 
Below, we suppress the common arguments $(\bv{k},\bv{q},-\bv{q})$ of the third-order kernels.

\subsubsection{Density auto-power spectrum} \label{subsubapp:13_integral_density_auto} 
The third-order redshift-space correction to the density field contains six terms, as shown in \eqs{eq:delg_rsd_3rd}{eq:def_DelZ3}. 
We list the corresponding symmetrized kernels at $(\bv{k},\bv{q},-\bv{q})$ in the same order: 
\begin{align}
    - \partial_\parallel u_\parallel^{(3)}:&\quad 
    &K_{u^{(3)}}
    &\equiv f\mu^2 G_3, 
    \label{eq:rsd_kernel_u3}\\
    - \partial_\parallel \left[u_\parallel^{(1)}\delg^{(2)}\right]:&\quad 
    &K_{u\delta^{(2)}}
    &\equiv \frac{b_2f\mu^2}{6}-\frac{b_1f\mu}{3}\left(1+\frac{k^2}{q^2}\right)x\h{q}_\parallel, 
    \label{eq:rsd_kernel_ud2}\\
    - \partial_\parallel \left[u_\parallel^{(2)}\delg^{(1)}\right]:&\quad 
    &K_{u^{(2)}\delta}
    &\equiv \frac{b_1f\mu}{3} \sum_{\sigma=\pm1} \left(\mu+\sigma\frac{q}{k}\h{q}_\parallel\right) \frac{k^2G_2(\bv{k},\sigma\bv{q})}{|\bv{k}+\sigma\bv{q}|^2}, 
    \label{eq:rsd_kernel_u2d}\\
    \frac{1}{2}\partial_\parallel^2\left[2u_\parallel^{(1)}u_\parallel^{(2)}\right]:&\quad 
    &K_{uu^{(2)}}
    &\equiv -\frac{f^2\mu^2}{3}\sum_{\sigma=\pm1}\left(\sigma\mu\frac{k}{q}\h{q}_\parallel + \h{q}_\parallel^2\right) \frac{k^2G_2(\bv{k},\sigma\bv{q})}{|\bv{k}+\sigma\bv{q}|^2}, 
    \label{eq:rsd_kernel_uu2}\\
    \frac{1}{2}\partial_\parallel^2\left[\left(u_\parallel^{(1)}\right)^2\delg^{(1)}\right]:&\quad 
    &K_{uu\delta}
    &\equiv -\frac{b_1f^2\mu^2}{6} \frac{k^2}{q^2} \h{q}_\parallel^2, 
    \label{eq:rsd_kernel_uud}\\
    - \frac{1}{6}\partial_\parallel^3\left[\left(u_\parallel^{(1)}\right)^3\right]:&\quad 
    &K_{uuu}
    &\equiv -\frac{f^3\mu^4}{6} \frac{k^2}{q^2} \h{q}_\parallel^2. 
    \label{eq:rsd_kernel_uuu}
\end{align}
The operators in Eqs.~\eqref{eq:rsd_kernel_ud2}, \eqref{eq:rsd_kernel_uud}, and \eqref{eq:rsd_kernel_uuu}
can be written as products of linear fields at the same position, with no inverse derivatives acting on the
products.  
Their contractions with $\delL$ therefore yield only zero-lag moments: 
\begin{align}
    3\PL(k)\int_\bv{q} K_{u\delta^{(2)}} \PL(q) 
    &= f\mu^2 \left[\frac{b_2}{2}\sigma_\mrm{L}^2 - b_1\left(\frac{\sigma_\mrm{L}^2}{3} + \sigma_{v,\mrm{L}}^2k^2\right)\right] \PL(k), 
    \label{eq:13_int_K_ud2}\\
    3\PL(k)\int_\bv{q} K_{uu\delta} \PL(q) 
    &= -\frac{1}{2}b_1f^2\mu^2\sigma_{v,\mrm{L}}^2k^2\PL(k), 
    \label{eq:13_int_K_uud}\\
    3\PL(k)\int_\bv{q} K_{uuu} \PL(q) 
    &= -\frac{1}{2}f^3\mu^4\sigma_{v,\mrm{L}}^2 k^2\PL(k),  
    \label{eq:13_int_K_uuu} 
\end{align}
where we used \eqs{eq:azm_avrg_qn}{eq:azm_avrg_qn2} and defined the zero-lag moments: 
\begin{align}
    \sigma_\mrm{L}^2 \equiv \int_\bv{q}\PL(q),\quad 
    \sigma_{v,\mrm{L}}^2 \equiv \frac{1}{3}\int_\bv{q} \frac{\PL(q)}{q^2}. 
    \label{eq:13_zero_lag_moments} 
\end{align}
The term proportional to $b_2\sigma_{\mrm{L}}^2/2$ in \eq{eq:13_int_K_ud2} is canceled by the contribution from the mean subtraction in 
$\delg^{(2)}\supset (b_2/2)(\delL^2-\sigma_\mrm{L}^2)$. 

For the remaining three terms, we use the real-space kernels in \eqs{eq:def_F3_13config}{eq:def_K_G2_3rd_13config} and the following identity obtained from the SPT recursion relations \citep{Bernardeau+2002:PT_review}: 
\begin{align}
    G_3(\bv{k},\bv{q},-\bv{q}) = F_3(\bv{k},\bv{q},-\bv{q}) + \frac{2}{15}K_{\mcal{G}_2}^{(3)}(\bv{k},\bv{q},-\bv{q}). 
\end{align}
Using these expressions and performing the angular integrations, we obtain: 
\begin{align}
    3\PL(k)\int_\bv{q} K_{u^{(3)}} \PL(q) 
    &= f\mu^2 \left[\frac12 P_m^{(13)} + \frac{2}{15}\mcal{F}_{\mcal{G}_2}\right], 
    \label{eq:13_int_K_u3}\\
    3\PL(k)\int_\bv{q} K_{u^{(2)}\delta} \PL(q) 
    &= b_1f\mu^2 \left[P_m^{(13)} - \frac{2}{15}\mcal{F}_{\mcal{G}_2} + \left(\frac{\sigma_\mrm{L}^2}{3} + k^2\sigma_{v,\mrm{L}}^2\right) \PL(k)\right], 
    \label{eq:13_int_K_u2d}\\
    3\PL(k)\int_\bv{q} K_{uu^{(2)}} \PL(q) 
    &= f^2\mu^2\left[\mu^2P_m^{(13)} + \frac{9+\mu^2}{60}\mcal{F}_{\mcal{G}_2}\right]. 
    \label{eq:13_int_K_uu2} 
\end{align}
The $b_1$-dependent zero-lag terms in \eqs{eq:13_int_K_ud2}{eq:13_int_K_u2d} cancel. 
From these and the real-space result in \eq{eq:pgg_13_real}, we recover \eq{eq:Pgg_13_rsd}.

\subsubsection{density--shape cross- and shape auto-power spectra}
\label{subsubapp:13_integral_shape} 
For the shape field in redshift space, there are three mapping terms, as shown in \eqs{eq:Sij_rsd_3rd}{eq:def_DelZ3ij}. 
The last term in \eq{eq:Sij_rsd_3rd} involves only a product of the linear fields: two velocity fields and one shape field. 
The corresponding kernel is 
\begin{align}
    K_{uuS,ij}(\bv{k},\bv{q},-\bv{q}) 
    &\equiv -\frac{b_Kf^2\mu^2}{6} \frac{k^2}{q^2} \h{q}_\parallel^2\mcal{D}_{ij}(\bv{k}) . 
    \label{eq:rsd_kernel_uuS} 
\end{align} 
Performing the angular integrations, we obtain the (13) contribution from this term
\begin{align}
    3\PL(k)\int_\bv{q} K_{uuS,ij}(\bv{k},\bv{q},-\bv{q}) \PL(q) 
    &= -\frac{1}{2}b_Kf^2\mu^2\sigma_{v,\mrm{L}}^2k^2\PL(k)\mcal{D}_{ij}(\bv{k}) , 
    \label{eq:13_int_K_uuS}
\end{align}
Thus, this term gives only a zero-lag contribution. 

Next, we consider the term containing $u_\parallel^{(2)}S_{ij}^{(1)}$, whose kernel is 
\begin{align}
    K_{u^{(2)}S,ij}
    &\equiv \frac{b_Kf\mu}{3} \sum_{\sigma=\pm1} \left(\mu+\sigma\frac{q}{k}\h{q}_\parallel\right) 
    \mcal{D}_{ij}(\hbv{q})
    \frac{k^2G_2(\bv{k},\sigma\bv{q})}{|\bv{k}+\sigma\bv{q}|^2}, 
    \label{eq:rsd_kernel_K_u2S_ij} 
\end{align}
Performing the $E$-mode projection and using \eqs{eq:azm_avrg_ME_qq}{eq:azm_avrg_nME_qqq}, we obtain 
\begin{align}
    \avrg{K_{u^{(2)}S}^E}_\varphi
    &\equiv
    M_{ij}^E \avrg{K_{u^{(2)}S,ij}}_\varphi \nonumber\\
    &= 
    \frac{b_Kf\mu^2(1-\mu^2)}{6} 
    \sum_{\sigma=\pm1} 
    \left(\mcal{L}_2(x) + \sigma \frac{q}{k}\mcal{L}_3(x)\right)
    \frac{k^2G_2(\bv{k},\sigma\bv{q})}{|\bv{k}+\sigma\bv{q}|^2} \nonumber\\
    &= 
    \frac{b_Kf\mu^2(1-\mu^2)}{6} 
    \left[
    6F_3 + \frac15K_{\mcal{G}_2}^{(3)} -\frac32K_{Q_1}^{(3)} + x\mcal{L}_3(x) + \frac{k^2}{q^2}x^2
    \right]. 
\end{align}
In the last line, we used \eq{eq:def_F3_13config}, \eq{eq:def_K_G2_3rd_13config}, and \eq{eq:def_K_Q1_3rd}. 
Therefore, the (13)-type loop contribution becomes 
\begin{align}
    3\PL(k)\int_\bv{q} K_{u^{(2)}S}^E \PL(q) 
    &= \frac{(1-\mu^2)}{2} b_Kf\mu^2 
    \left[
    P_m^{(13)} + \frac1{15}\mcal{F}_{\mcal{G}_2} -\frac12\mcal{F}_{Q_1} + \sigma_{v,\mrm{L}}^2k^2\PL(k)
    \right]. 
    \label{eq:13_int_K_u2S}
\end{align}

The kernel of the remaining correction term in \eq{eq:Sij_rsd_3rd} is 
\begin{align}
    K_{uS^{(2)},ij} 
    &= 
    \frac{f\mu k}{3} 
    \left[
    \frac{\mu}{k} K_{2,ij}(\bv{q},-\bv{q}) + \frac{q_\parallel}{q^2}K_{2,ij}(\bv{k},-\bv{q}) - \frac{q_\parallel}{q^2}K_{2,ij}(\bv{k},\bv{q})
    \right], \nonumber\\
    &= 
    \frac{b_{2,1}f\mu^2}{3}\mcal{D}_{ij}(\bv{q}) - 
    \frac{f\mu k}{3q}
    \sum_{\sigma=\pm1} \sigma
    \left[b_KF_2(\bv{k},\sigma\bv{q}) + \frac{1-x^2}{4}(b_{2,2}-b_{2,3})\right] 
    \h{q}_\parallel
    \mcal{D}_{ij}(\bv{k}+\sigma\bv{q}), \label{eq:rsd_kernel_K_uS2_ij}
\end{align}
where $K_{2,ij}$ is the second-order shape kernel defined in \eq{eq:shape_kernel_2nd} and we used the following relations: 
\begin{align}
    K_{Q_1,ij}(\bv{q},-\bv{q}) &= \mcal{D}_{ij}(\bv{q}),\quad 
    K_{Q_2,ij}(\bv{q},-\bv{q}) = K_{Q_3,ij}(\bv{q},-\bv{q}) = 0, \\
    \sum_{\sigma=\pm1} \sigma K_{Q_2,ij}(\bv{k},\sigma\bv{q})
    &=
    -\sum_{\sigma=\pm1} \sigma K_{Q_3,ij}(\bv{k},\sigma\bv{q})
    =
    \frac{1-x^2}{4}\sum_{\sigma=\pm1} \sigma \mcal{D}_{ij}(\bv{k}+\sigma\bv{q}). 
\end{align} 
The azimuthally averaged $E$-mode kernel then becomes 
\begin{align}
    \avrg{K_{uS^{(2)}}^E}_\varphi 
    &= 
    \frac{f\mu^2(1-\mu^2)}{6} b_{2,1} \mcal{L}_2(x) \nonumber\\
    &\quad-
    \frac{f\mu^2(1-\mu^2)}{6}
    \sum_{\sigma=\pm1} 
    \frac{k^2}{|\bv{k}+\sigma\bv{q}|^2}
    \left[b_KF_2(\bv{k},\sigma\bv{q}) + \frac{1-x^2}{4}(b_{2,2}-b_{2,3})\right] \nonumber\\
    &\quad\quad\quad\quad\quad\quad\quad\quad\quad\quad\quad\quad\quad\quad\quad
    \times
    \left[2\mcal{L}_2(x) + \sigma \left( \frac{q}{k}\mcal{L}_3(x) + \frac{k}{q}x \right)\right] \nonumber\\
    &= 
    \frac{f\mu^2(1-\mu^2)}{6} 
    \Bigg[
    b_{2,1} \mcal{L}_2(x) 
    - 
    \frac14 \left(b_K+\frac{7}{20}(b_{2,2}-b_{2,3})\right)
    \left(K_{\mcal{G}_2}^{(3)} -10 K_{Q_1}^{(3)}\right) \nonumber\\
    &\quad\quad\quad\quad\quad\quad\quad\quad
    - 
    b_K\left(x\mcal{L}_3(x) + \frac{k^2}{q^2}x^2\right)
    \Bigg].
\end{align}
Therefore, we obtain 
\begin{align}
    3\PL(k)\int_\bv{q} K_{uS^{(2)}}^E \PL(q) 
    = 
    &-\frac{f\mu^2(1-\mu^2)}{24} 
    \left(b_K+\frac{7}{20}(b_{2,2}-b_{2,3})\right) 
    \left(\mcal{F}_{\mcal{G}_2} -10\mcal{F}_{Q_1}\right) \nonumber\\ 
    &-
    \frac{b_Kf\mu^2(1-\mu^2)}{2} \sigma_{v,\mrm{L}}^2k^2\PL(k). 
    \label{eq:13_int_K_uS2} 
\end{align}
The zero-lag terms in \eqs{eq:13_int_K_u2S}{eq:13_int_K_uS2} cancel. 
From the three results in Eqs.~\eqref{eq:13_int_K_uuS}, \eqref{eq:13_int_K_u2S}, and \eqref{eq:13_int_K_uS2}, we finally obtain 
\begin{align}
    3\PL(k)\int_\bv{q}
    \Delta Z_3^E \PL(q)
    &=
    \frac{(1-\mu^2)}{2}f\mu^2
    \Bigg[
    b_K P_m^{(13)}
    - \frac{1}{60} \left(b_K+\frac74 (b_{2,2}-b_{2,3})\right) \mcal{F}_{\mcal{G}_2}
    \nonumber\\
    &\quad\quad\quad\quad+ 
    \frac{1}{3} \left(b_K + \frac78(b_{2,2}-b_{2,3})\right) \mcal{F}_{Q_1}
    - 
    \frac{b_Kf}{2}\sigma_{v,\mrm{L}}^2k^2\PL(k)
    \Bigg]. 
    \label{eq:13_int_DelZ3E}
\end{align}
Using this, the result in Appendix~\ref{subsubapp:13_integral_density_auto}, and the real-space results in \eqs{eq:delg3rd_delL_contraction}{eq:Sij3rd_delL_contraction}, we reproduce \eqs{eq:PgE_13_rsd}{eq:PEE_13_rsd}.

\subsubsection{UV limits and redshift-space counterterms}
\label{subsubapp:UVlim_3rd_kernels} 
At leading order in $k/q$, the contributions from UV modes, $q\gg k$, to the (13) loop functions are 
\begin{align}
    \left.P_m^{(13)}(k)\right|_\mrm{UV} 
    &= -\frac{61}{105} \left.\sigma_v^2\right|_\mrm{UV} k^2\PL(k), \\
    \left.\mcal{F}_{\mcal{G}_2}(k)\right|_\mrm{UV} 
    &= -\frac{32}{7} \left.\sigma_v^2\right|_\mrm{UV} k^2\PL(k), \\
    \left.\mcal{F}_{Q_1}(k)\right|_\mrm{UV} 
    &= \frac{16}{49} \left.\sigma_v^2\right|_\mrm{UV} k^2\PL(k), 
\end{align}
where we define the UV part of the zero-lag moment as 
\begin{align}
    \left.\sigma_v^2\right|_\mrm{UV} \equiv \frac{1}{3}\int_{\bv{q}, q > \Lambda} \frac{\PL(q)}{q^2}. 
    \label{eq:13_zero_lag_moments_UV} 
\end{align} 
For the density field in \eq{eq:tdelg_exp_cubic}, by substituting these limits to the contractions derived in Appendix~\ref{subsubapp:13_integral_density_auto}, we obtain the following
UV responses in Fourier space: 
\begin{alignat}{2}
    -\partial_\parallel \left[ u_\parallel \left[1+\delg\right] \right]: &
    &&
    -\left.\sigma_v^2\right|_\mrm{UV} f\mu^2
    \left(\frac9{10}-\frac{b_1}{35}\right)
    k^2 \delL(\bv{k}),\\
    \frac12 \partial_\parallel^2 \left[ u_\parallel^2 \left[1+\delg\right] \right]: &
    &\quad&
    -\left.\sigma_v^2\right|_\mrm{UV} f^2\mu^2
    \left(\frac{24}{35}+\frac{b_1}{2}
    +\frac{23}{35}\mu^2\right) 
    k^2 \delL(\bv{k}), \\
    -\frac16 \partial_\parallel^3 \left[ u_\parallel^3 \left[1+\delg\right] \right]: &
    &&
    -\left.\sigma_v^2\right|_\mrm{UV} \frac12f^3\mu^4
    k^2\delL(\bv{k}), 
\end{alignat} 
where we used Eqs.~\eqref{eq:13_int_K_ud2}, \eqref{eq:13_int_K_u3}, and \eqref{eq:13_int_K_u2d} for the first moment, 
\eqs{eq:13_int_K_uud}{eq:13_int_K_uu2} for the second, and \eq{eq:13_int_K_uuu} for the third. 
Therefore, these UV contributions are absorbed by the counterterms
\begin{align}
    \partial_\parallel
    \left[u_\parallel\left[1+\delg\right]\right]^\mrm{ctr}
    &\equiv
    c_{u\mrm{g}}\partial_\parallel^2\delL,
    \\
    \partial_\parallel^2
    \left[u_\parallel^2\left[1+\delg\right]\right]^\mrm{ctr}
    &\equiv
    c_{u^2\mrm{g},1}\partial_\parallel^2\delL
    +
    c_{u^2\mrm{g},2}
    \frac{\partial_\parallel^4}{\nabla^2}\delL,
    \\
    \partial_\parallel^3
    \left[u_\parallel^3\left[1+\delg\right]\right]^\mrm{ctr}
    &\equiv
    c_{u^3\mrm{g}}
    \frac{\partial_\parallel^4}{\nabla^2}\delL.
    \label{eq:ctr_density_moments}
\end{align} 
Including the real-space counterterm, $c_{\mrm{g},0}\nabla^2\delL$, we obtain 
\begin{align}
    \tdelg^\mrm{ctr} 
    &= 
    c_{\mrm{g},0} \nabla^2\delL + c_{\mrm{g},2} \partial_\parallel^2 \delL + c_{\mrm{g},4} \frac{\partial_\parallel^4}{\nabla^2} \delL, 
\end{align}
where we define 
\begin{align}
    c_{\mrm{g},2}
    \equiv
    -c_{u\mrm{g}}+\frac12c_{u^2\mrm{g},1}, \quad
    c_{\mrm{g},4}
    \equiv
    \frac12c_{u^2\mrm{g},2}-\frac16c_{u^3\mrm{g}}. 
\end{align}

For the shape field in \eq{eq:tSij_exp_cubic}, keeping the tensor indices in the kernels in \eqs{eq:rsd_kernel_K_u2S_ij}{eq:rsd_kernel_K_uS2_ij}, we first take their UV limits and retain terms up to $\mcal{O}(k^2/q^2)$: 
\begin{align}
    \left.K_{u^{(2)}S,ij}\right|_\mrm{UV}
    &= 
    \frac{f\mu}{3} 
    b_K\mcal{D}_{ij}(\bv{q})
    \left[
    x \h{q}_\parallel + \frac67(1-x^2)\left(\mu - 2x\h{q}_\parallel\right) \frac{k^2}{q^2} 
    \right], \label{eq:rsd_kernel_K_u2S_ij_UV} \\
    \left.K_{uS^{(2)},ij}\right|_\mrm{UV}
    &= 
    \frac{f\mu}{3} 
    \Bigg[
    b_{2,1}\mu \mcal{D}_{ij}(\bv{q}) 
    - b_Kx \h{q}_\parallel \left[\mcal{D}_{ij}(\bv{q}) + \frac{k^2}{q^2}\mcal{D}_{ij}(\bv{k})\right] \nonumber\\
    &\qquad- \left(\frac{20}7b_K + b_{2,2}-b_{2,3}\right) (1-x^2) \h{q}_\parallel 
    \left\{-x\mcal{D}_{ij}(\bv{q}) + \TF{\h{k}_{(i}\h{q}_{j)}}\right\}\frac{k^2}{q^2}
    \Bigg]. 
\end{align}
From these and \eq{eq:rsd_kernel_uuS}, using the angular averaging formulae in Appendix~\ref{subapp:azm_avrg_13}, we obtain the linear UV responses 
\begin{alignat}{2}
    -\partial_\parallel \left[ u_\parallel \Sij \right]: &
    &\quad&
    -\left.\sigma_v^2\right|_\mrm{UV} f 
    \Bigg[
    \left(
    \frac{97}{245}b_K
    -\frac8{35}(b_{2,2}-b_{2,3})
    \right)
    \mu^2\mcal{D}_{ij}(\bv{k})
    \nonumber\\
    &&&\quad\quad\quad\quad\quad\quad+
    \left(
    \frac{496}{245}b_K
    +\frac47(b_{2,2}-b_{2,3})
    \right)
    \mu\,\TF{\h{n}_{(i}\h{k}_{j)}}
    \Bigg]
    k^2\delL(\bv{k}),\\
    \frac12 \partial_\parallel^2 \left[ u_\parallel^2 \Sij \right]: &
    &&
    -\left.\sigma_v^2\right|_\mrm{UV} \frac12 f^2 \mu^2 b_K
    \mcal{D}_{ij}(\bv{k})
    k^2\delL(\bv{k}). 
\end{alignat}
Thus, the corresponding counterterms are 
\begin{align}
    \partial_\parallel \left[ u_\parallel \Sij \right]^\mrm{ctr}
    &= 
    c_{uS,1} \partial_\parallel^2 K_{\mrm{L},ij}
    + 
    c_{uS,2} \TF{\h{n}_{(i} \partial_{j)}} \partial_\parallel \delL, \label{eq:ctr_uS}\\
    \partial_\parallel^2 \left[ u_\parallel^2 \Sij \right]^\mrm{ctr}
    &= 
    c_{u^2S} \partial_\parallel^2 K_{\mrm{L},ij}, \label{eq:ctr_uuS}
\end{align} 
Including the real-space counterterm, $c_{S,0} \nabla^2 K_{\mrm{L},ij}$, we obtain 
\begin{align}
    \tSij^\mrm{ctr} 
    &= 
    c_{S,0} \nabla^2 K_{\mrm{L},ij} 
    + c_{S,2} \partial_\parallel^2 K_{\mrm{L},ij} 
    + c_{S,2}^\parallel \TF{\h{n}_{(i} \partial_{j)}} \partial_\parallel \delL, 
\end{align} 
where we defined the independent parameters as
\begin{align}
    c_{S,2}
    \equiv
    -c_{uS}+\frac12c_{u^2S}, \quad 
    c_{S,2}^\parallel 
    \equiv 
    - c_{uS,2}. 
\end{align}

\section{Azimuthal averages} \label{app:azm_avrg} 
Loop integrals involving galaxy shapes contain Cartesian indices carried by the internal loop momentum $\bv{q}$. 
In redshift space, even scalar density kernels contain LOS factors such as $q_\parallel=\bv{q}\cdot\hbv{n}$. 
We take the external momentum $\bv{k}$ as the polar axis and integrate analytically over the azimuthal angle $\varphi$ of $\bv{q}$. 
The remaining integrals can then be organized in terms of $q$ and $x\equiv\hbv{k}\cdot\hbv{q}$, which is convenient for numerical implementation. 
We summarize the azimuthal averaging formulae used in Appendices~\ref{app:details_real} and \ref{app:details_rsd}.
We use unit-vector formulae for the $(13)$ integrals and transverse-momentum formulae for the $(22)$ integrals. 
At fixed $k$, $q$, and $x$, we define the azimuthal average of $X$ by 
\begin{align}
    \avrg{X}_\varphi
    \equiv \int_0^{2\pi}\frac{\mrm{d}\varphi}{2\pi}X. \label{eq:def_azimuthal_average} 
\end{align}

\subsection{Products of unit vectors} \label{subapp:azm_avrg_13} 
The symmetric trace-free part of a product of unit vectors satisfies \citep[e.g.][]{Thorne1980:Multipole_GW} 
\begin{align}
    \avrg{\left\{\h{q}_{i_1}\cdots\h{q}_{i_\ell}\right\}^{\mrm{TF}}}_\varphi 
    = 
    \left\{\h{k}_{i_1}\cdots\h{k}_{i_\ell}\right\}^{\mrm{TF}} 
    \mcal{L}_{\ell}(x), \label{eq:int_of_TF_product}
\end{align}
where $\mcal{L}_{\ell}$ is the Legendre polynomial of degree $\ell$, and TF removes all traces. 
The first three cases are 
\begin{align}
    \avrg{\h{q}_i}_\varphi 
    &= \h{k}_i x, \label{eq:azm_avrg_qhat_rank1}\\
    \avrg{\h{q}_i\h{q}_j}_\varphi 
    &= \h{k}_i\h{k}_j \mcal{L}_2(x) + \frac{\delK_{ij}}{3}\left[1-\mcal{L}_2(x)\right], \label{eq:azm_avrg_qhat_rank2}\\
    \avrg{\h{q}_i\h{q}_j\h{q}_k}_\varphi 
    &= \h{k}_i\h{k}_j\h{k}_k \mcal{L}_3(x) 
    + \frac15\left( \delK_{ij}\h{k}_k + \delK_{ik}\h{k}_j + \delK_{jk}\h{k}_i\right)\left[x-\mcal{L}_3(x)\right]. \label{eq:azm_avrg_qhat_rank3}
\end{align}
Contracting these identities with $\hbv{n}$ and the projection tensor $M_{ij}^E$
defined in \eq{eq:def_M_E} gives 
\begin{align}
    \avrg{\h{q}_\parallel}_\varphi 
    &= \h{n}_i\avrg{\h{q}_i}_\varphi 
    = \mu x, \label{eq:azm_avrg_qn}\\
    \avrg{\h{q}_\parallel^2}_\varphi 
    &= \h{n}_i\h{n}_j\avrg{\h{q}_i\h{q}_j}_\varphi 
    = \mu^2 \mcal{L}_2(x) + \frac13 \left[1-\mcal{L}_2(x)\right], \label{eq:azm_avrg_qn2}\\
    M_{ij}^E \avrg{\mcal{D}_{ij}(\bv{q})}_\varphi  
    &= M_{ij}^E\avrg{\h{q}_i\h{q}_j}_\varphi 
    = \frac{1-\mu^2}{2} \mcal{L}_2(x), \label{eq:azm_avrg_ME_qq}\\
    M_{ij}^E \avrg{\h{q}_\parallel\mcal{D}_{ij}(\bv{q})}_\varphi  
    &= \h{n}_k M_{ij}^E\avrg{\h{q}_i\h{q}_j\h{q}_k}_\varphi 
    = \frac{(1-\mu^2)\mu}{2} \mcal{L}_3(x). \label{eq:azm_avrg_nME_qqq}
\end{align}

\subsection{Transverse components} \label{subapp:azm_avrg_22} 
We define the component of $\bv{q}$ transverse to $\bv{k}$ by
$q_{\perp i}\equiv\mcal{P}_{ij}(\bv{k})q_j$, where
$\mcal{P}_{ij}(\bv{k})=\delK_{ij}-\h{k}_i\h{k}_j$.
Thus $\bv{q}=qx\hbv{k}+\bv{q}_\perp$, and at fixed $q$ and $x$ all azimuthal dependence is carried by $\bv{q}_\perp$.
This component appears naturally in the shape kernel in \eq{eq:def_KQ3ij}.
For the $(22)$ integrals, the other internal momentum is $\bv{p}=\bv{k}-\bv{q}$, so $p_{\perp i}=-q_{\perp i}$.
We can therefore express both internal momenta using the same transverse vector.
The LOS component is $q_\parallel=qx\mu+\bv{q}_\perp\cdot\hbv{n}$.

The azimuthal average of products of $\bv{q}_\perp$ gives tensors that are isotropic in the plane perpendicular to $\bv{k}$. 
Thus, odd moments of $\bv{q}_\perp$ vanish, while the first two even moments are 
\begin{align}
    \avrg{q_{\perp i}q_{\perp j}}_\varphi
    &=\frac{q_\perp^2}{2}\mcal{P}_{ij}, \label{eq:azm_avrg_qperp_rank2}\\
    \avrg{q_{\perp i}q_{\perp j}q_{\perp k}q_{\perp l}}_\varphi
    &=\frac{q_\perp^4}{8}
    \left(
    \mcal{P}_{ij}\mcal{P}_{kl}
    +\mcal{P}_{ik}\mcal{P}_{jl}
    +\mcal{P}_{il}\mcal{P}_{jk}
    \right), \label{eq:azm_avrg_qperp_rank4}
\end{align}
where $q_\perp^2=q^2(1-x^2)$ is independent of $\varphi$. 
The coefficients follow by taking traces and using $\mcal{P}_{ii}=2$.\footnote{More generally, the average of $2r$ transverse components is 
\begin{align}
    \avrg{q_{\perp i_1}\cdots q_{\perp i_{2r}}}_\varphi
    =\frac{q_\perp^{2r}}{2^r r!}
    \sum_{\pi\in\mrm{Pair}(2r)}
    \prod_{(a,b)\in\pi}\mcal{P}_{i_a i_b},
\end{align}
where $\mrm{Pair}(2r)$ denotes the set of distinct pairings of the $2r$ indices. 
We need only the rank-two and rank-four cases, \eqs{eq:azm_avrg_qperp_rank2}{eq:azm_avrg_qperp_rank4}, in this work.} 
With these identities, we obtain the following contractions: 
\begin{align}
    \h{n}_i \h{n}_j \avrg{q_{\perp i}q_{\perp j}}_\varphi
    &= \frac{1-\mu^2}{2}q_\perp^2, \label{eq:nn_qperp_rank2}\\
    \h{n}_i \h{n}_j \h{n}_k \h{n}_l \avrg{q_{\perp i}q_{\perp j}q_{\perp k}q_{\perp l}}_\varphi
    &= \frac{3 (1-\mu^2)^2}{8} q_\perp^4, \label{eq:nnnn_qperp_rank4}\\
    M_{ij}^E \avrg{q_{\perp i}q_{\perp j}}_\varphi
    &= -\frac{1-\mu^2}{4}q_\perp^2, \label{eq:ME_qperp_rank2}\\
    \h{n}_i \h{n}_j M_{kl}^E \avrg{q_{\perp i}q_{\perp j}q_{\perp k}q_{\perp l}}_\varphi
    &= \frac{(1-\mu^2)(3\mu^2-1)}{16} q_\perp^4, \label{eq:nnME_qperp_rank4}\\
    M_{ij}^E M_{kl}^E \avrg{q_{\perp i}q_{\perp j}q_{\perp k}q_{\perp l}}_\varphi
    &= \frac{3\mu^4 - 2\mu^2 + 3}{32} q_\perp^4, \label{eq:MEME_qperp_rank4}\\
    M_{ij}^B M_{kl}^B \avrg{q_{\perp i}q_{\perp j}q_{\perp k}q_{\perp l}}_\varphi
    &= \frac{\mu^2}{8} q_\perp^4. \label{eq:MBMB_qperp_rank4}
\end{align}
The following parity-odd contractions vanish: 
\begin{align}
    M_{ij}^B \avrg{q_{\perp i}q_{\perp j}}_\varphi 
    &= 
    \h{n}_i \h{n}_j M_{kl}^B \avrg{q_{\perp i}q_{\perp j}q_{\perp k}q_{\perp l}}_\varphi
    = 
    M_{ij}^E M_{kl}^B \avrg{q_{\perp i}q_{\perp j}q_{\perp k}q_{\perp l}}_\varphi
    =0.
\end{align}

\section{Relation between the one-loop power spectrum and the TNS model}
\label{app:relation_TNS} 
Here, we derive \eq{eq:TNS_EPT_relation} for the deterministic part of the power spectra. 
For Gaussian initial conditions, by perturbatively expanding \eq{eq:TNS_cumulant} up to $\mcal{O}(\PL^2)$, we obtain 
\begin{align}
    \avrg{e^{j\Delta u_\parallel}XY}
    &\simeq
    \avrg{XY}_{\mrm{tree+1\mathchar`-loop}}
    +j\avrg{\Delta u_\parallel XY}_{\mrm{c,tree}}
    \nonumber\\
    &\quad
    +j^2
    \avrg{\Delta u_\parallel X}_{\mrm{tree}}
    \avrg{\Delta u_\parallel Y}_{\mrm{tree}}
    +\frac{j^2}{2}
    \avrg{(\Delta u_\parallel)^2}_{\mrm{tree}}
    \avrg{XY}_{\mrm{tree}}, 
    \label{eq:TNS_EPT_expansion}
\end{align}
where $X=X(\bv{x})$, $Y=Y(\bv{x}')$ and 
$\avrg{(\Delta u_\parallel)^2}_{\mrm{tree}}
= 
2f^2\sigma_{v,\mrm{L}}^2
-2\avrg{u_\parallel(\bv{x})u_\parallel(\bv{x}')}_{\mrm{tree}}$. 
Since the redshift-space kernels in Appendix~\ref{subapp:PT_kernel_rsd} are obtained by expanding the same exponential factor, the Fourier transform of \eq{eq:TNS_EPT_expansion} corresponds to the one-loop power spectrum in standard Eulerian perturbation theory, $\tP_{XY}^{\mrm{tree+1\mathchar`-loop}}$, by definition. 
Following Ref.~\cite{Taruya+2010:TNS}, we define the nonlinear Kaiser term and the $A$, $B$, and $C$ terms as the Fourier transforms of the terms in \eq{eq:TNS_EPT_expansion}:
\begin{align}
    K_{XY}(\bv{k})
    &\equiv
    \int_{\bv{r}}e^{-i\bv{k}\cdot\bv{r}}
    \avrg{XY}_{\mrm{tree+1\mathchar`-loop}},
    \\
    A_{XY}(\bv{k})
    &\equiv
    j\int_{\bv{r}}e^{-i\bv{k}\cdot\bv{r}}
    \avrg{\Delta u_\parallel XY}_{\mrm{c,tree}},
    \\
    B_{XY}(\bv{k})
    &\equiv
    j^2\int_{\bv{r}}e^{-i\bv{k}\cdot\bv{r}}
    \avrg{\Delta u_\parallel X}_{\mrm{tree}}
    \avrg{\Delta u_\parallel Y}_{\mrm{tree}},
    \\
    C_{XY}(\bv{k})
    &\equiv
    -j^2\int_{\bv{r}}e^{-i\bv{k}\cdot\bv{r}}
    \avrg{u_\parallel(\bv{x})u_\parallel(\bv{x}')}_{\mrm{tree}}
    \avrg{XY}_{\mrm{tree}}.
    \label{eq:TNS_KABC}
\end{align}
Using this notation, we obtain 
\begin{align}
    \tP_{XY}^{\mrm{EPT,1\mathchar`-loop}}
    =
    K_{XY}+A_{XY}+B_{XY}+C_{XY}
    -(kf\mu\sigma_{v,\mrm{L}})^2K_{XY}^{\mrm{tree}},
    \label{eq:EPT_in_KABC}
\end{align}
where $K_{XY}^{\mrm{tree}}$ is the tree part (linear Kaiser term) of $K_{XY}$. 

In \eq{eq:TNS_cumulant}, the one-point velocity dispersion appears as the overall damping factor. 
Expanding this factor gives the last term in \eq{eq:EPT_in_KABC}. 
The TNS prescription, on the other hand, keeps this factor unexpanded and replaces it with $D_{\mrm{FoG}}(kf\mu\sigma_v)$, as in \eq{eq:TNS_Gaussian_FoG}. 
Therefore, the power spectrum based on the TNS prescription is then written as
\begin{align}
    \tP_{XY}^{\mrm{TNS}}
    =
    D_{\mrm{FoG}}(kf\mu\sigma_v)
    \left[K_{XY}+A_{XY}+B_{XY}+C_{XY}\right].
    \label{eq:TNS_model_with_C}
\end{align}
Here we keep the $C$ term, which is omitted in the original TNS expression. 
From \eqs{eq:EPT_in_KABC}{eq:TNS_model_with_C}, we obtain \eq{eq:TNS_EPT_relation}. 

\section{Summary tables for the EFT model} \label{app:summary_table} 
We summarize the radial basis functions and coefficients defined in \eq{eq:PX_eft_tot} in Tables~\ref{tab:galaxy_rsd_coefficients}--\ref{tab:ia_rsd_BB_coefficients}. 
\clearpage
\begin{table}[p]
\centering
\begingroup
\fontsize{9}{11}\selectfont
\setlength{\arraycolsep}{3pt}
\setlength{\extrarowheight}{3pt}
\renewcommand{\arraystretch}{1.45}
$\begin{array}{@{}c|>{\displaystyle}c>{\displaystyle}c>{\displaystyle}c>{\displaystyle}c>{\displaystyle}c@{}}
\hline
\mcal{B}_i(k) & \mu^{0} & \mu^{2} & \mu^{4} & \mu^{6} & \mu^{8} \\ \hline
P_{\mrm L}
  & b_1^{2}
  & 2 b_1 f
  & f^{2}
  & 0
  & 0 \\[2pt]
\hline
P_m^{(22)}
  & b_1^{2}
  & 2 b_1 f \left(b_1 + 1\right)
  & f^{2} \left(b_1^{2} + 4 b_1 + 1\right)
  & 2 f^{3} \left(b_1 + 1\right)
  & f^{4} \\[2pt]
\mathcal I_{\delta,\delta^2}
  & b_1 b_2
  & b_2 f \left(b_1 + 1\right)
  & b_2 f^{2}
  & 0
  & 0 \\[2pt]
\mathcal I_{\delta,\mathcal G_2}
  & 2 b_1 b_{\mathcal G_2}
  & \frac{f}{14}\left[\begin{gathered}{}7 b_1 f \\ {}+8 b_1  \left(1 - b_1\right) \\ {}+28 b_{\mathcal G_2} \left(b_1 + 1\right)\end{gathered}\right]
  & \frac{f^{2}}{14}\left[\begin{gathered}{}- 8 b_1^{2} - b_1 + 28 b_{\mathcal G_2} \\ {}+7 f \left(b_1 + 1\right) + 8\end{gathered}\right]
  & \frac{f^{3}}{14}\left[\begin{gathered}{}- 9 b_1 + 7 f \\ {}+7\end{gathered}\right]
  & - \frac{f^{4}}{14} \\[2pt]
\mathcal I_{\delta^2,\delta^2}
  & \frac{b_2^{2}}{4}
  & 0
  & 0
  & 0
  & 0 \\[2pt]
\mathcal I_{\delta^2,\mathcal G_2}
  & b_2 b_{\mathcal G_2}
  & \frac{b_2 f \left(- 8 b_1 + 7 f + 8\right)}{28}
  & - \frac{b_2 f^{2}}{28}
  & 0
  & 0 \\[2pt]
\mathcal I_{\mathcal G_2,\mathcal G_2}
  & b_{\mathcal G_2}^{2}
  & \frac{b_{\mathcal G_2} f \left(- 8 b_1 + 7 f + 8\right)}{14}
  & \frac{f^{2}}{1568}\left[\begin{gathered}{}- 112 b_{\mathcal G_2} + 147 f^{2} \\ {}- 224 f \left(b_1 - 1\right) \\ {}+128 \left(b_1 - 1\right)^{2}\end{gathered}\right]
  & - \frac{f^{3}}{784}\left[\begin{gathered}{}- 16 b_1 \\ {}+63 f \\ {}+16\end{gathered}\right]
  & \frac{51 f^{4}}{1568} \\[2pt]
\mathcal I_{\mrm{RSD}}
  & 0
  & b_1^{2} f^{2}
  & - b_1 f^{2} \left(b_1 - 2 f\right)
  & - f^{3}  \left(2 b_1 - f\right)
  & - f^{4} \\[2pt]
\hline
P_m^{(13)}
  & b_1^{2}
  & 2 b_1 f \left(b_1 + 1\right)
  & f^{2}  \left(4 b_1 + 1\right)
  & 2 f^{3}
  & 0 \\[2pt]
\mathcal F_{\mathcal G_2}
  & 2 b_1 \left(b_{\mathcal G_2} + \frac{2 b_{\Gamma_3}}{5}\right)
  & \frac{f}{30}\left[\begin{gathered}{}9 b_1 f \\ {}+8 b_1  \left(1 - b_1\right) \\ {}+60 b_{\mathcal G_2} + 24 b_{\Gamma_3}\end{gathered}\right]
  & \frac{f^{2} \left(- 7 b_1 + 9 f + 8\right)}{30}
  & \frac{f^{3}}{30}
  & 0 \\[2pt]
k^2\sigma_{v,\mrm L}^2P_{\mrm L}
  & 0
  & - b_1^{2} f^{2}
  & - 2 b_1 f^{3}
  & - f^{4}
  & 0 \\[2pt]
\hline
k^2P_{\mrm L}
  & - 2 b_1 c_{\mrm g,0}
  & - 2 b_1 c_{\mrm g,2} - 2 c_{\mrm g,0} f
  & - 2 b_1 c_{\mrm g,4} - 2 c_{\mrm g,2} f
  & - 2 c_{\mrm g,4} f
  & 0 \\[2pt]
\hline
1
  & N_{\mrm{gg},0}
  & 0
  & 0
  & 0
  & 0 \\[2pt]
k^2
  & N_{\mrm{gg},2}
  & N_{\mrm{gg},2\mu}
  & 0
  & 0
  & 0 \\[2pt]
\hline
\end{array}$
\endgroup
\caption{Coefficients $c_{\mrm{gg}}^{(i,n)}$ in \eq{eq:PX_eft_tot}.}
\label{tab:galaxy_rsd_coefficients}
\end{table}

\clearpage
\begin{table}[p]
\centering
\begingroup
\fontsize{9}{11}\selectfont
\setlength{\arraycolsep}{3pt}
\setlength{\extrarowheight}{3pt}
\renewcommand{\arraystretch}{1.45}
$\begin{array}{@{}c|>{\displaystyle}c>{\displaystyle}c>{\displaystyle}c>{\displaystyle}c@{}}
\hline
\mcal{B}_i(k) & \mu^{0} & \mu^{2} & \mu^{4} & \mu^{6} \\ \hline
P_{\mrm L}
  & b_1 b_K
  & b_K f
  & 0
  & 0 \\[2pt]
\hline
P_m^{(22)}
  & b_1 b_K
  & b_K f \left(2 b_1 + 1\right)
  & b_K f^{2} \left(b_1 + 2\right)
  & b_K f^{3} \\[2pt]
\mathcal I_{\delta,\delta^2}
  & b_1 b_{2,1} + \frac{b_2 b_K}{2}
  & \frac{f \left(2 b_1 b_{2,1} + b_2 b_K + 2 b_{2,1}\right)}{2}
  & b_{2,1} f^{2}
  & 0 \\[2pt]
\mathcal I_{\delta,\mathcal G_2}
  & b_1 b_{2,1} - \frac{b_1 b_{2,2}}{4} + b_{\mathcal G_2} b_K
  & \frac{f}{28}\left[\begin{gathered}{}14 b_{2,1}  \left(3 b_1 + 2\right) \\ {}- 7 b_{2,2} \left(b_1 + 1\right) + 28 b_{\mathcal G_2} b_K \\ {}+b_K \left(- 9 b_1 + 7 f + 8\right)\end{gathered}\right]
  & \frac{f^{2}}{28}\left[\begin{gathered}{}42 b_{2,1} - 7 b_{2,2} \\ {}+b_K \left(- 9 b_1 + 7 f + 6\right)\end{gathered}\right]
  & - \frac{b_K f^{3}}{14} \\[2pt]
\mathcal I_{\delta^2,\mathcal G_2}
  & - \frac{b_1 b_{2,1}}{28} - \frac{b_2 b_{2,2}}{8}
  & \frac{f}{168}\left[\begin{gathered}{}- 62 b_1 b_{2,1} - 3 b_2 b_K \\ {}+7 b_{2,1} f - 6 b_{2,1}\end{gathered}\right]
  & - \frac{55 b_{2,1} f^{2}}{168}
  & 0 \\[2pt]
\mathcal I_{\mathcal G_2,\mathcal G_2}
  & - \frac{b_1 b_{2,1}}{28} - \frac{b_{2,2} b_{\mathcal G_2}}{4}
  & \frac{f}{672}\left[\begin{gathered}{}4 b_{2,1} \left(- 2 b_1 + 7 f - 6\right) \\ {}+6 b_{2,2}  \left(8 b_1 - 7 f - 8\right) \\ {}- 21 b_{2,3} f - 24 b_{\mathcal G_2} b_K\end{gathered}\right]
  & \frac{f^{2}}{4704}\left[\begin{gathered}{}48 b_1 b_K + 140 b_{2,1} \\ {}+42 b_{2,2} - 147 b_{2,3} \\ {}- 483 b_K f - 48 b_K\end{gathered}\right]
  & \frac{51 b_K f^{3}}{1568} \\[2pt]
\mathcal I_{\mrm{RSD}}
  & 0
  & b_1 b_K f^{2}
  & - b_K f^{2} \left(b_1 - f\right)
  & - b_K f^{3} \\[2pt]
\mathcal I_{\delta^2,Q_1}^{(0)}
  & \frac{\sqrt{6} b_2 b_{2,1}}{4}
  & 0
  & 0
  & 0 \\[2pt]
\mathcal I_{\mathcal G_2,Q_1}^{(0)}
  & - \frac{\sqrt{6} b_{2,1}  \left(5 b_1 - 7 b_{\mathcal G_2}\right)}{14}
  & \frac{\sqrt{6} b_{2,1} f \left(- 280 b_1 + 35 f - 72\right)}{336}
  & - \frac{35 \sqrt{6} b_{2,1} f^{2}}{48}
  & 0 \\[2pt]
\hline
P_m^{(13)}
  & b_1 b_K
  & b_K f \left(2 b_1 + 1\right)
  & 2 b_K f^{2}
  & 0 \\[2pt]
\mathcal F_{\mathcal G_2}
  & \begin{gathered}{}b_1 \bar b_{3,1} \\ {}+b_K \left(b_{\mathcal G_2} + \frac{2 b_{\Gamma_3}}{5}\right)\end{gathered}
  & \frac{f}{240}\left[\begin{gathered}{}- 7 b_1 \left(b_{2,2} - b_{2,3}\right) \\ {}+240 \bar b_{3,1} \\ {}+4 b_K \left(- 9 b_1 + 9 f + 8\right)\end{gathered}\right]
  & - \frac{7 f^{2} \left(b_{2,2} - b_{2,3}\right)}{240}
  & 0 \\[2pt]
\mathcal F_{Q_1}
  & b_1 \bar b_{3,2}
  & \frac{f}{24}\left[\begin{gathered}{}8 b_1 b_K + 7 b_1 \left(b_{2,2} - b_{2,3}\right) \\ {}+24 \bar b_{3,2}\end{gathered}\right]
  & \frac{f^{2}  \left(7 b_{2,2} - 7 b_{2,3} + 8 b_K\right)}{24}
  & 0 \\[2pt]
k^2\sigma_{v,\mrm L}^2P_{\mrm L}
  & 0
  & - b_1 b_K f^{2}
  & - b_K f^{3}
  & 0 \\[2pt]
\hline
k^2P_{\mrm L}
  & - b_1 c_{S,0} - b_K c_{\mrm g,0}
  & - b_1 c_{S,2} - b_K c_{\mrm g,2} - c_{S,0} f
  & - b_K c_{\mrm g,4} - c_{S,2} f
  & 0 \\[2pt]
\hline
k^2
  & N_{\mrm gS,2}
  & N_{\mrm gS,2\mu}
  & 0
  & 0 \\[2pt]
\hline
\end{array}$
\endgroup
\caption{Coefficients $c_{\mrm gE}^{(i,n)}$ in \eq{eq:PX_eft_tot}, after factoring out $A_{\mrm gE}=(1-\mu^2)/2$.}
\label{tab:ia_rsd_gE_coefficients}
\end{table}

\clearpage
\begin{table}[p]
\centering
\begingroup
\fontsize{9}{11}\selectfont
\setlength{\arraycolsep}{3pt}
\setlength{\extrarowheight}{3pt}
\renewcommand{\arraystretch}{1.45}
$\begin{array}{@{}c|>{\displaystyle}c>{\displaystyle}c>{\displaystyle}c@{}}
\hline
\mcal{B}_i(k) & \mu^{0} & \mu^{2} & \mu^{4} \\ \hline
P_{\mrm L}
  & \frac{b_K^{2}}{2}
  & 0
  & 0 \\[2pt]
\hline
P_m^{(22)}
  & \frac{b_K^{2}}{2}
  & b_K^{2} f
  & \frac{b_K^{2} f^{2}}{2} \\[2pt]
\mathcal I_{\delta,\delta^2}
  & b_{2,1} b_K
  & b_{2,1} b_K f
  & 0 \\[2pt]
\mathcal I_{\delta,\mathcal G_2}
  & \frac{b_K \left(4 b_{2,1} - b_{2,2}\right)}{4}
  & \frac{b_K f \left(42 b_{2,1} - 7 b_{2,2} - b_K\right)}{28}
  & - \frac{b_K^{2} f^{2}}{28} \\[2pt]
\mathcal I_{\delta^2,\delta^2}
  & - \frac{7 b_{2,1}^{2}}{36}
  & 0
  & 0 \\[2pt]
\mathcal I_{\delta^2,\mathcal G_2}
  & - \frac{b_{2,1}  \left(70 b_{2,1} + 21 b_{2,3} + 18 b_K\right)}{504}
  & - \frac{55 b_{2,1} b_K f}{168}
  & 0 \\[2pt]
\mathcal I_{\mathcal G_2,\mathcal G_2}
  & \frac{b_{2,1}^{2}}{18} - \frac{b_{2,1} b_{2,3}}{24} - \frac{b_{2,1} b_K}{28} + \frac{b_{2,2}^{2}}{32} + \frac{b_{2,3}^{2}}{64}
  & \frac{b_K f}{672}\left[\begin{gathered}{}20 b_{2,1} + 6 b_{2,2} - 21 b_{2,3} \\ {}- 42 b_K f\end{gathered}\right]
  & \frac{51 b_K^{2} f^{2}}{3136} \\[2pt]
\mathcal I_{\mrm{RSD}}
  & 0
  & \frac{b_K^{2} f^{2}}{2}
  & - \frac{b_K^{2} f^{2}}{2} \\[2pt]
\mathcal I_{\delta^2,Q_1}^{(0)}
  & - \frac{5 \sqrt{6} b_{2,1}^{2}}{36}
  & 0
  & 0 \\[2pt]
\mathcal I_{\mathcal G_2,Q_1}^{(0)}
  & - \frac{\sqrt{6} b_{2,1}}{1008}\left[\begin{gathered}{}140 b_{2,1} + 126 b_{2,2} - 21 b_{2,3} \\ {}+360 b_K\end{gathered}\right]
  & - \frac{35 \sqrt{6} b_{2,1} b_K f}{48}
  & 0 \\[2pt]
\mathcal I_{Q_1,Q_1}^{(0)}
  & \frac{35 b_{2,1}^{2}}{24}
  & 0
  & 0 \\[2pt]
\hline
P_m^{(13)}
  & \frac{b_K^{2}}{2}
  & b_K^{2} f
  & 0 \\[2pt]
\mathcal F_{\mathcal G_2}
  & \bar b_{3,1} b_K
  & \frac{- b_K f \left(4 b_K + 7 \left(b_{2,2} - b_{2,3}\right)\right)}{240}
  & 0 \\[2pt]
\mathcal F_{Q_1}
  & \bar b_{3,2} b_K
  & \frac{b_K f \left(8 b_K + 7 \left(b_{2,2} - b_{2,3}\right)\right)}{24}
  & 0 \\[2pt]
k^2\sigma_{v,\mrm L}^2P_{\mrm L}
  & 0
  & - \frac{b_K^{2} f^{2}}{2}
  & 0 \\[2pt]
\hline
k^2P_{\mrm L}
  & - b_K c_{S,0}
  & - b_K c_{S,2}
  & 0 \\[2pt]
\hline
k^2
  & \frac{N_{SS,2\Delta}}{2}
  & 0
  & 0 \\[2pt]
\hline
\end{array}$
\endgroup
\caption{Coefficients $c_-^{(i,n)}$ in \eq{eq:PX_eft_tot}, after factoring out $A_-=(1-\mu^2)^2/4$.}
\label{tab:ia_rsd_minus_coefficients}
\end{table}

\clearpage
\begin{table}[p]
\centering
\begingroup
\fontsize{9}{11}\selectfont
\setlength{\arraycolsep}{3pt}
\setlength{\extrarowheight}{3pt}
\renewcommand{\arraystretch}{1.45}
$\begin{array}{@{}c|>{\displaystyle}c>{\displaystyle}c@{}}
\hline
\mcal{B}_i(k) & \mu^{0} & \mu^{2} \\ \hline
\mathcal I_{\delta^2,\delta^2}
  & \frac{b_{2,1}^{2}}{9}
  & - \frac{b_{2,1}^{2}}{18} \\[2pt]
\mathcal I_{\delta^2,\mathcal G_2}
  & \frac{7 b_{2,1}^{2}}{72}
  & \frac{b_{2,1} \left(b_{2,1} - 6 \left(b_{2,3} + b_K f\right)\right)}{72} \\[2pt]
\mathcal I_{\mathcal G_2,\mathcal G_2}
  & - \frac{b_{2,1}^{2}}{72}
  & \frac{5 b_{2,1}^{2}}{72} - \frac{b_{2,1} \left(b_{2,3} + b_K f\right)}{12} + \frac{\left(b_{2,3} + b_K f\right)^{2}}{32} \\[2pt]
\mathcal I_{\delta^2,Q_1}^{(0)}
  & \frac{\sqrt{6} b_{2,1}^{2}}{18}
  & - \frac{\sqrt{6} b_{2,1}^{2}}{9} \\[2pt]
\mathcal I_{\mathcal G_2,Q_1}^{(0)}
  & \frac{\sqrt{6} b_{2,1}^{2}}{18}
  & \frac{- \sqrt{6} b_{2,1}  \left(8 b_{2,1} - 3 \left(b_{2,3} + b_K f\right)\right)}{72} \\[2pt]
\mathcal I_{Q_1,Q_1}^{(0)}
  & - \frac{b_{2,1}^{2}}{3}
  & \frac{5 b_{2,1}^{2}}{12} \\[2pt]
\hline
1
  & N_{SS,0}
  & 0 \\[2pt]
k^2
  & N_{SS,2}
  & N_{SS,2\mu} \\[2pt]
\hline
\end{array}$
\endgroup
\caption{Coefficients $c_{BB}^{(i,n)}$ in \eq{eq:PX_eft_tot}.}
\label{tab:ia_rsd_BB_coefficients}
\end{table}
\clearpage

\bibliographystyle{JHEP}
\bibliography{main}
\end{document}